\documentclass{emulateapj}

\usepackage{epsfig}
\usepackage{longtable}

\def \age{$\langle t\rangle_{\mbox{\scriptsize SFR}}$}
\begin{document}

\title{A Near-Infrared Spectroscopic Survey of K-selected Galaxies at \lowercase{$z$} $\sim$ 2.3: 
Comparison of stellar population synthesis codes and constraints
  from the rest-frame NIR
}
\author{Adam Muzzin\altaffilmark{1}, Danilo
  Marchesini\altaffilmark{1}, Pieter G. van Dokkum\altaffilmark{1}, Ivo
  Labb\'{e}\altaffilmark{2}, Mariska Kriek\altaffilmark{3}, \& Marijn Franx\altaffilmark{4}} 

\altaffiltext{1}{Department of Astronomy, Yale
  University, New Haven, CT, 06520-8101; adam.muzzin@yale.edu} 
\altaffiltext{2}{Hubble Fellow, Carnegie Observatories, 813 Santa
  Barbara Street, Pasadena, CA, 91101} 
\altaffiltext{3}{Department of Astrophysical Sciences, Princeton
  University, Princeton, NJ, 08544} 
\altaffiltext{4}{Leiden Observatory, Leiden University, PO Box 9513,
  2300 RA Leiden, Netherlands} 
\begin{abstract}
We present SED modeling of a sample of 34 K-selected galaxies
  at $z \sim$ 2.3.  These galaxies have  NIR spectroscopy that
  samples the rest-frame Balmer/4000\AA$ $ break as well as
  deep photometry in thirteen broadband filters.
New to our analysis is  IRAC data that extend the SEDs into the
  rest-frame NIR.   Comparing parameters determined
  from SED fits with and without the IRAC data we find that the
  IRAC photometry significantly improves the confidence intervals of  
  $\tau$, A$_{v}$, M$_{\mbox{\scriptsize star}}$, and SFR for
  individual galaxies, but does not systematically alter the mean
  parameters of the sample.   We use the IRAC data to assess how well
  current stellar population synthesis codes describe the rest-frame NIR SEDs of young galaxies
where discrepancies between treatments of the TP-AGB phase of stellar
evolution are most pronounced.  The models of Bruzual \&
Charlot (2003), Maraston (2005), and Charlot \& Bruzual (2008)  all successfully reproduce the SEDs of
our galaxies with $\leq$ 5\% differences in the quality of fit; however,  
the best-fit masses from each code differ systematically by as much as
  a factor of 1.5, and other parameters vary more, up to factors of 2-3.  
A comparison of best-fit stellar population parameters from different SPS codes, dust laws,
  and metallicities shows that the choice of SPS code is the largest systematic
  uncertainty in most parameters,  and that systematic uncertainties are typically larger than the formal random uncertainties.  The SED fitting confirms our previous result that galaxies with strongly suppressed star formation account for $\sim$ 50\%
  of the K-bright population at $z \sim$ 2.3; however, the uncertainty
  in this fraction is large due to systematic differences in the SSFRs
  derived from the three SPS models. 
\end{abstract}


\section{Introduction}
The observed flux from a galaxy as a function of wavelength or frequency, its spectral energy
distribution (SED), represents the integrated light of its stellar
populations and is a valuable tool for determining properties
such as its age, and current mass contained in stars.  Interpreting 
observations of galaxy SEDs requires creating synthetic spectra using our best
understanding of stellar spectra, stellar evolution, and dust
absorption.  This process, known as stellar population synthesis (SPS)
was first implemented by Tinsley (1967, 1972) and has subsequently been
refined into increasingly sophisticated models (e.g., Bruzual 1983;
Renzini \& Buzzoni 1986; Bruzual \& Charlot 1993; Worthey 1994; Fioc \&
Rocca-Volmerange 1997; Leitherer et al. 1999; Bruzual \& Charlot 2003;
Maraston 2005; Bruzual 2007; Conroy et al. 2008).  
\newline\indent
With the advent of  deep multiwavelength extragalactic surveys and the refinement of photometric
redshift techniques, large, statistically complete samples of galaxies
with broadband SEDs are now available for studying the evolution of
galaxy properties in a systematic way up to $z \sim$ 2-3 (e.g.,
F\"{o}rster Schreiber et al. 2004; Labb\'{e} et al. 2005; van Dokkum et
al. 2006; Papovich et al. 2006; Fontana et al. 2006; Daddi et al. 2007; Wuyts et
al. 2007; P\'{e}rez-Gonz\'{a}lez et al. 2008; Drory \& Alvarez 2008;
Marchesini et al. 2008, and many others).  The demand for SPS
codes that provide accurate synthetic SEDs for a given parameter set
is now greater than ever as they are $the$ tool for
extracting astrophysical information from these data.  
\newline\indent
Using the current generation of SPS models these studies have shown
that a large fraction of the stellar mass (M$_{\mbox{\scriptsize star}}$) in massive galaxies is
 assembled by $z \sim$ 2 (e.g., Fontana et al. 2006;
 P\'{e}rez-Gonz\'{a}lez et al. 2008;  Marchesini et al. 2008), and that a significant fraction of these
galaxies are already passively evolving (e.g., Labb\'{e} et al. 2005;
Papovich et al. 2006; Kriek et al. 2006).  
Unfortunately, the current SPS models do have significant challenges to overcome (see e.g., Conroy et
al. 2008), making it unclear how robust some of these results are
(see e.g., the discussion in Marchesini et al. 2008).  Furthermore, it is well known that stellar population parameters
derived from fitting broadband SEDs suffer from systematic uncertainties caused by the fact that
metallicity, the initial mass function (IMF), and dust extinction law  cannot be constrained
by broadband SEDs alone.  Instead, these 
parameters must be assumed {\it a priori} when creating the set of
synthetic SEDs used in the fitting.  More recently it has become clear that
the treatment of SPS itself (e.g., the choice of isochrones, stellar
libraries, and integration method) can
result in synthetic spectra that look strikingly different for a similar set of input
parameters (e.g., Maraston 2005, Maraston et al. 2006; Bruzual 2007;
Conroy et al. 2008) and that the choice of SPS method is now an
additional systematic error in the interpretation of galaxy SEDs.  
\newline\indent
One of the major challenges for SPS remains to be a complete treatment of the thermally-pulsating
asymptotic giant branch (TP-AGB) stars.  This population of stars
are notoriously challenging to model because they  
have complex physics such as thermal
pulses, evolving chemical compositions, rapid mass-loss
rates, and self-obscuring dust shells (see e.g., Marigo et al. 2008, and
references therein).
Furthermore, their lifetimes are short which means they are rarely
found in clusters and therefore the majority of empirical spectra come from field
stars of unknown metallicity (e.g., Lan\c{c}on \& Wood 2000, Lan\c{c}on \& Mouchine 2002).  Nevertheless, they cannot be
ignored in SPS because they are 
bright in the rest-frame NIR and can dominate the total NIR luminosity
of young stellar populations with ages between 0.2 to 2.0 Gyr (e.g.,
Maraston 2005, Bruzual 2007).  The
treatment of TP-AGB stars in SPS modeling is now a key issue in the study
of stellar masses, because the rest-frame NIR also traces a galaxy's old stellar
population and therefore represents its integrated star formation
history (SFH).  If galaxies undergo multiple
bursts of star formation (SF) at widely spaced epochs, the rest-frame
NIR becomes the critical wavelength range for determining stellar masses.  For young
galaxies at high redshift, SPS codes with different treatments of
TP-AGB evolution can produce stellar masses and ages that
systematically differ by
roughly a factor of 2 for identical SEDs (e.g., Maraston et al. 2006;
Kannappan \& Gawiser 2007;
Wuyts et al. 2007; Bruzual 2007). 
Indeed, based on discrepancies between kinematically-derived, and photometrically-derived stellar masses of galaxies at $z
\sim$ 1, van der Wel et al. (2006) have even argued that the stellar masses
of young, early-type galaxies are better determined by SED fits that
exclude the rest-frame NIR.
\newline\indent
Recent papers by Maraston et al. (2006) and Bruzual (2007) have suggested
that $Spitzer$ observations could be important for resolving some of the
differences between SPS codes, particularly the TP-AGB treatment.  
IRAC  observes at 3.6$\micron$, 4.5$\micron$,
5.8$\micron$ and 8.0$\micron$, effectively the rest-frame NIR of galaxies at $z \sim$ 2.
At this redshift the universe is  young, $\sim$ 3 Gyr, and
therefore the majority of observed galaxies contain the young stellar
populations that should be dominated by TP-AGB stars; however, they  have not yet acquired an underlying old
stellar population that must be accounted for when studying TP-AGB
effects in the rest-frame NIR in low redshift galaxies (e.g., Riffel et
al. 2008). Using a sample of 7 galaxies at 1.4 $< z <$ 2.7 from the GOODS survey,
Maraston et al. (2006) have argued that their models produce 
better fits to young galaxies than the Bruzual
\& Charlot (2003, hereafter BC03) models and this is caused by their empirically
calibrated fuel consumption treatment of
the TP-AGB phase.  Using the next generation of the
BC03 code that includes improved evolutionary tracks for TP-AGB stars
from Marigo et al. (2008), Bruzual (2007) have fit the same 7 galaxies and
showed that they now obtain parameters closer to those in Maraston et
al. (2006). 
\newline\indent
In Kriek et al. (2006; 2007; 2008a, hereafter K08) we presented SED modeling of a
sample of 36 K-selected galaxies at $z \sim$ 2.3 that was based on
extensive NIR spectroscopy covering the JHK bands obtained from
GNIRS on Gemini, as well as broadband photometry.  These data constrain the
galaxy's SEDs from the rest-frame UV to optical and K08 compared fits with and without the NIR spectroscopy to
determine how parameters determined from fitting broadband data
alone compare
with those determined with the higher resolution spectroscopy and
spectroscopic redshifts.  Recently, we have obtained IRAC observations
of this sample  and  this allows us to extend the SED modeling into the
rest-frame NIR.  With deep photometry in 13 broadband filters
(UBVRIz$^{\prime}$JHK+IRAC) as well as high-resolution NIR spectroscopy that
covers the rest-frame Balmer/4000\AA~break, this sample of galaxies are
likely to have the best-constrained SEDs of young massive galaxies at $z \sim$ 2.3 
until more powerful optical and mid-infrared spectroscopic capabilities
become available from ground-based 20-30m telescopes and JWST,
respectively.
\newline\indent
In this paper we explore how well the properties of these galaxies can
be derived from their SEDs, with a focus on the rest-frame NIR.
Specifically, we concentrate on four issues, 1) {\it Improvement in
  Constraints and Systematic Errors:}
Does IRAC data improve the constraints on stellar
population parameters for $z \sim$ 2.3 galaxies, or does it increase the
uncertainties because the models fail to correctly reproduce the
rest-frame NIR SED?  Are there systematic
errors in the stellar population parameters determined without the rest-frame NIR
data?
2) {\it Best-possible constraints:} Ignoring systematic effects such as metallicity, dust law, IMF, and choice of SPS
code, what are the best-possible constraints on the stellar population
parameters of individual galaxies we can expect from their
SEDs alone? 3) {\it Comparison of SPS Codes:} How well do the most-used SPS codes describe the SEDs
of young stellar populations in the rest-frame NIR? Including the IRAC data, do any of the
codes provide better fits to the data?  What are the systematic differences in
parameters for individual galaxies using various SPS codes, dust laws,
and metallicities, and which of these effects is most significant?  4)
{\it Are passive galaxies really passive?}  Approximately 50\% of the
galaxies in our sample show no emission lines in their spectra.  Do
all of these galaxies have strongly suppressed star formation, or are some dusty star
forming galaxies with enough dust to obscure the emission-line regions?
\newline\indent
This paper is organized as follows.  In $\S$ 2 we briefly review the observational
data used in our analysis.  In $\S$ 3 we introduce the
 SPS codes used in our fitting, and discuss our fitting and error estimation
methods.  In $\S$ 4 we evaluate the
constraints on stellar population parameters when including the rest-frame NIR data.  In
$\S$ 5 we compare the quality of fits from the various SPS codes and
compare the systematic changes in parameters from these codes.  We
assess the systematic effects from the various assumptions of dust law
and metallicity compared to SPS code in $\S$ 6, and in $\S$ 7 we
examine what fraction of the galaxies have
strongly suppressed star formation based on their SEDs.   We conclude with a
summary in $\S$ 8
\newline\indent
Throughout this paper we assume a $\Omega_{m}$ = 0.3,
$\Omega_{\Lambda}$ = 0.7, H$_{0}$ = 70 km s$^{-1}$ Mpc$^{-1}$ cosmology.
\section{Data}
\subsection{The Galaxy Sample}
The 34 galaxies\footnote{The original sample contains 36 galaxies;
  however, two were not observed as part of the IRAC program, see $\S$
  2.2} used in this study are those with NIR spectroscopic
observations performed by Kriek et al. (2006; 2007), K08.  Galaxies were selected
for the sample based on their K-band magnitude (K $<$ 19.7,
Vega magnitude) and that they were likely to lie in the 2.0 $< z <$ 2.7
redshift range based on their photometric redshift
probability distribution as derived from broadband UBVRIz$^{\prime}$JHK 
photometry.  As discussed in K08, Mann-Whitney and
Kolmorov-Smirnov tests show that this sample is
representative of the distribution of the $z_{\mbox{\scriptsize
  phot}}$, J-K, R-K, and (U-V)$_{rest}$ colors of a mass-limited sample at
2 $< z <$ 3.  It may be less representative of
a K-bright subsample of the 2 $< z <$ 3 population because the
overall K-bright population tends to have a lower median redshift and narrower redshift
distribution than the spectroscopic sample (K08).  Small biases like
  these should not affect the analysis in this paper.  Approximately 20\%
(7/34) of the sample have spectroscopic redshifts 1.5 $< z <$ 2.0.
These galaxies may not necessarily be representative of the stellar
populations of a mass-limited sample in this
redshift range; however, we include them in this analysis.  
\subsection{Photometric Data} 
The UBVRIz$^{\prime}$JHK photometry for
galaxies in the  fields SDSS1030, CW1255, HDFS1, and HDFS2,
(hereafter, the MUSYC fields) is derived from the catalogues
presented in Gawiser et al. (2006) and Quadri et al. (2007).  
Photometry in the same bands for galaxies in the ECDFS field are derived from the catalogues presented
in Taylor et al. (2008).  
The  IRAC photometry for galaxies in the MUSYC fields is presented in
Marchesini et al. (2008) and we refer to that paper for a detailed
discussion of the data reduction and catalogue creation. 
The IRAC data for the ECDFS field was obtained as part of the SIMPLE
survey and is discussed in Damen et al. (2008).  We have added a systematic error of
10\% in quadrature to the IRAC photometric errors to account for zero
point uncertainties and the color-dependent flat fielding errors.
\newline\indent  
A few of the galaxies in the K08 sample were located close to the edge, or
completely out of the region covered by the IRAC data.  SDSS1030-101
is missing 3.6$\micron$ and 5.8$\micron$ data, whereas SDSS1030-1839
is missing 4.5$\micron$ and 8.0$\micron$ data.  These galaxies still
have data in the complementary IRAC bands so we retain them in the sample.
The galaxies SDSS1030-301 and SDSS1030-1813 were near the edge of the field and due to the
dither pattern have only a few frames of data in the
IRAC bands.  Although they are detected, the lack of data make background subtraction and cosmic
ray rejection difficult so we remove them from the
sample.  Removal of these galaxies reduces the sample to 34 of the 36 galaxies presented in K08.  
\subsection{NIR Spectroscopic Data}
The reduction of the NIR spectroscopic data used in our SED modeling was discussed in detail in
Kriek et al. (2006).  Briefly, the data were obtained with the GNIRS
instrument on Gemini-South using the 0.675$''$-wide
slit and the 32 l/mm grating in cross-dispersed mode which provided  a
resolving power of R $\sim$ 1000.  Spectroscopic redshifts were determined
for 17/34 galaxies that had detected emission lines.  For the
remaining 17 galaxies that did not have detectable emission lines 
K08 determined a spectroscopic continuum redshift using SED modeling
of the NIR spectra and broadband photometry.  In this
analysis we use the low-resolution binned spectra constructed by K08.
Each data point in the binned spectra contains 80 good pixels from the
observed spectrum, corresponding to $\sim$ 400\AA~observed frame.  Absolute flux calibration of the 
spectra is performed using the broadband JHK magnitudes.
\section{SED Fitting and Stellar Population Parameters}
\subsection{Stellar Population Models}
At present, there are several well-tested SPS codes publically available.  Codes such as PEGASE (Fioc \&
Rocca-Volmerage 1997), Bruzual \& Charlot (2003, hereafter BC03),
Maraston (2005, hereafter M05),
Starburst99 (Vazquez \& Leitherer 2005), and Charlot \& Bruzual (2008, in
preparation, hereafter CB08) all provide synthetic
spectra for SED modeling.  
\newline\indent
As pointed out by M05, the PEGASE, Starburst99, and
BC03 models use similar stellar spectral libraries, isochrones, and
treatment of the TP-AGB phase of stellar evolution.  Therefore, in our
fitting we  use the BC03 models assuming they are  representative of this
class.  We also perform fits using the M05 code which uses the fuel
consumption approach as an alternative to isochrone synthesis for modeling stellar evolution off the main
sequence.  Lastly, we  perform fits with the CB08 code which is similar
to the BC03 code except that it also includes  updated
evolutionary tracks for TP-AGB stars from Marigo et al. (2008).  Taken
together these
three codes span the different treatments of TP-AGB stars currently available.
\subsection{Fitting Method}
The fitting procedure used in our analysis is analogous to the one
outlined in Kriek et al. (2006) and K08.  The photometric and
spectroscopic data are fit to models with 34  exponentially
declining SFHs (parameterized by $\tau$) ranging from 0.01 to 20.0
Gyr.  We fit 45 different ages ($t$) ranging from 0.01 to 13 Gyr, but only allow
those that are less than the age of the universe at the redshift of
the galaxy.  For the majority of comparisons in this paper we assume solar
metallicity and a Calzetti et al. (2000) dust law\footnote{A comparison of fits using other
  dust laws and metallicities is presented in $\S$ 6} as our ``control'' model.
The V-band attenuation (A$_{v}$) is allowed to range between 0 and 4 mag in increments of 0.1.  When fitting
galaxies without emission lines we also adopt $z$ as a free parameter
and fit in increments of $\delta$z = 0.01.  
\newline\indent
To maintain continuity with the fits presented in K08 we adopt a Salpeter (1955)
IMF.  Although it is widely accepted that the Kroupa (2001) and
Charbrier (2003) IMFs are likely to be better choices for a universal IMF; we note that the
galaxies in our sample are young and the main sequence turnoff mass is much
larger than the mass at which these IMFs differ, therefore, the
choice between these IMFs does not significantly affect the shape of
the synthetic SEDs, only the normalization of the M$_{\mbox{\scriptsize star}}$.  Given
an average age of $\sim$ 1 Gyr, the M$_{\mbox{\scriptsize star}}$ computed for our sample
can be converted to a Charbrier or  Kroupa IMF simply by multiplying by scaling factors of 0.57 and 0.63,
respectively.  Our values of M$_{\mbox{\scriptsize star}}$ are also
corrected to account for gas recycled to the interstellar medium from
supernovae and planetary nebula.  Therefore, our M$_{\mbox{\scriptsize
    star}}$ corresponds to living stars plus stellar remnants, but do not
include recycled gas.
\newline\indent
For each SPS SED, we find a normalization ($X$)
by minimizing the function
\begin{equation}
\chi_{r}^2 = \frac{1}{N_{DOF}} \sum_{i = 1}^{N_{filters}} \frac{(X\cdot
  f_{model(\tau,t,A_{v},z),i} - f_{obs,i})^{2}}{\delta
  f_{obs,i}^2},
\end{equation}
where $N_{DOF}$ is the number of degrees of freedom, f$_{obs,i}$ is the observed flux in the $i^{th}$ filter,
$\delta$f$_{obs,i}$ is the error in the observed flux, and f$_{model(\tau,t,A_{v},z),i}$
is the flux of model SED which is a function of $\tau$, $t$, A$_{v}$,
and $z$.  We then construct a $\chi_{r}^2$-surface as a function of the
synthetic SPS model parameters $\tau$, $t$, A$_{v}$, and $z$.  The best
fit SPS model is determined by finding 
the location of the minimum on the $\chi_{r}^2$ surface.
\newline\indent
The 1$\sigma$ confidence intervals in the model parameters are
determined using the Monte Carlo method suggested by Papovich et al. (2001) and K08.  In these simulations 
the photometric and spectroscopic data are perturbed randomly within
their uncertainties and then the SPS models are fit again.  From
200 simulations per galaxy we find the value of $\chi_{r}^2$
that encompasses the $\chi_{r}^2$ from 68\% of the Monte Carlo simulations.  Returning to the
original $\chi_{r}^2$ fitting surface, locations with $\chi_{r}^2$ below this
value are considered to be allowable within the 1$\sigma$ errors.
The strength of this method is that by using the original $\chi_{r}^2$-surface
the correlations between errors in the parameters are preserved.  
\section{Improvement in Stellar Population Parameters from Rest-Frame NIR Data}
\indent
In this section we examine the importance of IRAC data
for constraining the stellar population parameters of $z \sim$ 2.3
galaxies from their SEDs without considering systematic effects such
as the choice of metallicity, dust law, IMF and SPS code.  We test for
potential systematic differences in parameters that result from
fitting with and without constraints on the rest-frame NIR SED, as well as how much (if at all) the rest-frame NIR
improves the uncertainties in the stellar population parameters.  
Throughout this
section we use the BC03 $\tau$-models with solar metallicity, the Calzetti et al. (2000) dust law, and a
Salpeter IMF as our ``control'' model for the comparisons.  
\subsection{Comparison of SED Fits With and Without IRAC Data}
\indent
In Table 1 we list the best fit SED parameters and 1$\sigma$
uncertainties determined from fits to  
the UBVRIz$^{\prime}$ broadband data and NIR spectroscopy (i.e., without the IRAC data,
hereafter we refer to this as U$\rightarrow$z$^{\prime}$+NIRspec), as well as
those from fits with the UBVRIz$^{\prime}$, NIR
spectroscopy and the IRAC data (hereafter
U$\rightarrow$8$\micron$+NIRspec).  Table 1 also contains the
parameter \age, defined as the SFR-weighted mean age of the
stellar population (see F\"{o}rster Schreiber et al. 2004).  The
parameter $t$ in Table 1 is the time since the onset of SF; however,
in $\tau$-models galaxies
are continually forming stars and therefore \age~
defines the mean age of the stellar population that dominates the
light of the galaxy.  Throughout this paper we use $\langle t\rangle_{\mbox{\scriptsize
    SFR}}$ as the metric of the age of the galaxies.
\newline\indent
K08 also determined SED parameters for this sample of galaxies using
the U$\rightarrow$z$^{\prime}$+NIRspec data (see their Table 2).  
Some of our U$\rightarrow$z$^{\prime}$+NIRspec fit parameters are
identical to those determined by K08; however, most of the 
parameters and uncertainties have changed by small amounts.  These
minor differences occur for several reasons.  Firstly, our grid of $\tau$ and $t$ is larger
and the spacing is somewhat different than the K08 grid.  Secondly, some of the broadband
fluxes have changed slightly in our newer photometric
catalogues.  Thirdly, we use a new fitting code which finds the
SED normalizations using a different algorithm than the K08 code.  This leads to small numerical
differences in the $\chi_{r}^2$'s.  Frequently the difference between
the minimum $\chi_{r}^2$ of the K08 best fit and our best fit is $<$ 0.1\%;
however, the best fit 
parameters can be significantly different.  There are a few galaxies that have large differences in
parameters; however, the 1$\sigma$ confidence intervals from our fits and the K08 fits still
span the same range of values for these galaxies.
Due to these  changes 
we use the parameters and uncertainties  for the U$\rightarrow$z$^{\prime}$+NIRspec data
determined using our fitting code for consistency of
comparison.  
\newline\indent  
In Figures 1 and 2 we plot F$_{\lambda}$ as a function of observed
wavelength for the photometric data as well as the best fitting SEDs.
Broadband photometry is plotted in red and the binned NIR
spectroscopic data is plotted in cyan.  The
SEDs fit with the U$\rightarrow$z$^{\prime}$+NIRspec data are plotted in grey,
and the SEDs fit with the U$\rightarrow$8$\micron$+NIRspec data are plotted in black.  The fit
parameters are  listed in the panels of Figures 1 and 2 for ease of comparison.  The broadband JHK fluxes have been plotted for
reference, but were not used in the fitting.  
\newline\indent
Examination of Figures 1 and 2
shows that for $\sim$ 50\% of the galaxies the SEDs fit without the
IRAC photometry are still consistent with the IRAC photometry at $<$ 
1$\sigma$.  Comparing the best fit SED parameters we find that  for 27/34
galaxies (79\%)  the $z$, $\tau$, $t$, A$_{v}$, M$_{\mbox{\scriptsize star}}$, SFR, and
\age~determined without IRAC data all agree with those fit with
IRAC data within the 1 $\sigma$ confidence intervals.  This
demonstrates that for most galaxies any systematic changes in the fits
that occur due to the inclusion of the IRAC data are smaller than the
formal confidence intervals (we explore this in more detail in $\S$ 4.3).  For
the remaining 7/34 galaxies (1030-1531, 1030-2728, 1256-519, 1256-1967,
HDFS2-509, ECDFS-6842, ECDFS-12514) there is a mixed 
range of agreement between the black and grey SEDs and the best 
fit parameters.
\newline\indent
The largest difference in best fit parameters as well as SED shape
is for the galaxy 1256-519, which provides a qualitative  illustration of
the importance of the rest-frame NIR photometry for constraining stellar population
parameters. Without the IRAC data galaxy 1256-519 is best fit as a young, dusty star forming
galaxy ($t$ = 0.3 Gyr, A$_{v}$ = 3.2, SFR $\sim$ 400 M$_{\odot}$
yr$^{-1}$), but including the IRAC data it is best fit as an old\footnote{Throughout this paper we use the term ``old'' when referring to stellar populations that are $\gtrsim$ 1 Gyr old.  In the local universe this age would be considered young; however, at $z \sim$ 2.3 it is nearly a maximally old stellar population.  Therefore, all galaxies in our sample are implicitly ``young'', and we use the terms ``young'' and ``old'' in their relative sense.}, moderately-dusty
quiescent galaxy  ($t$ = 2.5 Gyr, A$_{v}$ = 0.9, SFR
$\sim$ 3 M$_{\odot}$ yr$^{-1}$).  
As Figure 1 illustrates, there is little difference between these two SEDs in the
rest-frame optical; however, their fluxes in the rest-frame NIR are
quite different.  A young-and-dusty population is much brighter in the
rest-frame NIR than an old-and-quiescent population. Once the IRAC data is
included it is clear that the galaxy has an SED 
that is consistent with an old-and-quiescent population.  
\newline\indent
The importance of the rest-frame NIR for distinguishing
young-and-dusty populations from old-and-quiescent populations has
already been discussed by previous authors, e.g., Labb\'{e} et al. (2005), Papovich et al. (2006),
Wuyts et al. (2007) and Williams et al. (2008).  These studies have suggested that simple
color-color diagrams could be an efficient method for separating these types.  
Galaxy 1256-519 clearly demonstrates how critical the rest-frame NIR
is; even with spectroscopic redshifts and high-resolution spectrophotometry
near the Balmer/4000\AA~break, it can still be difficult to robustly distinguish
young-and-dusty from old-and-quiescent populations without additional data
on the rest-frame NIR SED.
\begin{figure*}
\plotone{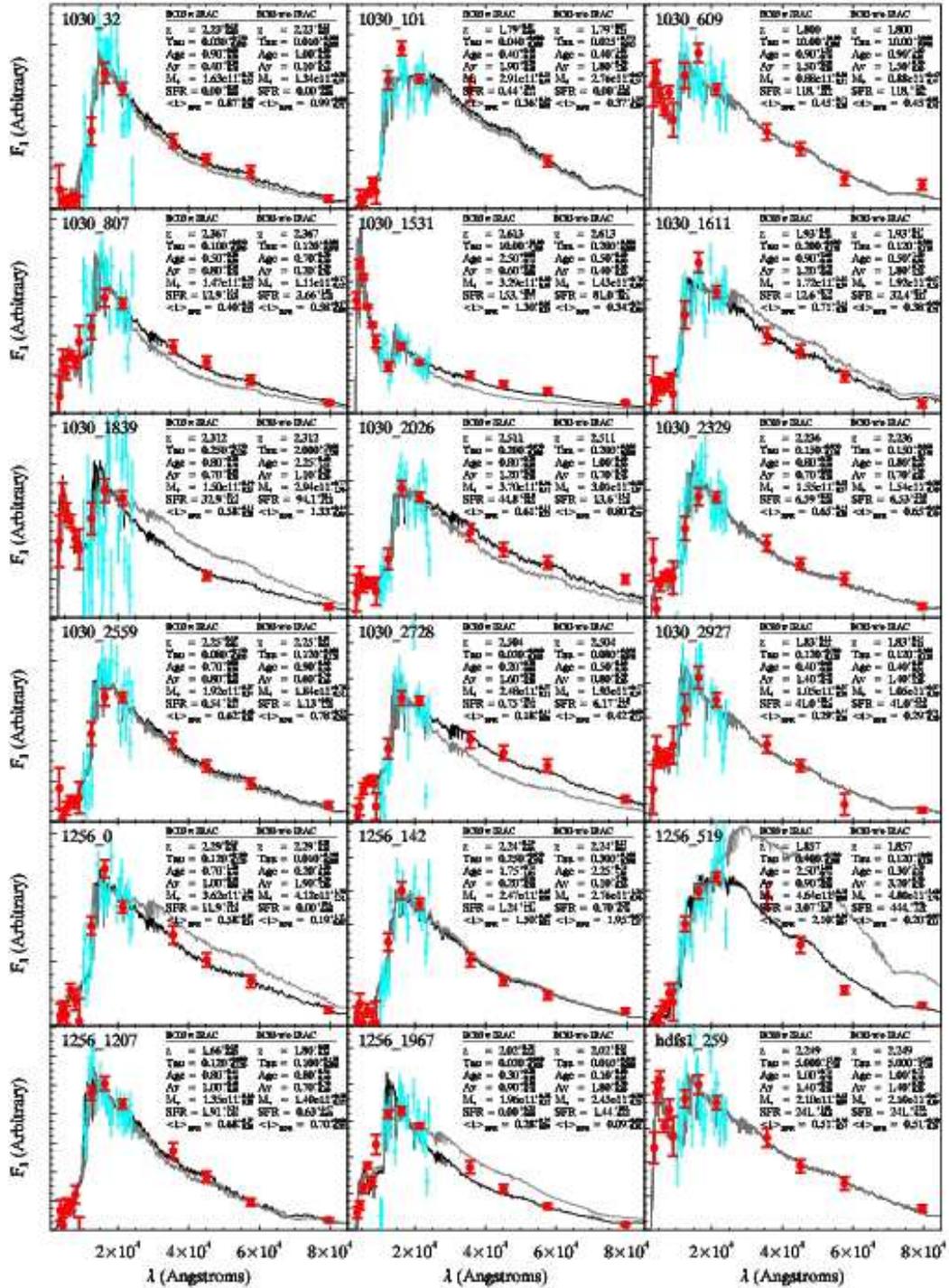}
\caption{\footnotesize SEDS fit using BC03 models without IRAC data (grey SED) and
 with IRAC data (black SED). The red points are the broadband
 photometry and the cyan points are the binned NIR spectroscopy.
}
\end{figure*}
\begin{figure*}
\plotone{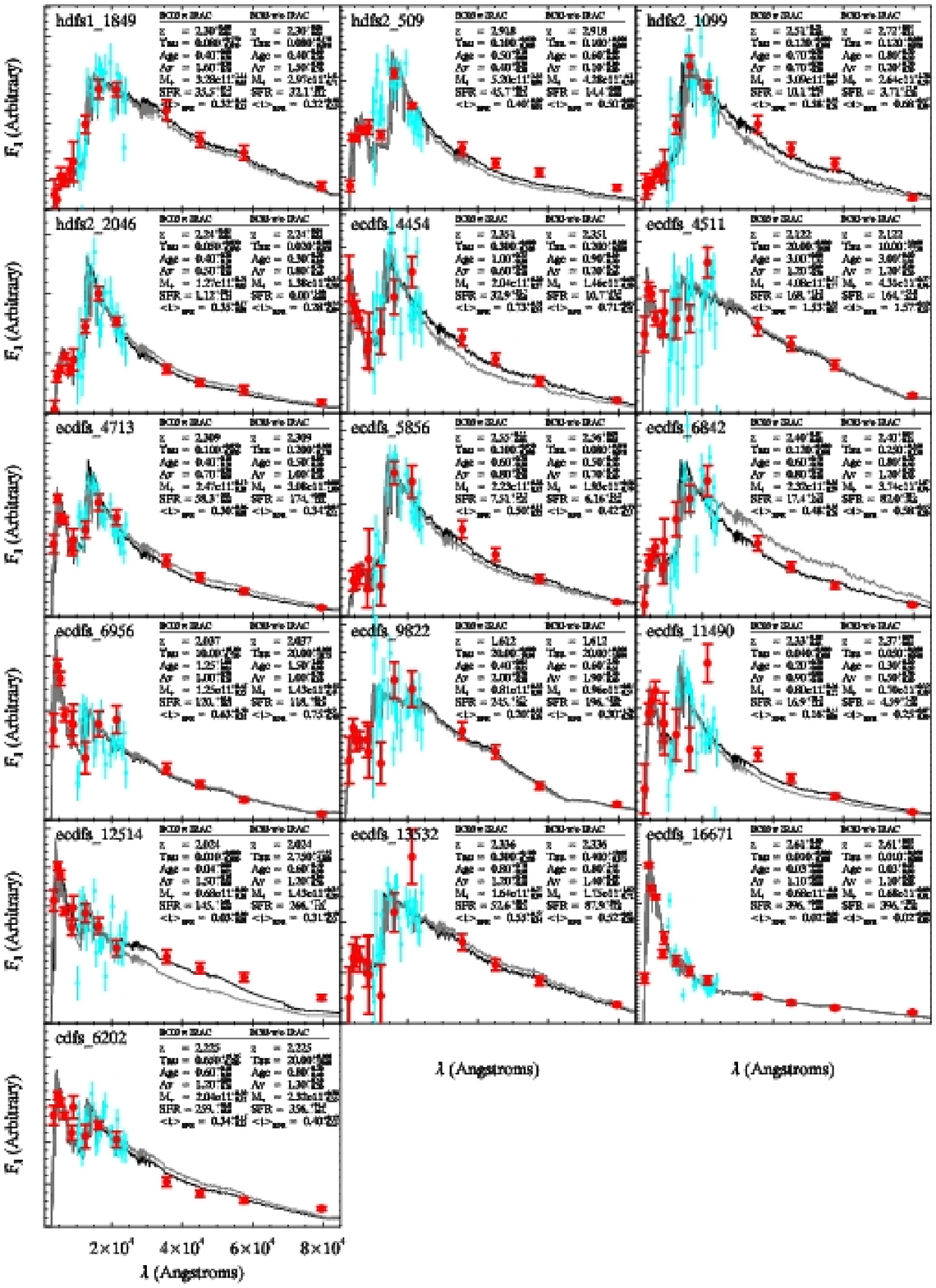}
\caption{\footnotesize As Figure 1. }
\end{figure*}
\subsection{Improvement in Photometric Redshifts with IRAC data}
\indent
Accurate determination of the stellar population parameters of $z
\sim$ 2 galaxies with broadband data requires high quality
photometric redshifts ($z_{\mbox{\scriptsize phot}}$).   Using this
spectroscopic sample K08 showed
that the $z_{\mbox{\scriptsize phot}}$'s determined using the
Rudnick et al. (2001; 2003) code, which employs the
Coleman et al. (1980) and Kinney et al. (1996) templates, were systematically overestimated by $\Delta
z$/(1 + $z$) = 0.08, where $\Delta$$z$ =
($z_{\mbox{\scriptsize phot}}$ - $z_{\mbox{\scriptsize spec}}$), and had a scatter of 0.13 in the same units.
Although random errors in $z_{\mbox{\scriptsize phot}}$ can be overcome with larger samples, systematic errors can pose significant
problems, particularly when studying  the evolution of the stellar
mass density or luminosity density (see, e.g., K08, van Dokkum et al. 2006).  It
has been suggested by some authors that the 1.6$\micron$ bump feature
present in the SEDs of evolved stellar populations could be a useful
$z_{\mbox{\scriptsize phot}}$ indicator (e.g., Simpson \& Eisenhardt 1999;
Sawicki 2002).  This feature falls in the IRAC bands for galaxies at
1.5 $< z <$ 3.0 and given the large scatter and systematic offset between the
$z_{\mbox{\scriptsize phot}}$ and the spectroscopic redshifts ($z_{\mbox{\scriptsize spec}}$) seen by K08, it is worth
investigating whether deep IRAC data might improve
the $z_{\mbox{\scriptsize phot}}$ of $z \sim$ 2.3 galaxies.  
We test this by determining $z_{\mbox{\scriptsize phot}}$ using the broadband photometry with and
without the IRAC data and comparing these to the $z_{\mbox{\scriptsize spec}}$. 
\newline\indent
The $z_{\mbox{\scriptsize phot}}$'s are computed using the EAZY
photometric redshift code (Brammer et al. 2008).  We use the standard EAZY
v1.0 template set which is determined using nonnegative matrix
factorization of SED models from the P\'{E}GASE code (Fioc \&
Rocca-Volmerange 1997).  For comparison, the $z_{\mbox{\scriptsize phot}}$ are  fit using three
datasets, the U$\rightarrow$K photometry, the U$\rightarrow$4.5$\micron$
photometry and the U$\rightarrow$8.0$\micron$ photometry.  The
U$\rightarrow$4.5$\micron$ photometry is tested separately from the
U$\rightarrow$8.0$\micron$ photometry because in practice the
5.8$\micron$ and 8.0$\micron$ bands are often avoided when fitting
$z_{\mbox{\scriptsize phot}}$ for large samples of galaxies (e.g., Brodwin et al. 2006;
Marchesini et al. 2008) because at present most
template sets do not include the PAH features that fall in these
bandpasses at $z <$ 0.7.  We omit the galaxies 1030-101 and 1030-1839 when
comparing the $z_{\mbox{\scriptsize phot}}$  because they  only have
photometry in two of the IRAC bands.
\newline\indent
In Figure 3 we plot  $z_{\mbox{\scriptsize phot}}$
vs. $z_{\mbox{\scriptsize spec}}$ for the three datasets.  For
galaxies without emission lines the $z_{\mbox{\scriptsize spec}}$ is
the best-fit redshift using the U$\rightarrow$8$\micron$+NIRspec data
in $\S$ 4.1, including the appropriate error bar.   Following K08,
galaxies in the ECDFS field, which has
much lower S/N  photometry in the JHK bands than the other MUSYC fields are plotted
with open grey circles.  Comparing the $z_{\mbox{\scriptsize phot}}$ fit with the
U$\rightarrow$K photometry to the $z_{\mbox{\scriptsize spec}}$ we find a systematic offset of 0.17 in
$\Delta z$/(1 + $z$) and a scatter of 0.12.  This offset
and scatter are similar, but not identical to the ones measured by K08; however, they
used the Rudnick et al. (2001; 2003) code 
rather than EAZY, so we do not expect perfect agreement.  Most of the systematic offset in the $z_{\mbox{\scriptsize phot}}$
is caused by the ECDFS galaxies (hereafter the ``wide''
sample, following the convention from K08).  Removing those from the sample (hereafter the ``deep''
sample) the systematic
offset is  0.00 and the scatter is 0.05, comparable to the offset and scatter of 0.03 and 0.08 measured by K08.
We list  the offsets and scatters measured for all datasets in Table 2.
\newline\indent
When the $z_{\mbox{\scriptsize phot}}$ are computed with the
U$\rightarrow$4.5$\micron$ photometry, there is no improvement in
the offset and scatter.  
Surprisingly, adding the  3.6$\micron$ and 4.5$\micron$ data does not
even improve the $z_{\mbox{\scriptsize phot}}$ of the wide sample, which
has the lowest S/N near-IR photometry.  When we fit with the entire
U$\rightarrow$8$\micron$ data the offset and scatter in the overall
sample improve to 0.14
and 0.11, respectively, and this overall improvement is caused
primarily by improvement in the wide
sample where the systematic offset is reduced by $\sim$ 0.1 in
$\delta$$z$.  For the deep sample the offset and scatter are 0.02 and
0.05, comparable to what is found with just the U$\rightarrow$K
photometry.  
\newline\indent
These comparisons show that the IRAC photometry does modestly
improve the $z_{\mbox{\scriptsize phot}}$ of galaxies at $z
\sim$ 2, but it does so only for galaxies with marginal S/N  photometry in the
optical and NIR bandpasses, and only
when all four IRAC channels are used.  For
galaxies with good S/N in the optical and NIR bandpasses, there is no evidence for an
improvement in the photometeric redshifts when IRAC data is
included.  This is best illustrated by comparing the $z_{\mbox{\scriptsize phot}}$ of the deep sample, fit
without the IRAC data to the wide sample, fit including all four IRAC
channels.  For the deep sample without IRAC data the offset and scatter are 0.00 and
0.05, respectively, and for the wide sample with IRAC data they are 0.35 and 0.17, respectively.  Clearly deep NIR data are  more
useful than IRAC data when determining the $z_{\mbox{\scriptsize
    phot}}$'s of
$z \sim$ 2.3 galaxies.  This suggests that for
this sample, the constraints on the $z_{\mbox{\scriptsize phot}}$ are dominated
by the location of the Lyman, Balmer or 4000\AA~breaks in the SEDs, and that constraints from the 1.6$\micron$ bump
are much weaker.  The Lyman, Balmer and 4000\AA~breaks are sharper
features than the 1.6$\micron$ bump, and the
optical and NIR bands are closer spaced in wavelength than the IRAC
bands, and therefore it is not surprising that this is the case. 
\newline\indent
It is worth pointing out that although the $z_{\mbox{\scriptsize phot}}$
of the 1.6 $< z <$ 2.9 spectroscopic sample are not significantly improved
with the IRAC data, this does not necessarily mean that IRAC data will not improve 
$z_{\mbox{\scriptsize phot}}$ for a different sample.  Our
K-selected sample is comprised primarily of massive
galaxies, which are frequently red with strong Balmer/4000\AA~breaks.  Blue galaxies
with a weak Balmer break and the Lyman break blueward of the
observed U-band ($z <$ 2.5) may see larger improvements in
$z_{\mbox{\scriptsize phot}}$ when IRAC data is included.  
\begin{figure}
\begin{center}
\includegraphics[scale = 1.5]{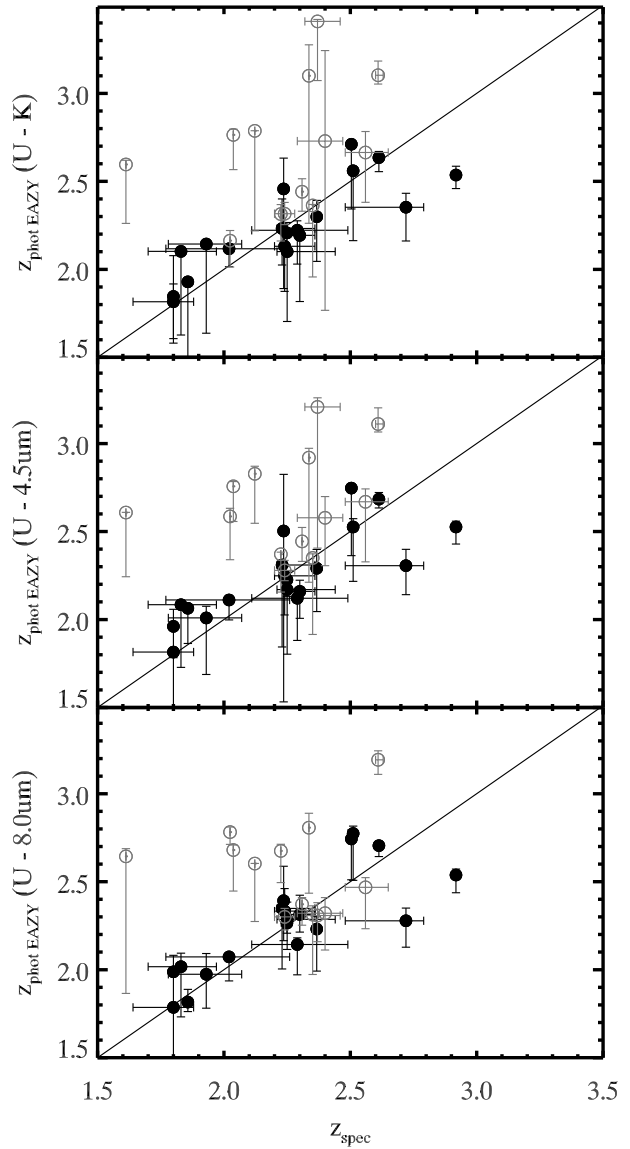}
\end{center}
\caption{\footnotesize Top panel:  EAZY $z_{\mbox{\scriptsize phot}}$ calculated
  using the U$\rightarrow$K photometry vs. $z_{\mbox{\scriptsize spec}}$. Galaxies in the ECDFS, which has much poorer S/N JHK
  photometry than the MUSYC deep fields are plotted as open grey circles.  Middle
  panel: Same as top panel but for $z_{\mbox{\scriptsize phot}}$'s determined
  with U$\rightarrow$4.5$\micron$ photometry.  Bottom Panel: Same as
  top panel but for $z_{\mbox{\scriptsize phot}}$'s determined with
  U$\rightarrow$8.0$\micron$ photometry.  The IRAC data does not
   improve the $z_{\mbox{\scriptsize phot}}$ of galaxies with good S/N
  JHK photometry.  It does reduce the systematic overestimate of $z_{\mbox{\scriptsize phot}}$ for galaxies
  with poorer S/N JHK photometry, but only by 0.1 in $\delta$$z$, and only when all
  four IRAC channels are used.  This shows that high S/N JHK photometry is significantly more
  valuable than IRAC data when determining the $z_{\mbox{\scriptsize phot}}$ of
  galaxies at $z \sim$ 2.3.}
\end{figure}
\subsection{Reduction of Systematic Errors in Stellar Population Parameters with IRAC Data}
\indent
Examination of the SEDs in $\S$ 4.1 showed that when the IRAC data were
included in the fits to the U$\rightarrow$z$^{\prime}$+NIRspec
data, the stellar population parameters of most galaxies remain
consistent within the 1$\sigma$
uncertainties.  Although consistent, the differences may be systematic,
which could change the mean stellar population parameters of the sample.
In this section we quantitatively check
for systematic differences in the mean parameters with the inclusion of
the IRAC data.  
\newline\indent
In Table 3 we list the mean value of the parameters determined with
different subsets of the data compared to the mean value determined using the
full U$\rightarrow$8$\micron$+NIRspec data set.  Specifically, we
compare parameters from fits to the broadband photometry and NIR
spectroscopy without the IRAC data
(U$\rightarrow$z$^{\prime}$+NIRspec, i.e., the same data as used in
the fits by K08), broadband photometry including the IRAC data
(U$\rightarrow$8$\micron$), and broadband photometry without IRAC data
(U$\rightarrow$K).  When fitting the broadband data we leave redshift
as a free parameter in order to preserve 
systematic differences that may result from either random or
systematic errors in 
$z_{\mbox{\scriptsize phot}}$.  The uncertainties listed in Table 3 are
standard errors of the mean ($\sigma$/$\sqrt{N}$).   More details of
the uncertainties and histograms that show the shape of the distributions are
presented as an Appendix.
\newline\indent
Table 3 shows that changes in the mean values of M$_{\mbox{\scriptsize star}}$,
$\tau$, \age, A$_{v}$, and SFR determined without the IRAC data 
(U$\rightarrow$z$^{\prime}$+NIRspec) compared to those determined
with the IRAC data (U$\rightarrow$8$\micron$+NIRspec) are all
consistent within the statistical accuracy achievable with our sample,
approximately 10\%.  This demonstrates that not only are the
parameters determined by K08 consistent with those determined with the
IRAC data, but that none of the differences serve to change the mean parameters
of the sample in a systematic way.
\newline\indent
The mean values of most  parameters determined using the broadband data are
also consistent with those determined using the
U$\rightarrow$8$\micron$+NIRspec data.  The only exception is
M$_{\mbox{\scriptsize star}}$, which appears to be overestimated by a factor of $\sim$
1.3 (0.11 dex) when only
U$\rightarrow$K photometry is used in the fitting.  This systematic overestimate
was also seen by K08 when they compared the M$_{\mbox{\scriptsize star}}$ from fits to the
U$\rightarrow$K data to those from the U$\rightarrow$z$^{\prime}$+NIRspec
data.  Given that  M$_{\mbox{\scriptsize star}}$ determined with both the
U$\rightarrow$8$\micron$+NIRspec and
U$\rightarrow$z$^{\prime}$+NIRspec data are consistent, our fits
support the K08 conclusions.  As discussed in K08, the
systematic effect on M$_{\mbox{\scriptsize star}}$ can lead to an overestimate of the
number of massive galaxies at high redshift, and hence an
underestimate of the evolution of the stellar mass density.
\newline\indent
Interestingly, it appears that the systematic overestimate of
M$_{\mbox{\scriptsize star}}$ is removed when the IRAC data is included in the SED
fits to the broadband data.  Elsner et al. (2008) also saw a systematic reduction in M$_{\mbox{\scriptsize star}}$ when
including IRAC data in SED fitting of broadband photometry of $z \sim$ 2 galaxies.  Their M$_{\mbox{\scriptsize star}}$ was systematically reduced by a factor of 1.6, which is similar to our factor of 1.3. Shapley et al. (2005) and Wuyts et
al. (2007)  made the same
comparison; however, neither found that adding IRAC data caused a
significant systematic change in the mean M$_{\mbox{\scriptsize star}}$ of their sample.
Shapley et et al. (2005) did suggest 
that the M$_{\mbox{\scriptsize star}}$ of a subsample of their galaxies was systematically
reduced when  IRAC data was included in the fitting.
\newline\indent
Why does including the IRAC data remove the systematic overestimate of
M$_{\mbox{\scriptsize star}}$ in our sample?  In $\S$ 4.2 we showed that the $z_{\mbox{\scriptsize
    phot}}$ were
systematically overestimated for galaxies in the ECDFS field, and that
the IRAC data  reduced this systematic by $\sim$ 0.1 in
$\Delta$$z$/(1 + $z$).  If we remove the ECDFS galaxies and compare the mean
values of M$_{\mbox{\scriptsize star}}$ from the U$\rightarrow$K photometry again, we find that the systematic overestimate of M$_{\mbox{\scriptsize star}}$ is
reduced to a factor of 1.08 $\pm$ 0.10.  This suggests that it is primarily the
systematic overestimate of the $z_{\mbox{\scriptsize
    phot}}$, caused by low S/N
JHK photometry that is responsible for the overestimate of
M$_{\mbox{\scriptsize star}}$.  Given that Wuyts et al. (2007) already had high S/N JHK
photometry, this may explain why they saw no change in M$_{\mbox{\scriptsize star}}$ when 
the IRAC data was included in the fitting.  Both the Elsner et
al. (2008) and Shapley et al. (2005) studies use the same redshifts
for their fits with and without IRAC data, which suggests that their
differences may be more data-specific, i.e., depend on the
combination of broadband filters.  Shapley et al. (2005) suggest it
may be caused by contamination of their observed K$_{s}$-band fluxes from H$\alpha$
emission.
\newline\indent
In summary, our comparison shows that when comparing stellar population parameters determined with various permutations
of the data, the only parameter that suffers a systematic bias at the
$>$ 10\% level is  M$_{\mbox{\scriptsize star}}$ when determined
using the U$\rightarrow$K photometry.    This bias can be removed by
including either NIR
spectroscopy or IRAC data in the fitting because both  improve the $z_{\mbox{\scriptsize
    phot}}$.  None of the other parameters, $\tau$, \age, A$_{v}$, or SFR show evidence
for systematic differences at $>$ 10\% even without using IRAC data or NIR
spectroscopy in the fitting.
\subsection{Improvement of Uncertainties in Stellar Population
  Parameters with IRAC Data}
\indent
Although the IRAC data do reduce the systematic errors in $z_{\mbox{\scriptsize phot}}$
and M$_{\mbox{\scriptsize star}}$ for our sample, there
are no changes in the mean values of the other stellar
population parameters.  This suggests that the IRAC data do not
improve the accuracy for the majority of stellar population parameters.  In this
section we examine if including the IRAC photometry improves the
uncertainties in the parameters determined for individual galaxies. 
\newline\indent
A simple test is to compare the parameter uncertainties  computed using the
Monte Carlo method.  While potentially informative,
such a comparison requires the uncertainties themselves to be robust, even in the case of sparse data.   A 
direct way to test for an improvement is to measure
the distribution of parameters computed without IRAC data relative to the
distribution computed with IRAC data.  The rms scatter of this distribution
is a metric of the average uncertainty in a parameter for the sample, independent of
the method used to estimate the uncertainties for individual galaxies. 
\newline\indent
Measuring a relative distribution of parameters requires a
comparison sample.  Since we do not have independent knowledge of
the parameters without measurement uncertainty, we will assume that the parameters determined using  the
U$\rightarrow$8$\micron$+NIRspec data are likely to be the most
precise, and compute the standard deviation ($\sigma$) of parameters determined
without the IRAC data relative to these.  A more detailed discussion of
these comparisons, as well as histograms of the distributions are presented in
the Appendix.  In Figure 4 we graphically summarize the result of
the comparisons for the parameters M$_{\mbox{\scriptsize star}}$, $\tau$, \age,
A$_{v}$, and SFR, computed using the U$\rightarrow$K,
U$\rightarrow$8$\micron$, U$\rightarrow$z$^{\prime}$+NIRspec, and
U$\rightarrow$8$\micron$+NIRspec data.
\newline\indent
The left panel shows the $\sigma$ of the logarithm of the parameter distributions 
computed using each subset of the data plotted as a function of the
$\sigma$ of the parameter distribution from the
U$\rightarrow$8$\micron$+NIRspec data.  In order to compare parameters with different units we use
the fractional scatter in each parameter ($\sigma$/mean).  
Reading from left to right along the X-axis of Figure 4 shows which parameters
are determined with the best precision using the
U$\rightarrow$8$\micron$+NIRspec data.  
\newline\indent
The Y-axis of Figure 4 shows the logarithm of the scatter in parameters determined using
the U$\rightarrow$K, U$\rightarrow$8$\micron$, and
U$\rightarrow$z$^{\prime}$+NIRspec data which are plotted as green diamonds,
blue squares, and red circles, respectively.  The one-to-one relation
is plotted as a dotted line.  If a data point  lies on that line it indicates
that the precision of that parameter, with the given subset of data,
is as good as the precision of that parameter attainable with the
entire dataset.  
\newline\indent
Figure 4 shows some interesting trends, perhaps the most unsurprising of
which is that the uncertainties in parameters determined using  the
U$\rightarrow$K data are always the
largest.  Adding either the NIR spectroscopy, the IRAC data, or the
combination of the two significantly improves the
uncertainties in all parameters.  
If we compare the uncertainties in parameters determined with the
U$\rightarrow$8$\micron$ data (blue squares) to those determined with
the U$\rightarrow$z$^{\prime}$+NIRspec (red circles) we can assess which parameters
are most improved with each type of data, as well as the size of
the improvement.  
\newline\indent
Examining the constraints from the U$\rightarrow$8$\micron$ data, it
is clear that the IRAC data is most useful for constraining the
M$_{\mbox{\scriptsize star}}$, A$_{v}$, and SFR.  Interestingly, adding
the NIR spectroscopy
in combination with the IRAC data does not further improve these
parameter estimates.  The U$\rightarrow$z$^{\prime}$+NIRspec data
shows that the NIR spectroscopy is most useful for constraining the \age~ and $\tau$
of the models;  however, the uncertainty in both of these parameters can
be further improved by including the IRAC data.  The NIR spectroscopy
also improves the M$_{\mbox{\scriptsize star}}$ significantly, but including the IRAC data makes the
constraints only $\sim$ 5\% better.
\newline\indent
Why do the different types of data  affect the parameters in these ways?  How
can IRAC data improve SFR estimates, when it primarily traces the old
stellar population?   
Although young-and-dusty and old-and-quiescent systems are difficult to distinguish with data that 
covers only the rest-frame UV to optical SED, they are 
separable once rest-frame NIR data is available (e.g., $\S$ 4.1, Labb\'{e} et
al. 2005; Williams et al. 2008).  This is not because the NIR
wavelength range itself is particularly valuable, but because it completes the coverage of the
part of a galaxy's SED that is dominated by light from stellar photospheres.  In
turn this  improves the constraints on the A$_{v}$, because A$_{v}$
changes the broad shape of the stellar SED.  In the $\tau$-models the
SFR is not a quantity that is fit for, instead it is inferred from the
number of $e$-folding times ($t$/$\tau$), scaled by the M$_{\mbox{\scriptsize star}}$, and
corrected for the A$_{v}$.  Improved constraints on A$_{v}$  lead
to better SFRs, and hence  IRAC data  actually improves
estimates of SFRs from SED fitting.
\newline\indent
The NIR spectroscopy adds both a $z_{\mbox{\scriptsize spec}}$, as well as
high-resolution information on the SED near the Balmer/4000\AA~break.  As
shown by K08, it is the high resolution information near the Balmer/4000\AA~
break that drives the improvement in both $t$ and $\tau$.
Interestingly, the IRAC data still helps to improve the constraints on
these parameters further.  Most likely this is because it traces the old
stellar population, which in tandem with the optical data constrains
the ratio of old stars to young stars, and hence the number of
$e$-folding times, which is directly related to both $t$ and $\tau$.
\newline\indent
Perhaps the most remarkable result from Figure 4 is that for our sample, the
parameters M$_{\mbox{\scriptsize star}}$ and SFR are as well-determined from
the  U$\rightarrow$8$\micron$ photometry and $z_{\mbox{\scriptsize
    phot}}$, as they are from the U$\rightarrow$8$\micron$+NIRspec
data and $z_{\mbox{\scriptsize spec}}$.  This suggests that deep
broadband photometry that extends into the rest-frame NIR can
potentially provide unbiased high-quality estimates of the
M$_{\mbox{\scriptsize star}}$ and SFR for massive galaxies at 2 $< z <$ 3.  
NIR spectroscopy is extremely valuable  for getting
accurate rest-frame colors using $z_{\mbox{\scriptsize spec}}$  (e.g.,
K08), identifying AGN (e.g., Kriek et al. 2007), and to provide an independent check on
SFRs from emission line fluxes (e.g., Erb et al. 2006); however, it
does not significantly improve the
constraints on M$_{\mbox{\scriptsize star}}$ or SFR from the SED fits if IRAC data is available.
\subsection{How Well Can We Constrain Stellar Populations Parameters with SEDs?}
\indent
Given that our current sample of galaxies has the best-constrained SEDs
of $z \sim$ 2.3 galaxies currently available, we can estimate the best-possible
constraints on stellar population parameters that can be
determined from SED fitting ignoring systematic effects such as choice
of metallicity, IMF, and dust law.
Assuming that the BC03 models with solar metallicity, a Salpeter IMF, and a Calzetti
et al. (2000) dust law create an appropriate set of template SEDs,
the X-axis of Figure 4 shows that the SED fitting best-constrains the parameters M$_{\mbox{\scriptsize star}}$, A$_{v}$,
and \age.  Using the U$\rightarrow$8$\micron$+NIRspec data,
M$_{\mbox{\scriptsize star}}$ can be determined to $\pm$ 0.12 dex, \age~ to $\pm$
0.26 dex, and A$_{v}$ to $\pm$ 0.3 mag.  Interestingly, when only
broadband U$\rightarrow$8$\micron$  data is used, the
constraints on these parameters are similar, 
$\pm$ 0.11 dex, $\pm$ 0.36 dex, and $\pm$ 0.3 mag, for M$_{\mbox{\scriptsize star}}$,
\age, and A$_{v}$, respectively. 
\newline\indent
The $\tau$ and SFR, are  not as well-determined as the other parameters.  With the
entire U$\rightarrow$8$\micron$+NIRspec data set the scatter in $\tau$ and SFR
are  factors of 2.5 and 2.8, respectively.  The
uncertainty in the SFR with the broadband data alone is still a factor
of 2.8; however,  the uncertainty in $\tau$ is much worse, a factor of 4.6.  .
\newline\indent
Overall, this exercise shows that the best constrained parameters of
individual galaxies in our sample from
SED fitting are M$_{\mbox{\scriptsize star}}$, \age~, and
A$_{v}$.  
As we will demonstrate in the $\S$ 5,
these uncertainties are now small enough that they are comparable to the level of the systematic 
errors caused by the uncertainty in SPS code, metallicity, dust law
and IMF.
\begin{figure*}
\plotone{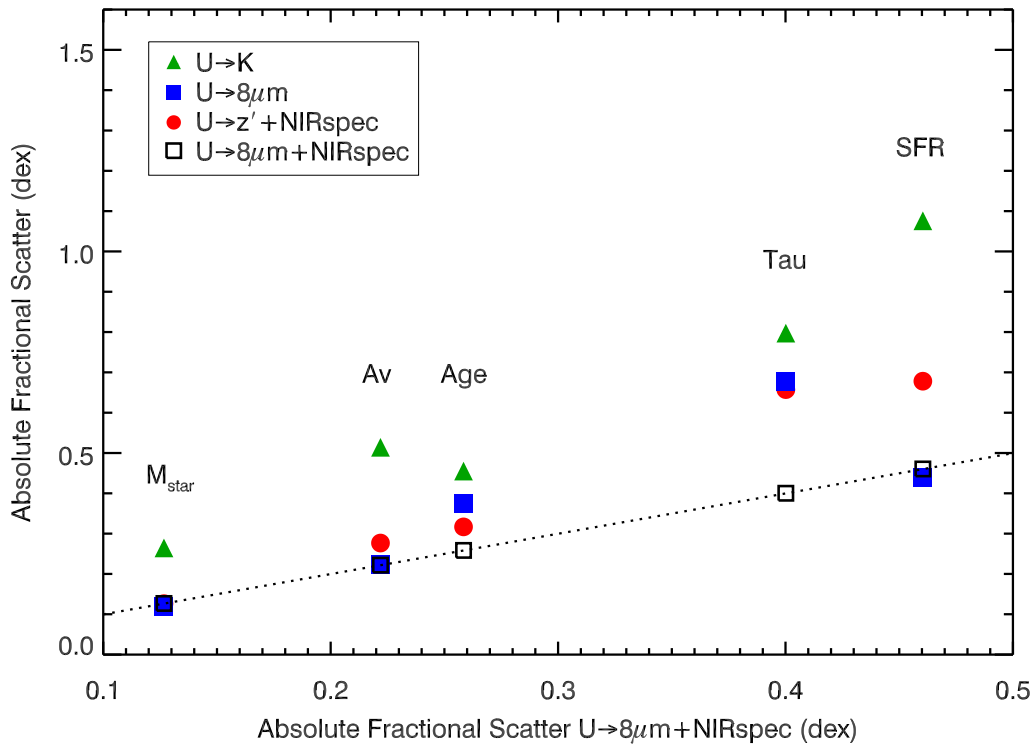}
\caption{\footnotesize The logarithm of the absolute
  fractional scatter in parameters,
  Log($\sigma$/mean), determined with various subsets of the data
  (U$\rightarrow$K, U$\rightarrow$8$\micron$,
  U$\rightarrow$z$^{\prime}$+NIRspec) plotted versus the
  logarithm of the absolute fractional scatter determined with the entire dataset
  (U$\rightarrow$8$\micron$+NIRspec).  The dotted line has a slope of
  unity.  Reading along the X-axis, the best constrained parameters are M$_{\mbox{\scriptsize star}}$,
  A$_{v}$, and \age,  which have rms scatters of $\sim$
  0.12 - 0.25 dex.  The $\tau$ and SFR are not as well constrained from the
  SED alone, and have a scatter of a factor of $\sim$ 2-3.}
\end{figure*}
\section{Comparison of SPS Codes}
In this section we fit the full U$\rightarrow$8$\micron$+NIRspec photometric
dataset to a grid of $\tau$-models from the SPS codes of BC03, M05, and CB08.  We examine the
overall quality of the fits based on the $\chi_{r}^2$ statistic as well as
the residuals from the mean SEDs to test whether
a particular set of models provides a better description of the data.  Given the different treatment of the TP-AGB phase of stellar
evolution in the M05 and CB08 models compared to the BC03 models and
the larger rest-frame NIR fluxes this produces, we
compare fits with and without the IRAC data as a test of how well 
the models describe the SEDs in the rest-frame NIR.  We also examine
the systematic differences in the stellar population parameters
determined with the different codes and their implications for studies
of the stellar populations of distant galaxies.
Throughout this comparison we assume solar metallicity, a Salpeter
IMF, and the Calzetti et al. (2000) dust law as a ``control'' model.
\subsection{Comparison of Models When Excluding IRAC}
Our first comparison of the models is made using the
U$\rightarrow$z$^{\prime}$+NIRspec data which effectively spans the UV through optical wavelength range
for the galaxies in our sample.
The parameters determined from these fits are listed in Table 1.  
In the top panels of Figure 5 we plot the $\chi_{r}^2$'s of the fits to the M05 and CB08 models
against those from the fits to the BC03 models.
Figure 5 shows that there is a large range of $\chi_{r}^2$ values for
the fits; however, the majority of the systems, 26/34 (76\%) are reasonably well
described by the models, having $\chi_{r}^2$ $<$ 2.  This shows that even with the high-resolution
photometric data available from the NIR spectroscopy, simple $\tau$-models from
all three SPS codes can still successfully reproduce the SEDs of the
majority of $z \sim$ 2 galaxies.  
\newline\indent 
Comparison of the $\chi_{r}^2$ values shows that the BC03 models
have the lowest $\chi_{r}^2$ for  9/34 galaxies (26\%), the M05 models
have the lowest $\chi_{r}^2$ for 15/34 galaxies (44\%), and the CB08 models
have the lowest $\chi_{r}^2$ for 10/34 galaxies (29\%).  This demonstrates
that within our sample there is no statistically significant
evidence that one of the models provides the best fits more frequently
than the others.  Summing the $\chi_{r}^2$
values for all 34 galaxies results in total
$\chi_{r}^2$'s of 57.72, 60.54, and 58.02 for the BC03, M05, and
CB08 models, respectively.  These imply an average $\chi_{r}^2$ of 1.70 $\pm$ 0.03, 1.78 $\pm$
0.03, and 1.71 $\pm$ 0.03 for the BC03, M05, and CB08 models,
respectively, where the quoted errors are standard errors of the mean.  This suggests that the BC03 and CB08 models
may provide modestly better fits on average to the rest-frame UV
through optical SEDs than the M05 models; however, the majority of
this difference is caused by a few galaxies that are fit much better
by the BC03 and CB08 models (e.g., galaxy HDFS2-509 and 1256-1967) 
\newline\indent
It is worth noting that in Figure 5 the $\chi_{r}^2$ values of the BC03 and CB08 models are nearly identical
for all galaxies, whereas there is a larger scatter in the
$\chi_{r}^2$'s between the BC03 models and the M05 models.  This
confirms that the  SEDs in the BC03 and CB08 models appear to be nearly
identical in the rest-frame UV through optical wavelength range.  In $\S$ 5.2 we will
show that the fits to the rest-frame UV to NIR using the BC03 and CB08 models are quite different;
and the current comparison makes it clear that those differences must be
driven exclusively by the rest-frame NIR part of the SED.
\newline\indent
Overall, based on the average $\chi_{r}^2$, the BC03 and CB08 models perform slightly better than the M05
models at fitting the rest-frame UV to optical SEDs, but the
difference is small ($\sim$ 5\% in $\chi_{r}^2$) and only significant
at  $\sim$ 1.5$\sigma$ within our sample of 34 galaxies.
\begin{figure*}
\plotone{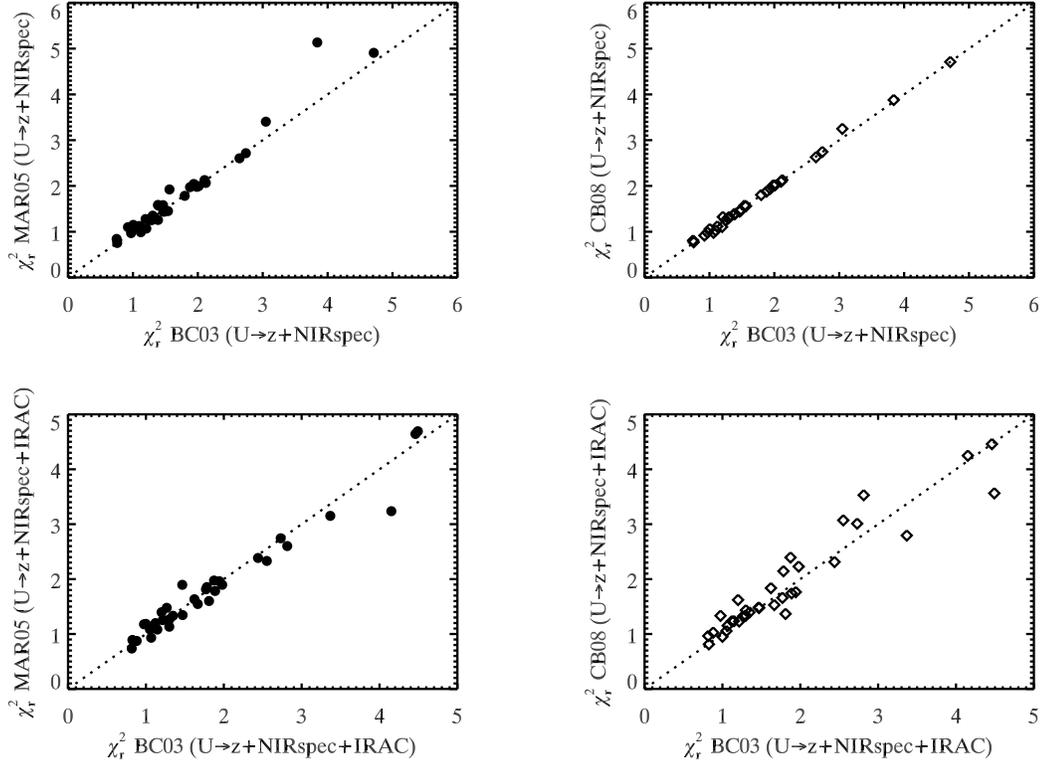}
\caption{\footnotesize Top Panels: Comparison of $\chi_{r}^2$ values
  for fits without the IRAC data included using models from the M05
  code vs. the BC03 code (left) and the CB08 code vs. the
  BC03 code (right).  The $\langle \chi_{r}^2\rangle$ values are 1.70 $\pm$ 0.03, 1.78 $\pm$
  0.03, and 1.71 $\pm$ 0.03 for the
  BC03, M05, and CB08 models, respectively.  This suggests that on
  average the BC03 and CB08 models may provide slightly better fits to the rest-frame UV to optical
  of young galaxies than the M05 models.  
 Bottom Panels: Comparison of $\chi_{r}^2$ values
  for fits including the IRAC data using models from the M05 code
  vs. the BC03 code (left) and  the CB08 code vs. the
  BC03 code (right).  The $\langle \chi_{r}^2\rangle$ values are 1.79 $\pm$ 0.03, 1.77 $\pm$
  0.03, and 1.85 $\pm$ 0.03 for the BC03, M05, and CB08 models,
  respectively.  Based on this comparison, all three
  codes perform equally well at fitting the rest-frame UV to NIR SEDs of young galaxies.}
\end{figure*}
\subsection{Comparison of Models When Including IRAC}
We next compare how well the models describe the entire rest-frame UV to NIR SED of
the galaxies by fitting the U$\rightarrow$8$\micron$+NIRspec data.
The  best fit SEDs from each SPS code are plotted in units of
F$_{\nu}$ vs. $\lambda$ in Figures 6 and
7.  The black line, green line,
and blue line are the best fits using the BC03, M05, and CB08 models,
respectively.    
The best fit stellar population parameters, their
uncertainties, and the $\chi_{r}^2$ of the best fits are listed in Table
1.  We plot the $\chi_{r}^2$'s of the fits to the M05 and CB08 models
against those from the fits to the BC03 models in the bottom panels of
Figure 5.
\newline\indent
Comparison of the $\langle \chi_{r}^2\rangle$'s in the top and bottom
panels Figure 5 shows that the variance in
$\chi_{r}^2$'s between models is significantly larger when the rest-frame NIR is included in the fitting;
however, the frequency with which each model provides the best fit is
still similar.  The BC03 models have the lowest $\chi_{r}^2$ for 11/34 (32\%) of the
galaxies, the M05 models have the lowest $\chi_{r}^2$ for 12/34 (35\%)
of the galaxies, and the CB08 models have the lowest $\chi_{r}^2$ for 11/34 (32\%)
of the galaxies.  
\newline\indent
Totaling the $\chi_{r}^2$ for all fits with each SPS codes gives
values of 62.80, 62.11, and 64.72 for the BC03, M05, and CB08 models,
respectively.  
Converting the total $\chi_{r}^2$'s to a $\langle \chi_{r}^2\rangle$
per galaxy for our sample gives values of 1.85 $\pm$ 0.03, 1.83 $\pm$
0.03, and 1.90 $\pm$ 0.03 for the BC03,
M05, and CB08 models, respectively, where the errors are the standard
error of the mean.  Interestingly, when the rest-frame NIR is
included, the M05 models describe the SEDs slightly better than both the BC03
and CB08 models on average.  The difference is not  significant when comparing the
BC03 and M05 models; however, it is significant at $\sim$ 1.5$\sigma$ when
comparing the M05 and CB08 models.  Maraston et al. (2006) found
similar results using a sample of 7 galaxies at 1.4 $< z <$ 2.7
selected from the GOODS survey.  For their sample the $\langle \chi_{r}^2\rangle$ was 1.38 $\pm$ 0.10 for fits with the M05 models, and 1.51 $\pm$ 0.10
for fits with the BC03 models, where we have computed the error bars
based on the standard error of the mean.  Their results also show a small, but not
statistically significant, advantage for the M05 models compared to the
BC03 models when the entire rest-frame UV to NIR SED is fit.
\newline\indent
If the M05 models had the largest  $\langle \chi_{r}^2\rangle$ when
fitting the rest-frame UV to optical SEDs of the galaxies, but the
smallest  $\langle \chi_{r}^2\rangle$ when the
rest-frame NIR data is included, it  suggests that they may describe
the rest-frame NIR for these galaxies better than both the BC03 and CB08
models.   We test this by
plotting the $\chi_{r}^2$ values obtained when fitting
using only the U$\rightarrow$z$^{\prime}$+NIRspec data versus those obtained when
fitting the U$\rightarrow$z$^{\prime}$+NIRspec+IRAC data for all three SPS codes in Figure 8.  Galaxies that
have much poorer fits when the rest-frame NIR is included will lie to
the right of the dotted line in Figure 8.  
\newline\indent
Before we discuss the $\chi_{r}^2$ values from fits with and without the
rest-frame NIR data, it is worth noting that the fraction of massive galaxies that host an AGN at $z \sim$ 2 is
much higher than at lower redshift (e.g. Kriek et al. 2007, Daddi et
al. 2007).  The presence of a
dusty AGN can cause significant emission at rest-frame NIR wavelengths
(e.g., Donley et al. 2008, and references therein),
and therefore can ``contaminate'' the stellar SED in the rest-frame NIR.
Such systems are now being identified at high redshift using the
presence of 
excess emission at  $\lambda$ $>$ 1.6$\micron$, rest-frame.  Galaxies
with a dusty AGN are frequently referred to as power-law galaxies (PLGs)
because the strong NIR emission at $\lambda$ $>$ 1.6$\micron$ from the AGN overpowers the stellar
emission removing the turnover in the SED at 1.6$\micron$ from the
peak of the stellar emission, and making the SED appear more like a power-law.  Given
that the SPS models do not include an AGN component, we expect any
galaxies identified as PLGs will be fit much more poorly when the
rest-frame NIR is included.  Poor fits to these galaxies in the
rest-frame NIR probably does not reflect a deficiency with the models
and therefore we need to identify any PLGs
within our sample.
\newline\indent
We identified PLGs by fitting the IRAC data for each
galaxy to a power-law of the form $f_{\nu} \propto \nu^{\alpha}$.
Galaxies fit with $\alpha$ $>$ -0.5 are considered PLGs, and therefore
likely candidates for hosting a dusty AGN (e.g., Alonso Herrero et
al. 2006, Donley et al. 2007).  The properties of these galaxies and
their overlap with emission-line AGN (Kriek et al. 2007) will be presented in
more detail, and including MIPS 24$\micron$ observations, in a future paper.
Here we only discuss them as they are relevant to SED fitting in the
rest-frame NIR.  
\newline\indent
The PLGs are plotted as open red
circles in Figure 8.  Interestingly, not only do most of the PLGs
have significantly poorer fits in the rest-frame NIR than the average
galaxy, they also have significantly poorer fits in the
rest-frame UV to optical than the average galaxy.   This suggests that if these
systems do have a dusty AGN, the AGN may also emit radiation blueward
of rest-frame 1.6$\micron$ and ``contaminate'' this part of the SED as
well.
\newline\indent
Excluding the PLGs, we can see that for the M05 models, very few
galaxies have significantly higher $\chi_{r}^2$ values when the
rest-frame NIR data are included in the fit.  For the BC03 and CB08 models, there
are some non-PLG systems where the fits are much worse when the
rest-frame NIR data is included; however, there are only a handful of such
galaxies.  The most discrepant galaxies in the BC03 models are
1030-1531, 1256-519, and 1256-1967.  Galaxy 1256-519 has a poor fit
with the IRAC data in all models; however,  it is notable that
1030-1531 and 1256-1967  have $more$ flux in the rest-frame NIR than
predicted by the BC03 models and are better described by the M05
and CB08 models.  
Interestingly, these two galaxies also have similar ages 
(0.47$^{+0.09}_{-0.07}$, and 0.45$^{+0.07}_{-0.16}$ Gyr) from the M05
  fits, and are right in the middle of the age range were emission
  from TP-AGB
  stars is near the maximum for a range of metallicities
  (e.g. M05) suggesting that the different treatment of these stars in
  the M05 and CB08 models is the cause of the improved fit.    
Still, it is possible that these two galaxies could  be modeled using a
composite of young and old bursts with the BC03 code (e.g., Yan et
al. 2004), and with only two non-PLG galaxies
in the sample that would require such modeling to explain the
SED, we suggest that there is only mild evidence at best from the $\chi_{r}^2$'s that the SEDs from
the BC03 code require significant changes in the rest-frame NIR in order to match the
observed SEDs of our galaxies.  
\newline\indent
This can also be seen in Figure 9 where we plot the residuals from
the SEDs fit using models from the three SPS codes, corrected to the rest-frame.  The solid line represents a
running average of the 30 nearest points.  The running average 
suggests that the BC03 models may underpredict the mean flux of the
average galaxy in the wavelength range 1.0 - 1.8$\micron$; however,
the difference is at most, 5-10\%, conceivably within the allowable
range of the uncertanties in the IRAC photometry.  The residuals from the M05 and CB08 models do not
show the same trend,  suggesting that they may describe the
rest-frame NIR SEDs of these galaxies slightly better than the BC03 models.
\newline\indent
In summary, these comparisons show a few key results.  Most
importantly, based on the frequency of lowest $\chi_{r}^2$ and the
$\langle \chi_{r}^2\rangle$ for all galaxies, we conclude that there is no
significant evidence that any of the three  SPS codes we tested describe the complete rest-frame UV to
NIR SEDs of massive $z \sim$ 2.3 galaxies significantly better than the others.
Comparison of the $\langle \chi_{r}^2\rangle$'s from fits with and without the rest-frame NIR data suggest
that the BC03 and CB08 models may describe the rest-frame UV to optical
SEDs of the galaxies slightly better than the M05 models, but that
the latter may fit the rest-frame NIR better than the former.  Still, the
differences in the $\langle \chi_{r}^2\rangle$ between the models are always $<$
5\% and are only significant by $\sim$ 1.5$\sigma$ at most.  The
residuals from the fits also seem to show that the M05 and CB08 models
describe the rest-frame NIR slightly better than the BC03 models, but again, the difference
is at most 5-10\%, and within the photometric errors.  These comparisons
suggest that it will be extremely difficult to refine the
models based on the observed SEDs of young stellar populations at high
redshift.
Mostly likely such constraints  will  have to come from other
avenues such as NIR spectroscopy of young stellar populations in the
local universe (e.g., Riffel et al. 2008) or  improved theoretical
understanding of post main sequence stellar evolution (e.g., Marigo et
al. 2008), or both.
\begin{figure*}
\plotone{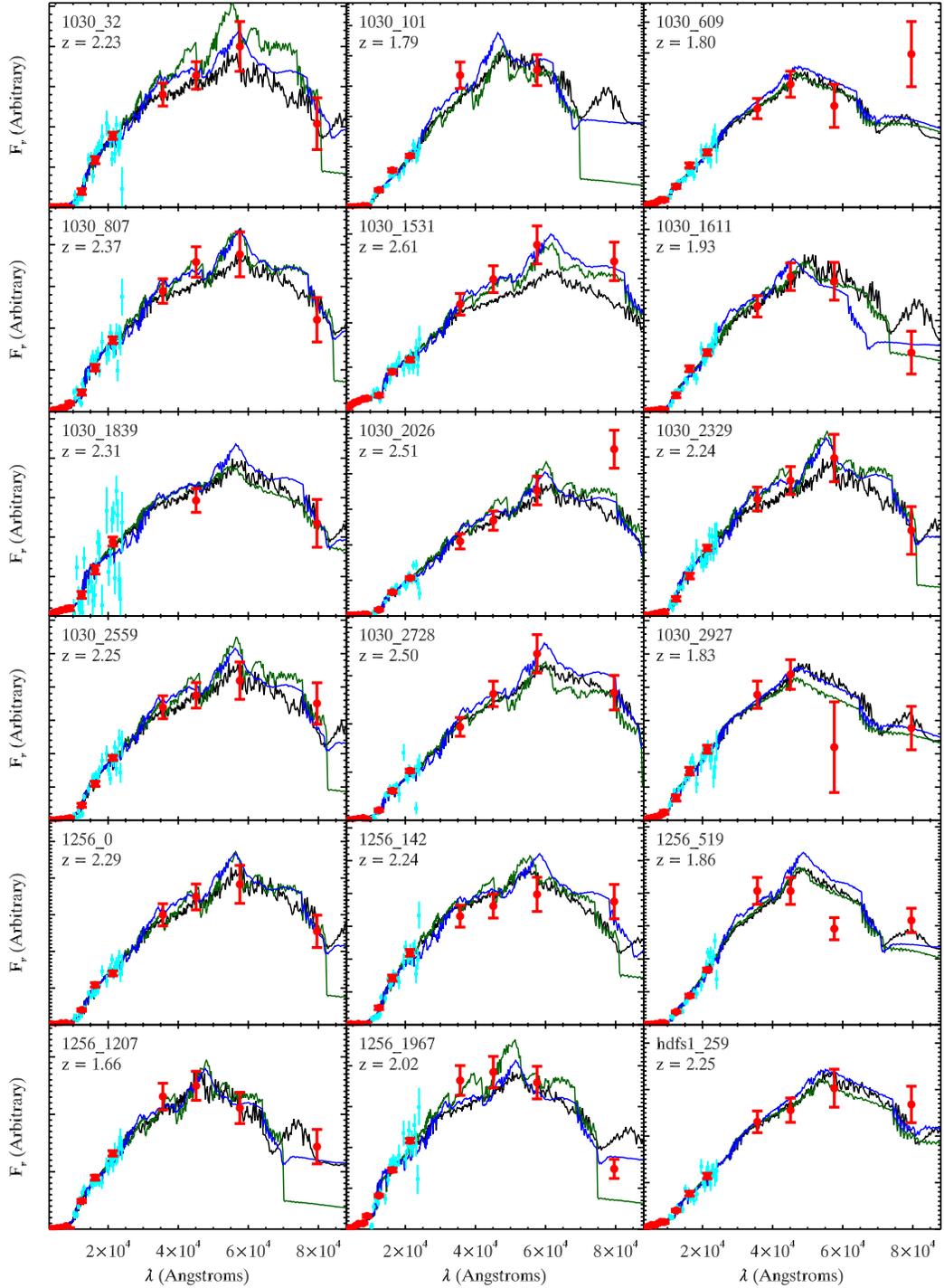}
\caption{\footnotesize SEDs  fit with BC03 models (black), M05 models
  (green) and CB08 models (blue), plotted as F$_{\nu}$ vs. Log($\lambda$). Red points are the broadband
  photometric data and cyan points are the binned NIR spectroscopy.}
\end{figure*}
\begin{figure*}
\plotone{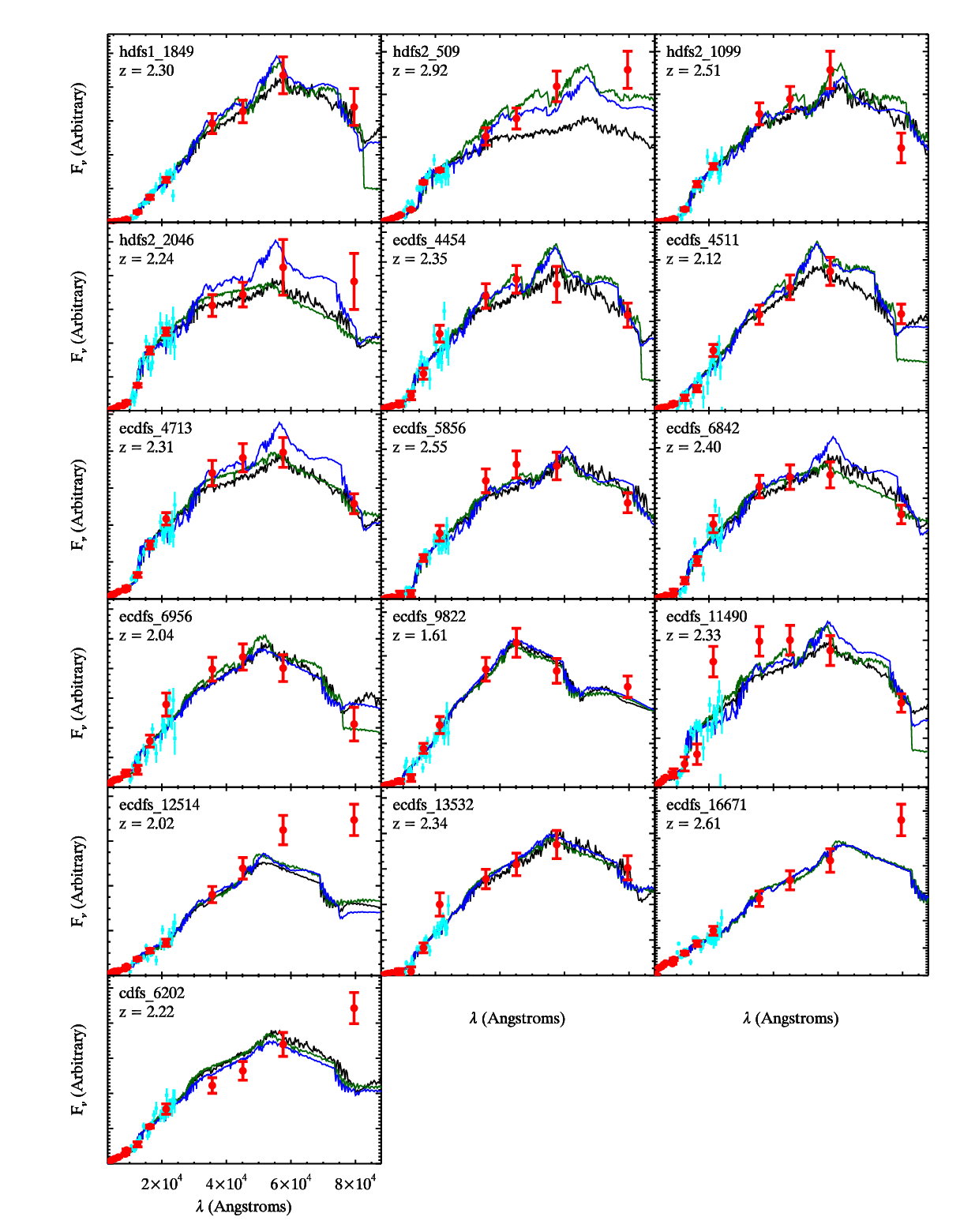}
\caption{\footnotesize As Figure 6. }
\end{figure*}
\subsection{How do best-fit parameters vary between SPS codes?}
Although the three codes describe the overall shape of the SEDs of
the galaxies equally well, the parameters of the best fit SEDs from
the codes imply significantly different stellar
populations for the same galaxies.  In Figure 10 we plot the SED parameters
determined from the U$\rightarrow$8$\micron$+NIRspec data using the M05 and CB08 models against those determined
using the BC03 models as solid circles and open diamonds, respectively.
\begin{figure*}
\plotone{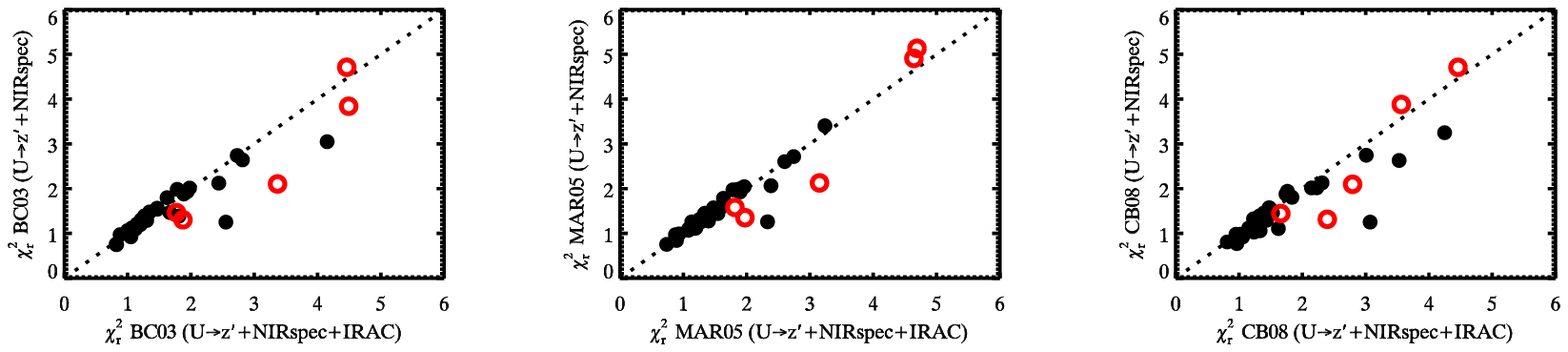}
\caption{\footnotesize  Left Panel: Comparison of $\chi_{r}^2$ values
  for fits with and without the IRAC data using the BC03 models.  Middle
  Panel: Same as  left panel but for fits with the M05 models.
  Right Panel:  Same as  left panel but for fits with the
  CB08 models.  Open red circles represent galaxies whose SEDs
  follow a power law in the IRAC bands of the form f$_{\nu}$ $\propto$ $\nu^{\alpha}$, where $\alpha$ $<$ -0.5.
  These galaxies are commonly defined as ``power-law galaxies'' and
  likely have the rest-frame NIR dominated by emission from dust
  heated by an AGN.  The M05 models show the least increase in $\chi_{r}^2$
  when the rest-frame NIR data is included in the fitting; however,
  the BC03 and CB08 models have only a few non-power-law galaxies that
  have notably worse fits when the rest-frame NIR data is used in the fitting. }
\end{figure*}
\subsubsection{Differences in Stellar Populations Between M05 and
  BC03 Models}
The mean and median ratio of stellar population parameters determined with the M05 and BC03 models are listed in Table 4.
The most significant difference between these models
is in the median M$_{\mbox{\scriptsize star}}$.  From Figure 10 we can see that the median M$_{\mbox{\scriptsize star}}$ from the M05 models is 0.63
that of the BC03 models, and appears to be a systematic offset,
independent of the mass of the galaxy.   If we divide the galaxies
into two samples, those with detected emission lines in the NIR
spectrum (hereafter EL-galaxies) and
those without detectable emission lines (hereafter NEL-galaxies) the
systematic change in M$_{\mbox{\scriptsize star}}$ for both of
these groups is 0.63 and 0.69, respectively, nearly identical to the median
difference of 0.63 for the whole sample.  This offset in M$_{\mbox{\scriptsize star}}$ is similar to 
those measured by Maraston et al. (2006) and Wuyts et al. (2007) who
saw ratios of 0.58 and 0.72, respectively. 
\newline\indent
Comparing the best fit values of $\tau$ between the models shows that
the M05 models prefer slightly lower values, they have a median
$\tau$ that is
0.75 that of the BC03 $\tau$.  This difference is driven mostly by
the EL-galaxies which have a $\tau$ of 0.66 times the BC03 value;
whereas the NEL-galaxies have a $\tau$ 0.83 times the BC03 value.  This suggests that in the M05 models,
the SFH  of EL-galaxies is burstier than for the BC03
models; however, we note that $\tau$ is one of the poorest constrained
parameters when fitting the SEDs alone.  
\newline\indent
If we compare the \age~ of the M05 models to the BC03 models
we find that they are lower by a factor of 0.65.  Again, this ratio is
similar to the ratios of 0.58 and 0.51 found by Maraston et al. (2006)
and Wuyts et al. (2007), respectively.  However, unlike M$_{\mbox{\scriptsize star}}$, the age differences are particularly pronounced
between the EL- and NEL-galaxies.  The ratio of ages of the NEL-galaxies
between the M05 models and BC03 models is 1.03, but for the
EL-galaxies the ratio is 0.42.  The age of the stellar population is
closely tied to the SFH, as the shape of the SED is primarily determined by the
number of e-folding times of the SFH, which is effectively the ratio
of $t$/$\tau$.  Given that there are differences in $\tau$ between the models, a
similar trend in $t$ and \age~ is expected in order to preserve the number of e-folding
times.   
\newline\indent
Due to these differences in $\tau$ and \age~ we 
also expect a significant difference in the SFRs for EL- and
NEL-galaxies.  The median ratio of SFRs is 1.12; however,
it is 1.58 and 0.33 for EL- and NEL-galaxies, respectively (the clear
trend can also be seen in Figure 11).  In the M05 models NEL-galaxies have significantly less SF than
the BC03 models would predict, whereas EL-galaxies are more active
than the BC03 models would predict, leading to a more bimodal galaxy
distribution, if galaxies are classified by SFR.
\newline\indent
The final parameter we consider is the A$_{v}$ of the
galaxies.  The median $\Delta$A$_{v}$ for the sample (where
$\Delta$A$_{v}$ $\equiv$ A$_{v,M05}$-A$_{v,BC03}$) is 0.0; however,
$\Delta$A$_{v}$ = 0.1 for the EL-galaxies, and $\Delta$A$_{v}$ = -0.4
for the NEL-galaxies.  The M05 models require far less dust
to fit the NEL-galaxies' SEDs, and to first order this
explains why the NEL-galaxies have smaller M$_{\mbox{\scriptsize star}}$ when fit with the
M05 models, despite the fact that they have similar $\tau$ and
\age~.  Maraston et al. (2006) also showed that when fitting
without dust the M05 models produce fits with significantly lower
$\chi_{r}^2$  than the BC03 models, suggesting they require less
dust to reproduce the SEDs of their sample as well.    
\newline\indent
In summary, galaxies fit with the M05 models have a median values of 0.63$\cdot$M$_{\mbox{\scriptsize star}}$, 0.65$\cdot$\age~, and
0.75$\cdot$$\tau$ with similar A$_{v}$s and SFRs when compared to
galaxies fit with the BC03 models, consistent with earlier studies.  However, breaking the
sample up into EL- and NEL-galaxies shows more interesting trends.  In
the M05 models
EL-galaxies have shorter timescales for SF and younger ages and this leads to
SFRs that are higher by a factor of 1.58.  Conversely, for NEL-galaxies the values
of $\tau$ and \age~ are fairly similar between the M05 models
and the BC03 models; however the A$_{v}$ is 0.4 mag less in the M05 models, and the SFRs
are only 0.33 of the BC03 value.  Overall, the M05 models produce a
more strongly bi-modal population of galaxies at $z \sim$ 2.3.
EL-galaxies have a shorter-timescale for SF, and larger SFRs; whereas
NEL-galaxies have less dust and are more quiescent.
\begin{figure}
\plotone{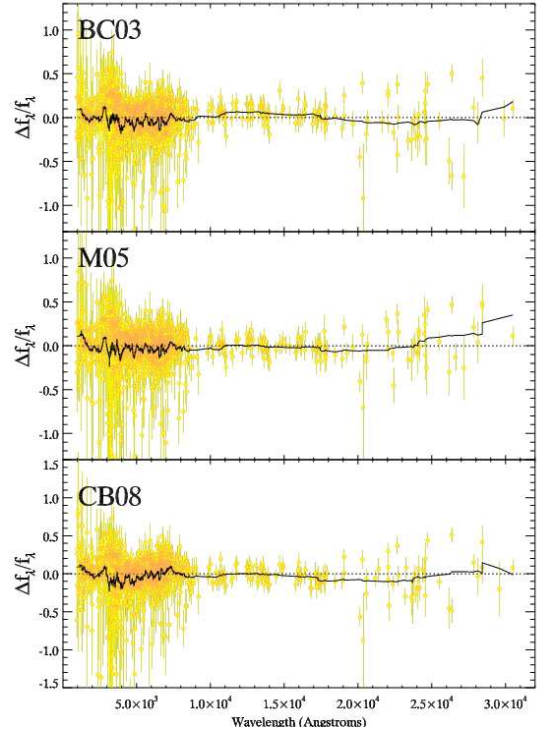}
\caption{\footnotesize Residuals from the SED fits using the BC03,
  M05, and CB08 models corrected to the rest-frame.  The residuals
  are the fractional residual where $\Delta f_{\lambda}/f_{\lambda} \equiv
  f_{obs}-f_{model}/f_{obs}$.  The black line represents a running
  average of the nearest 30 data points.  The running average
  suggests that the M05 and CB08 models may describe the SEDs of the
  galaxies better in the range 1.0 x 10$^{4}$ \AA~$< \lambda <$ 1.8 x
  10$^{4}$ \AA; however, the improvement is $\sim$ 5-10\%, well
  within the photometric errors.}
\end{figure}
\subsubsection{Differences in Stellar Populations from the CB08 Models}
\indent
Parameters determined from the CB08 models also have marked
differences from those determined with the BC03 models.  Similar to
the M05 models, M$_{\mbox{\scriptsize star}}$ from the CB08 models is lower by a factor of
0.74.  Dividing these into the EL- and NEL-galaxy categories we find
that for the EL-galaxies M$_{star,CB08}$/M$_{star,BC03}$ = 0.77 and
NEL-galaxies M$_{star,CB08}$/M$_{star,BC03}$ = 0.70.  This demonstrates that
similar to the M05 fits, the differences in M$_{\mbox{\scriptsize star}}$ are not a
strong function of SED type and are a systematic offset between the
models.  Of interest, this difference in M$_{\mbox{\scriptsize star}}$ is less dramatic
than the factor of 0.32 found by Bruzual (2007) when fitting the
galaxies used in Maraston et al. (2006); however, they compared
M$_{\mbox{\scriptsize star}}$ using fits without including dust extinction.  Clearly, once dust is
included in the fitting the differences in M$_{\mbox{\scriptsize star}}$ are not as
significant.  
\newline\indent
The median $\tau$ is similar between the CB08 and BC03 fits;
however, like the M05 fits, its dependence is strongly bimodal.  In the CB08 fits, the
EL-galaxies have $\tau$'s which are 0.66 that of the BC03 $\tau$'s.
This behavior is similar to the M05 EL-galaxies, which also have
shorter SF timescales.  The NEL-galaxies actually have larger
$\tau$'s in the CB08 models compared to the BC03 models by a factor of
1.25.  
\newline\indent
Unlike the M05 models, the \age~of galaxies in the CB08 models are
actually older than the BC03 models, with a median value 1.24 times larger.
This dependence is again strongly bimodal, with the EL-galaxies having
a \age~0.96 times the BC03 models, but the NEL-galaxies are
older, with \age$_{,CB08}$  = 1.46\age$_{,BC03}$.
Interestingly, this gives the same relative behavior as the M05
models, i.e., the EL-galaxies get younger {\it relative} to the
NEL-galaxies; however, with the CB08 models it is the NEL-galaxies
that get older while the EL-galaxies stay the same age. In the M05 models, the EL-galaxies
get younger, but the NEL-galaxies stay the same age.  
\newline\indent
The median extinction is lower in the CB08 models compared to the BC03
models, being $\Delta$A$_{v}$ = -0.4 for all galaxies.  The effect is
stronger for NEL-galaxies, which have $\Delta$A$_{v}$ = -0.6, compared
to the EL-galaxies which have $\Delta$A$_{v}$ = -0.2.
\newline\indent
Overall, the net effect of using the CB08 models compared to the BC03
models is that  the mean M$_{\mbox{\scriptsize star}}$ of galaxies is lower by a
factor of 0.74, and perhaps more importantly, the galaxy population
becomes more bimodal in several parameters.
This is similar to the result from the M05 models; however, the
bimodalism is driven by movement in different types of galaxies.  In the CB08 models compared to the BC03
models EL-galaxies have lower $\tau$'s, but similar \age~'s,
giving them a slightly burstier SFH.  This is different than the M05
models which are both significantly younger and have even smaller
$\tau$'s.  In the CB08 models the NEL-galaxies have both longer
$\tau$'s, older ages, and surprisingly, less dust.  In the M05
models, the NEL-galaxies had similar $\tau$'s, and \age~ as the
BC03 models.
\newline\indent
If we divide the galaxy
population by SFR in both the M05 and CB08 model it is more bimodal than in the BC03 models.  For the M05
models, the change is dominated by the EL-galaxies, which are notably
younger and bustier with higher SFRs than their BC03 counterparts.  For
the CB08 models, the bimodality is driven by the NEL-galaxies which
become older, with longer $\tau$ and smaller SFRs than their BC03
counterparts. 
\begin{figure*}
\plotone{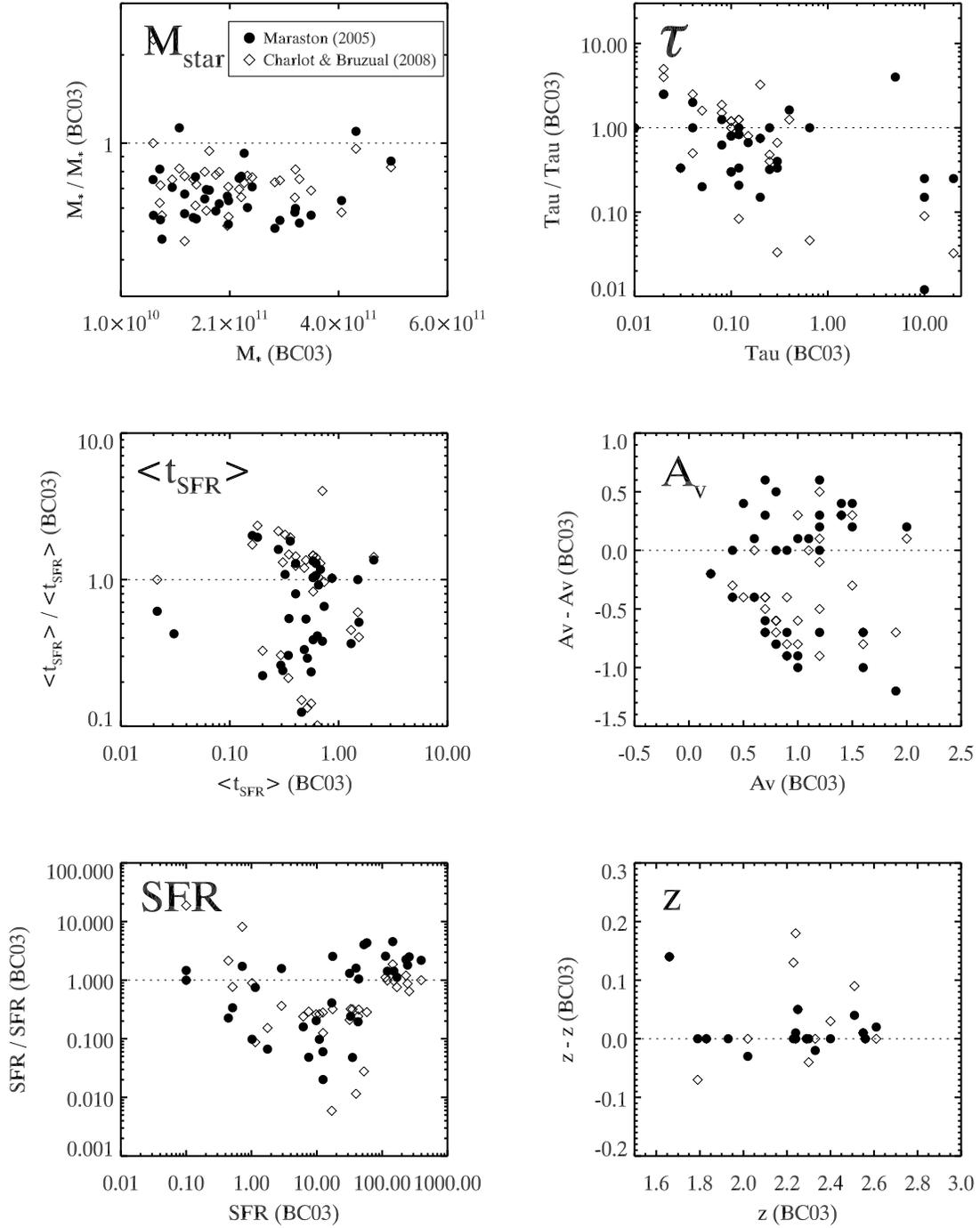}
\caption{\footnotesize Ratio of best-fit CB08 (open triangles)
  and M05 (filled circles) SED parameters to the best fit BC03 SED
  parameters as a function of the BC03 parameters.  The median M05
  fit is 0.63 of M$_{\mbox{\scriptsize star}}$, and 0.65 of \age~compared to the BC03
  fits, while the $z$, SFR, and A$_{v}$ are similar.  The median CB08
  fit is 0.74 of M$_{\mbox{\scriptsize star}}$, and 1.24 of \age, and has 0.4 mag
  less A$_{v}$ compared to the BC03 fits, while the $z$, SFR are similar }
\end{figure*}
\begin{figure*}
\plotone{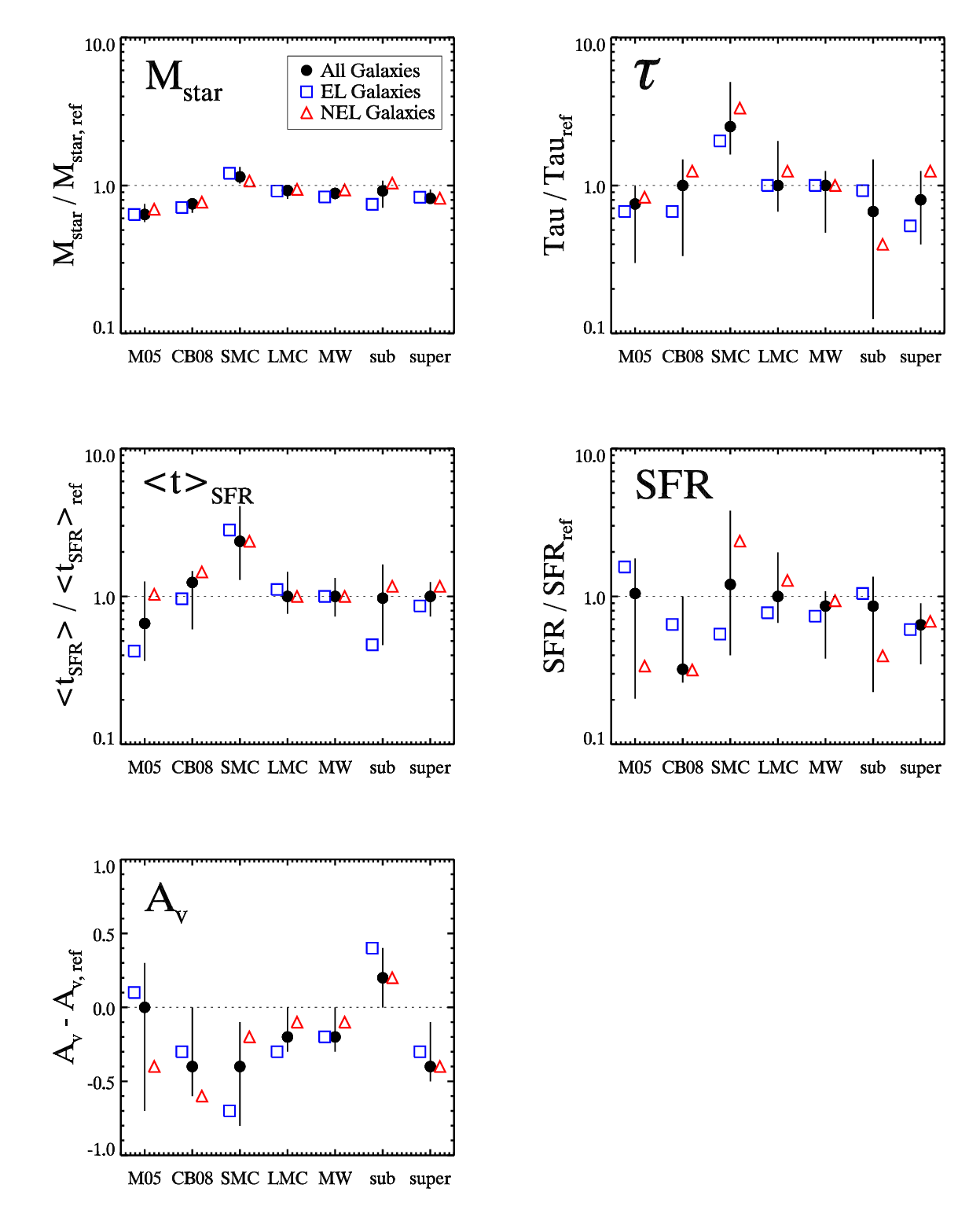}
\caption{\footnotesize Difference in the median value of parameters
  fit using the various SPS codes, dust laws, and metallicities (see
  text) compared to the ``control'' model.  Open blue squares
  represent galaxies with detectable emission lines and open red
  triangles represent galaxies without detectable emission lines.  The filled
  black circles are the median for all galaxies.  The lines denote the range that contains 50\% of the galaxies.  Most of the largest  systematic offsets are from the SPS codes, and their most significant effect
  is on M$_{\mbox{\scriptsize star}}$, and \age, the two parameters best determined by
  the SED fitting.}
\end{figure*}
\section{Systematic Effects on Stellar Population Parameters from
  Metallicity and Dust Law}
As we have shown in the previous section, SED fits using the current generation
of SPS codes provide stellar populations parameters that suffer from significant systematic effects.  In this section we quantify how large
these systematic effects are relative to other known systematics such as the choice of dust
law and metallicity.
\newline\indent
We allow three dust laws to redden the SEDS:  the Milky
Way (MW) and Large Magellanic Cloud (LMC) dust laws determined by
Fitzpatrick et al. (1986),  and the Small Magellanic Cloud (SMC) dust law determined by Pr\'{e}vot et
al. (1984).  We also incorporate two new metallicities into the fits, a
subsolar (Z = 0.2 Z$_{\odot}$), and supersolar (Z =
2.5 Z$_{\odot}$) population.  For the BC03 
models these are the m42, and m72 stellar populations, respectively.  
\newline\indent
In an attempt to isolate variables we use the BC03 models as our control
SPS code. When comparing dust laws we assume solar metallicity as our
control metallicity, and when comparing metallicities we assume a Calzetti et
al. (2000) dust law as the control dust law.   It is possible that using a
different control model will affect some of these  comparisons;
however, the large number
of potential permutations allowed in choosing the control model (there are 36
possible combinations of SPS code, dust law, and metallicity)  require us to chose only one in order to
make the interpretation of systematic effects tractable.
Moreover, the most commonly used models for determining
stellar population parameters thus far have been the BC03 models with solar
metallicity and the Calzetti dust law, so systematic changes in parameters
relative to this model are probably the most interesting for
understanding the implications of assumptions made in previous work.
\newline\indent
We first examine the effect of using the various dust laws on the
parameter estimates.  The ratios of parameters determined from fits using the SMC, MW,
and LMC dust laws compared to those from the BC03\_m62+Calzetti
control model are listed in Table 5 and are plotted in
Figure 11.  The open red triangles and open blue squares represent the medians of the NEL-galaxies and
EL-galaxies, respectively, and the black points denote the
medians of the whole sample.  The error bars on the black points denote the range spanned by 50\% of the sample.   Figure 11 shows that when parameters
determined using the LMC and MW dust laws are compared to those determined using the Calzetti law, the only
parameter that is significantly different is the
A$_{v}$.  Parameters such as M$_{\mbox{\scriptsize star}}$, $\tau$, \age~,
and SFR only change systematically by of order $\sim$ 10\%, and therefore appear to be robust regardless of whether the MW,
LMC, or Calzetti dust laws are used.  Furthermore,
the changes do not appear to depend on the
EL- and NEL-galaxy classification.  As pointed out by F\"{o}rster Schreiber
et al. (2004), the lack of significant changes in parameters between
these three dust laws is not surprising because they are fairly
similar in shape.  The
only notable difference is that the Calzetti law lacks the
2175\AA~absorption feature and that the ratio of total to
selective absorption is slightly higher,
R$_{v}$ = 4.05, compared to R$_{v}$ = 3.1 for the MW and LMC laws.
\newline\indent
Using the SMC dust law does cause some significant changes to several
parameters.  The median $\tau$ and \age~ are larger by a factors
of 2.5, and 2.8, respectively, and the median A$_{v}$ is lower by 0.4
mag.  The median SFR is similar; however, this results from a
coincidental cancellation caused by the NEL-galaxies having SFRs a factor of $\sim$ 2 higher, and EL-galaxies having
SFRs a factor of $\sim$ 2 lower.   
The SMC dust law has by far the strongest attenuation of all four dust laws at $\lambda$ $<$
is 2000\AA~rest-frame and therefore it requires stellar populations with
more rest-frame UV flux  to describe the SEDs.  This leads to a
preference for longer timescales for SF and older ages so that there
is a   contribution from both a young and old population simultaneously.
Because of the shape of the SMC dust law, the fits
prefer intrinsically redder stellar populations (i.e., those with
older ages and longer $\tau$) and less dust, rather than short bursts with
more dust.  The preference for older ages and less dust would probably diminish if
we used the SMC law with a subsolar metallicity population, which is
intrinsically bluer and would allow for more dust and shorter
timescales for SF.
\newline\indent
Next we examine the effect on the parameters caused by varying the
metallicity while holding the SPS model and dust law fixed.  The
median values of the parameters determined from those fits compared to the control models fits are listed in Table 6 and are
also plotted in Figure 11.  The choice of metallicity affects more parameters
than does the choice of dust law; however, it is still a
modest effect.  The $\tau$ of the stellar populations
is lower for both metallicities; however, the dependence for EL- and
NEL-galaxies is reversed.  The A$_{v}$ depends on metallicity in the predictable way, with subsolar
metallicity requiring more dust, and supersolar requiring less.
\newline\indent
Perhaps what is most interesting about the variations in parameters
caused by choice of dust law and metallicity is that other than the SMC dust law, their net systematic effect on M$_{\mbox{\scriptsize star}}$, and \age~, the two parameters that
are the best-determined using SED fitting ($\S$ 4.5) is fairly small, $\sim$ 10-20\%.
They do have larger effects on $\tau$ and SFR; however, these are the parameters that are
most poorly determined  using the SED fitting.
\newline\indent
One  disconcerting aspect of Figure 11 is that it shows clearly that it is the SPS codes
that have the largest systematic effects on most parameters, and
most importantly,  that {\it the largest systematic changes occur on the
parameters that are best determined by the SED fitting}: M$_{\mbox{\scriptsize star}}$,
\age~, and A${v}$.  So, despite the fact that using the best
 data currently available we can  constrain the M$_{\mbox{\scriptsize star}}$,
\age~, and A$_{v}$ of individual galaxies to $\sim$ 0.1 dex, 0.3 dex, and $\pm$ 0.3 mag,
respectively ($\S$ 4.5), it appears that systematic errors in these
parameters from the various SPS codes dominate over random errors for all three
parameters as they  cause systematic uncertainties of roughly $\pm$ 0.3 dex in M$_{\mbox{\scriptsize star}}$ and \age~, and
$\pm$ 0.4 mag in A$_{v}$.  This suggests than unless these
systematic discrepancies between the SPS models are resolved, more
multiwavelength data on the stellar component of the galaxy
SED will not provide better constraints on the evolution of the stellar mass density in the universe, or the peak
epoch of the formation of stars in the universe.  Longer wavelength
data that probes emission from dust (e.g., MIPS, ALMA) will almost
certainly improve the SFR estimates; however,
parameters determined from the
stellar SED are now  limited by systematic errors.  Adding to the challenge is the analysis
in $\S$ 5.2 that shows that even with the best constrained SEDs of
young stellar populations currently available we will not be able to
``fit out'' these systematic problems, as all three models describe the SEDs
equally well.
\section{Implications: Confirmation of the Existence of a Substantial
  Population of Quiescent Galaxies at $z \sim$ 2.3}
\indent
Using this sample of galaxies Kriek et al. (2006) have argued that $\sim$
50\% of K-selected galaxies at $z \sim$ 2.3 have strongly suppressed star
formation based on their lack of observed emission lines and the fact that the rest-frame UV through optical SED was best described by an old quiescent population.
Although they had the high-resolution NIR spectroscopy with which to
model the stellar populations, Kriek et al.\ lacked the rest-frame NIR data
from IRAC, without which it can be challenging to distinguish between young-and-dusty and old-and-quiescent populations
(e.g., Labb\'{e} et al. 2005; Papovich et al. 2006;
Wuyts et al. 2007, $\S$ 4).  Although unlikely, they could not completely rule out the possibility that some fraction of the systems
without emission lines could be highly extincted star forming galaxies rather than old and
quiescent galaxies, and that the lack of emission lines and red optical
colors are the result of significant dust obscuration.
Here we examine the stellar populations from the SED fits of the
EL and NEL galaxies to verify that the
SEDs of the galaxies without emission lines are still consistent with them being systems with strongly suppressed star formation.
\newline\indent
In the top panels Figure 12 we plot the specific star formation rate
(SSFR), defined as SFR/M$_{\mbox{\scriptsize star}}$, 
as a function of M$_{\mbox{\scriptsize star}}$ for the EL galaxies (blue) and
NEL galaxies (red).  Candidate AGNs selected using the emission-line
diagnostics in Kriek et al. (2007) as well as PLGs ($\S$ 5.2) are
indicated with an open diamond.  As we showed in  $\S$ 6, the
SPS codes have the largest systematic effects on  M$_{\mbox{\scriptsize star}}$ and
SFR, and therefore we plot three versions of the diagram in Figure 12, one each for values
determined using the three SPS codes.  
\newline\indent
It is clear from Figure 12 that the majority of galaxies without
emission lines are found at the lowest SSFRs, whereas those with emission
lines are found at the highest SSFRs.  There are some notable
model-dependent differences, with the M05 and CB08 models implying that the population
of galaxies at $z \sim$ 2.3 has a more bimodal distribution of SSFRs, whereas the
BC03 models suggest more of a continuum in SSFRs.  This is apparent
in the bottom panels of Figure 12, where we plot histograms of the
SSFRs.  
\newline\indent
If we adopt the Kriek et al. (2008b) definition of quiescent systems as those that have SSFRs $<$ 0.05 Gyr$^{-1}$ (i.e., those that
will increase their M$_{\mbox{\scriptsize star}}$ by $<$ 5\% in the next Gyr if the SFR
remains at the same value), then we find that 13, 17, and 22 of the 34 galaxies
(38\%$\pm$12\%, 50\%$\pm$15\%, 65\%$\pm$18\%) would be classified as quiescent based on the SED parameters from the
BC03, M05, and CB08 models, respectively.  These numbers are
in good agreement with the 45$^{+18}_{-12}$\% determined by Kriek et
al. (2006) based on the fraction of galaxies without emission lines and  SED fitting.
Despite the good agreement, there are a few subtle issues in
the emission line classification.  There are 16 galaxies
without emission lines,  and of these, 10, 12, and 15 of these would be considered
quiescent based on their SED parameters from fits to the BC03,
M05, and CB08 models, respectively.  Therefore, 
the majority of systems without emission lines are quiescent;
however, there is a non-negligible population ($\sim$ 6\% - 33\%) of
galaxies without emission lines that do have some ongoing
obscured star formation ranging from $\sim$ 10-850 M$_{\odot}$ yr$^{-1}$.  Coincidentally, these
few galaxies are canceled out in the overall quiescent fraction by a small fraction of 
emission line galaxies that are likely to be quiescent based on their SED, but probably have an AGN
component that is responsible for the emission lines.
\newline\indent
Taken together, these results suggest that the classification of $z
\sim$ 2 galaxies into star forming and
quiescent categories based on the presence/lack of emission lines and without any information on their SED is probably correct
for the majority of systems ($\sim$ 70-90\%); however, there is some cross
contamination between the two categories from dusty star forming galaxies without emission
lines, and quiescent galaxies with an emission line AGN.  Within our
sample these two populations appear to be small, and
 equally abundant and therefore cancel each other when estimating the
quiescent fraction.  As a result, the fraction of quiescent galaxies
is roughly the same using the SED fits or the emission line
classification.  Our sample is fairly small, so it is not clear if the
same cancellation would occur for larger samples.
\newline\indent
Although the contamination in the emission line classification
suggests we should appeal to the SED fits for a more robust
classification, that method is not completely ``clean'' either: the
quiescent fraction ranges by roughly a factor of 2 depending on which SPS code is used to
model the SEDs.  Considering this systematic effect and using the 1$\sigma$ confidence
intervals from the models, we can conclude that the fraction of K-selected galaxies at
$z \sim$ 2.3 that are classified as quiescent based on their SEDs ranges from as low as 26\% to as high as 83\% at 68\%
confidence, in good agreement with the value determined from the
emission line fraction by Kriek et al. (2006).  Clearly this
uncertainty is large; however, even within our modest sample of 34
galaxies, the statistical uncertainties are already smaller than the
systematic uncertainty from the SPS codes.
\newline\indent
The histograms in Figure 12 suggest a
more bimodal distribution of SSFRs when using the M05 and CB08 models
compared to the BC03 models.  Kriek et al. 2008b have shown that there is
already a developed red sequence in the (U-B)$_{rest}$
vs. M$_{\mbox{\scriptsize star}}$ plane for this sample of galaxies
and that most of the galaxies on the red sequence have low SSFRs as
inferred from SED fits using the BC03 models.  It is enticing to
prefer SPS models that show the same bimodality in SSFRs as in
rest-frame color; however, given that the current sample is small, and the
random uncertainties in the SSFRs are large (factor of 2-3) it is
better to defer such a comparison to a larger sample of galaxies
(e.g., van Dokkum et al. 2009). It will be interesting to see how 
the SSFRs of galaxies in the larger sample relate to their rest-frame
colors, which will be a much more accurately determined quantity using
those data.
\begin{figure*}
\plotone{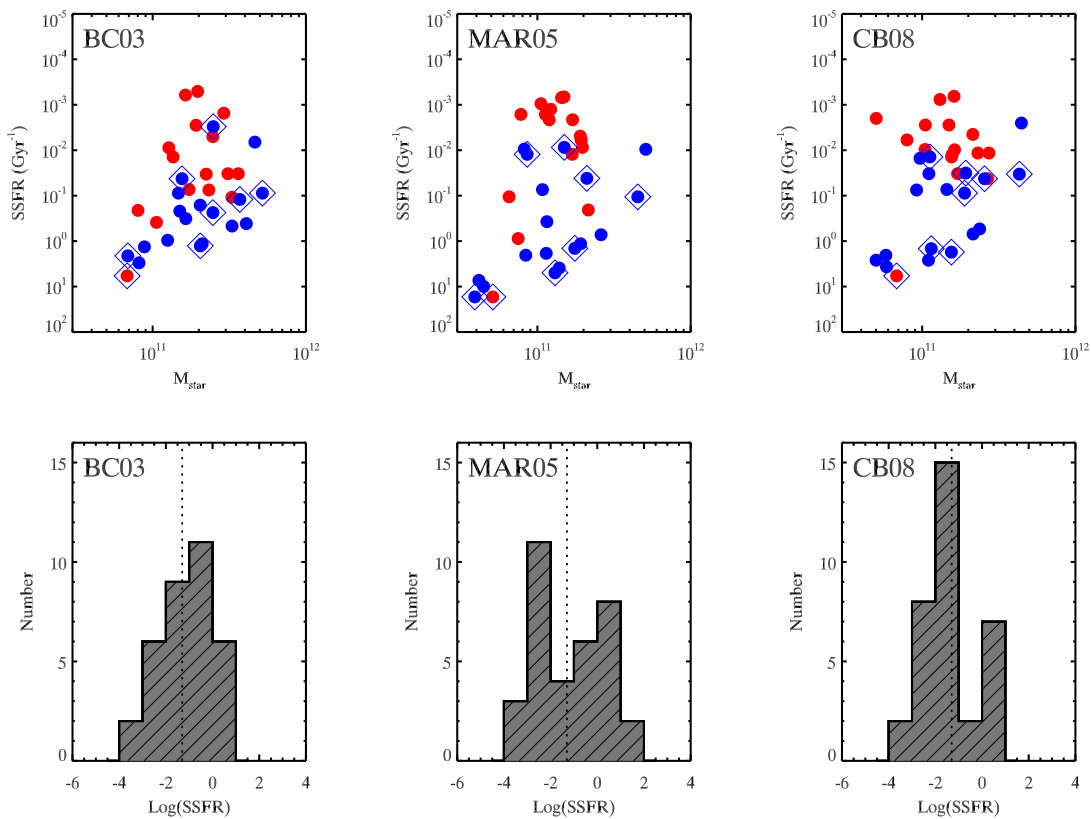}
\caption{\footnotesize Top Panels: Specific star formation rate
  (SSFR $\equiv$ SFR/M$_{\mbox{\scriptsize star}}$) as a function of
  M$_{\mbox{\scriptsize star}}$ determined from fits to 
  the BC03, M05, and CB08 models.  Galaxies with detected
  emission lines are plotted in blue, those without detectable
  emission lines are plotted in red.  Candidate AGN (see text) are
  indicated using open diamonds.  The three SPS models produce
  significantly different distributions in this parameter space;
  however, galaxies without emission lines typically have much lower
  SSFRs than those with emission lines.
  Bottom Panels: Histograms of the SSFR determined using models from
  the three SPS codes.  The dashed line represents our definition of a
  quiescent galaxy (SSFR $<$ 0.05 Gyr$^{-1}$).  Based on this
  definition, 38\%$\pm$12\%, 50\%$\pm$15\%, 65\%$\pm$18\% of our
  sample would be consider quiescent using the BC03, M05, and CB08
  models, respectively.}
\end{figure*}
\section{Summary}
In this paper we performed SED fitting on a sample of 34 K-selected
galaxies at $z \sim$ 2.3.  With NIR spectroscopy  as well as
  deep photometry in thirteen broadband filters, these galaxies have arguably the best constrained
SEDs of $z >$ 2 galaxies currently available. With this sample we studied the
importance of the rest-frame NIR in determining stellar population
parameters, the systematic differences between the current
generation of SPS codes, as well as quantify the systematic effects in determining
stellar population parameters of young galaxies from the choice of SPS
code, dust law, and metallicity.  Our
main results are summarized below:
\newline
1. Using the BC03 models with solar metallicity, the Calzetti dust
law, and a Salpeter IMF as a control model, fits with and without the
IRAC data show that including the rest-frame NIR in the SED fitting
of young stellar populations
provides improved constraints on nearly all stellar population
parameters.  In particular, $\tau$, A$_{v}$, and SFR are improved by
factors of $\sim$ 1.5 - 2.0, and M$_{\mbox{\scriptsize star}}$ is improved by a factor of 1.3.
\newline
2.  The IRAC data is particularly helpful when only broadband data (no
spectroscopy) is available at other wavelengths.
In particular, constraints on M$_{\mbox{\scriptsize star}}$,
SFR, and A$_{v}$ improve by factors of 4, 2.5, and 0.5 mag, respectively.
\newline
3. IRAC data only modestly improves the photometric redshifts of 1.6
$< z <$ 2.9 galaxies.  The improvement only occurs for galaxies with
low S/N JHK photometry, and only when all four IRAC bands are included
in fitting.  Our comparisons show that deep JHK data is far more
valuable than IRAC data when determining  the $z_{\mbox{\scriptsize
    phot}}$ of galaxies at 1.6 $< z <$ 2.9.
\newline
4. Neglecting systematic effects from choice of SPS code, dust law,
metallicity, and IMF, and using our entire photometric dataset (U$\rightarrow$8$\micron$+NIRspec), we find
that the best possible constraints on M$_{\mbox{\scriptsize star}}$, \age, and
A$_{v}$ of individual galaxies in our sample from SED fitting are
$\pm$ 0.12 dex,
$\pm$ 0.26 dex, and $\pm$0.3 mag, respectively.  Even with these well-constrained
SEDs, the $\tau$ and SFR of
individual galaxies are still uncertain to factors of 2-3.  Using only the broadband data
(U$\rightarrow$8$\micron$) as well as $z_{\mbox{\scriptsize phot}}$, the constraints on M$_{\mbox{\scriptsize star}}$, \age, and
A$_{v}$ are $\pm$ 0.11 dex, $\pm$0.36 dex, and $\pm$0.3 mag, nearly as good as when the NIR
spectroscopy is included in the fitting.  
\newline
5. Comparison of the quality of fits provided by the BC03, M05, and
   CB08 models using several methods shows that all three provide
   equally good descriptions of the data.  There is tenuous evidence
   that the BC03 and CB08 models describe the rest-frame UV through
   optical SED slightly better than the M05 models; however, the latter models may provide a slightly better
   description of the rest-frame NIR SED.  Taken together, there is
   little evidence that one set of models provides an overall better
   description of the SEDs of young massive galaxies based on the
   quality of fit.  Furthermore, given that the differences in the
   predicted fluxes of the best-fit models at all wavelengths is $<$5\%, it is unlikely that  the current
   generation of data on the SEDs of young, massive galaxies can
   inform the models enough to aid in their improvement.
\newline
6. Comparison of the systematic changes in stellar population
   parameters determined using the three SPS codes shows that when we
   fit with the M05 models the M$_{\mbox{\scriptsize star}}$
   and \age~of galaxies go down by factors of 0.63 and 0.65,
   respectively relative to the BC03 fits.  Galaxies
   fit with the CB08 models have 0.75 the M$_{\mbox{\scriptsize star}}$ of the BC03 fits,
   but are on average older by a factor of 1.24.   Notably, the
   changes in most parameters appear to
   depend on whether galaxies have detectable emission lines.  Because
   they affect the SFRs rates of SF and NEL-galaxies differentially, both
   the M05 and CB08 models provide an overall stellar population that
   has ``burstier'' episodes of star formation than the BC03 models,
   as well as suggest a galaxy population that is more bimodal in terms of the
   overall SSFRs.
\newline
7. Comparison of systematic changes in stellar population parameters
   caused by choice of dust law,  metallicity, and SPS code show that
   the latter are the most significant of the three.  The systematic
   effects on M$_{\mbox{\scriptsize star}}$, and \age, the best determined
   parameters from the SED fitting, from different dust laws and metallicities are generally
small, of the order of 10-20\% on M$_{\mbox{\scriptsize star}}$, and \age.  Conversely, the
   systematic effects from the SPS code on these parameters are most significant.  These systematic effects are larger than the
   random errors from the SED fitting and suggest that further
   progress in determining the evolution of the stellar mass density
   and the peak epoch of SF in the universe will require significant
   improvements in the models.
\newline
8. Comparison of the SSFRs determined using the SED fitting for
galaxies with and without detectable emission lines shows that most
galaxies without emission lines are quiescent (66\% to 96\%
depending on SPS model); however, the remainder are likely to star
forming galaxies that have obscured emission lines.  In our sample, the
obscured emission line population is roughly equal to the population
of galaxies that are quiescent based on their SED, but $have$
detected emission lines, probably from
an AGN.  Therefore, despite some small cross contamination of quiescent and
active galaxies in the emission line
and non-emission line categories, respectively, the overall fraction of K-bright, $z \sim$ 2.3 galaxies that would be considered
quiescent based on their SED is $\sim$ 50\% (26\% - 83\% depending on
SPS model), nearly identical to the
fraction without emission lines.
\acknowledgements
We are grateful to C. Maraston for providing stellar population synthesis models in binary format, and S. Charlot for providing the unpublished CB08 stellar population synthesis models.  We thank the members of the MUSYC collaboration for their contribution to this research.
A.M. acknowledges postdoctoral support from the National Science and
   Engineering Research Council (NSERC) from a PDF fellowship.  D.M. is supported by
NASA LTSA NNG04GE12G. The authors acknowledge support from NSF CAREER
AST-0449678, and Spitzer/JPL grants RSA 1277255, RSA 1282692, and RSA 1288440.

Based on observations obtained at the Gemini Observatory, which is operated by the
Association of Universities for Research in Astronomy, Inc., under a cooperative agreement
with the NSF on behalf of the Gemini partnership: the National Science Foundation (United
States), the Science and Technology Facilities Council (United Kingdom), the
National Research Council (Canada), CONICYT (Chile), the Australian Research Council
(Australia), Ministerio da Ciencia e Tecnologia (Brazil) and SECYT (Argentina)

This work is based in part on observations made with the Spitzer Space
Telescope, which is operated by the Jet Propulsion Laboratory, California
Institute of Technology under a contract with NASA.

\begin{appendix}
\section{Determination of Systematic and Random Errors in Stellar
  Population Parameters}
In Figure 13 we plot histograms that show the distributions of
parameters determined with the U$\rightarrow$K,
U$\rightarrow$z$^{\prime}$+NIRspec, and U$\rightarrow$8$\micron$ data
relative to those determined with the
U$\rightarrow$8$\micron$+NIRspec data.  The mean and median of each
distribution is listed in the upper right of each panel.  These demonstrate
the systematic differences in the parameters with various datasets
relative to the U$\rightarrow$8$\micron$+NIRspec data, and are the
parameters listed in Table 1. 
\newline\indent
\begin{figure*}
\plotone{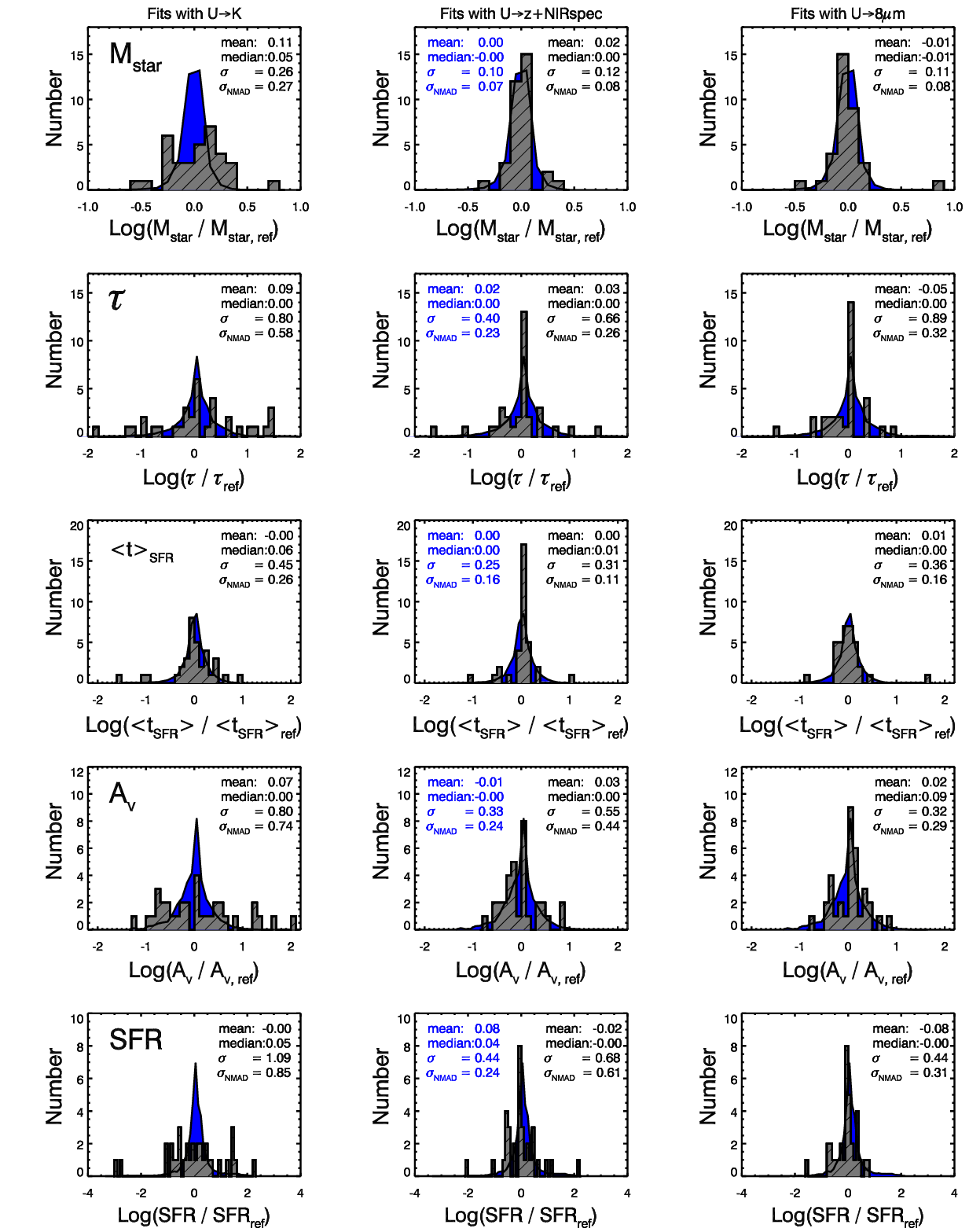}
\caption{\footnotesize Left Panels: Comparison of stellar population
  parameters determined using only the broadband U$\rightarrow$K photometry to
  those determined using all of the data (U$\rightarrow$8$\micron$+NIRspec). Middle Panels: 
  As the left panels, but comparing parameters from broadband plus NIR
  spectroscopy (U$\rightarrow$z$^{\prime}$+NIRspec) to those
  determined with U$\rightarrow$8$\micron$+NIRspec.  Right Panels: As left panels but
  comparing parameters from broadband optical and IRAC photometry
  (U$\rightarrow$8$\micron$) to those determined with U$\rightarrow$8$\micron$+NIRspec.
  The blue curves in each panel are the spread in the distribution
  that would be expected based on the Monte Carlo errors determined
  using the U$\rightarrow$8$\micron$+NIRspec data.  If the parameters
  determined from a subset of the data have the same accuracy and precision as those determined
  from the entire U$\rightarrow$8$\micron$+NIRspec data set, the blue
  and grey histograms should be similar.  The differences in the
  accuracy of the parameters using the various datasets can be determined by comparing the
  $\sigma$ or the $\sigma_{\mbox{\scriptsize NMAD}}$ accross the different datasets.  The addition of rest-frame NIR data from
  IRAC provides significant improvement in most parameters,
  particularly compared to those determined from U$\rightarrow$K photometry alone. 
  The IRAC data provides better constraints than the NIR spectroscopy on the parameters M$_{\mbox{\scriptsize star}}$, A$_{v}$, and SFR; whereas the NIR spectroscopy provides better constraints on $\tau$ and \age.  }
\end{figure*}
The standard deviation ($\sigma$) of the distributions are listed in the upper right of
each panel in Figure 13.  These serve as an indication of the scatter between the
best fit parameters, even though some of the distributions are
non-Gaussian.  Standard deviations are sensitive to large outliers and
therefore we also compute the normalized median absolute
deviation ($\sigma_{\mbox{\scriptsize NMAD}}$)  to compare if the $\sigma$ is
representative of the true scatter in parameters or is dominated by a few large outliers.
For a Gaussian error distribution, $\sigma_{\mbox{\scriptsize NMAD}}$ = rms.
\newline\indent
Because the histograms are ratios of parameters computed with
different datasets, the scatter is caused by the combination of errors in both datasets.  To determine whether the given data improves the
quality of the fits we must assess how much of the scatter is caused by
uncertainties in U$\rightarrow$K, U$\rightarrow$z$^{\prime}$+NIRspec, or U$\rightarrow$8$\micron$ fits and how much is
intrinsic to the SED fitting with the comparison sample (U$\rightarrow$8$\micron$+NIRspec).
The intrinsic contribution is estimated by simulating a new distribution of the
U$\rightarrow$8$\micron$+NIRspec fit parameters based on the error estimates listed
in Table 1.  We generate this distribution by drawing 200 new values of each parameter, for
each galaxy, using the uncertainties and compare the new parameters to
the original ones.  The resulting ``expected'' distributions are plotted in blue in
Figure 13 with the measured mean, median, $\sigma$ and $\sigma_{\mbox{\scriptsize NMAD}}$
also listed in blue.  These values
are the scatters used on the X-axis of Figure 13.  Where the grey and blue histograms differ in mean, median, or sigma, the
IRAC data provides additional constraints.  
The main results from the histograms in Figure 13 are summarized in $\S$ 4.3 and $\S$ 4.4.
\end{appendix}

\LongTables
\begin{deluxetable}{lllllllrll}
\LongTables
\tabletypesize{\footnotesize}
\scriptsize
\tablecolumns{10}
\tablecaption{SED Modeling Parameters}
\tablewidth{7.3in}
\tablehead{\colhead{Name} & \colhead{ Model } & \colhead{ $z$ } &  \colhead{ age } &
\colhead{$\tau$} & \colhead{A$_{v}$} & \colhead{M$_{*}$} &
\colhead{SFR} & \colhead{$<t_{SFR}>$} & \colhead{$<\chi_{r}^2>$}
\\ 
\colhead{} & \colhead{} & \colhead{} & \colhead{Gyr} & \colhead{Gyr} & \colhead{mag} &
\colhead{10$^{11}$M$_{\odot}$} & \colhead{$M_{\odot}$ yr$^{-1}$} & \colhead{Gyr}
\\
\colhead{(1)}& \colhead{(2)}& \colhead{(3)}& \colhead{(4)}&
\colhead{(5)}& \colhead{(6)} &
\colhead{(7)}& \colhead{(8)} & \colhead{(9)}
}
\startdata
1030-32 &  BC03 (no irac) & 2.23$^{+0.13}_{-0.02}$ & 1.00$^{+0.00}_{-0.20}$ & 0.01$^{+0.09}_{-0.00}$ & 0.1$^{+0.4}_{-0.1}$ & 1.34$^{+0.30}_{-0.24}$ & 0.0$^{+0.0}_{-0.0}$ & 0.99$^{+0.00}_{-0.21}$ & 1.56\\
 ... &  BC03 & 2.23$^{+0.13}_{-0.03}$ & 0.90$^{+0.60}_{-0.20}$ & 0.03$^{+0.12}_{-0.02}$ & 0.4$^{+0.3}_{-0.4}$ & 1.63$^{+0.25}_{-0.25}$ & 0.0$^{+0.5}_{-0.0}$ & 0.87$^{+0.60}_{-0.20}$ & 1.46\\
 ... &  MAR05 (no irac) & 2.23$^{+0.13}_{-0.03}$ & 1.00$^{+0.25}_{-0.20}$ & 0.05$^{+0.10}_{-0.04}$ & 0.0$^{+0.1}_{-0.0}$ & 1.16$^{+0.12}_{-0.20}$ & 9.6$^{+0.2}_{-9.6}$ & 0.95$^{+0.15}_{-0.18}$ & 1.92\\
 ... &  MAR05 & 2.23$^{+0.02}_{-0.14}$ & 0.90$^{+2.10}_{-0.10}$ & 0.01$^{+0.39}_{-0.00}$ & 0.0$^{+0.0}_{-0.0}$ & 1.05$^{+1.44}_{-0.13}$ & 0.0$^{+0.4}_{-0.0}$ & 0.89$^{+1.71}_{-0.13}$ & 1.89\\
 ... &  CB08 (no irac) & 2.36$^{+0.00}_{-0.14}$ & 0.90$^{+0.10}_{-0.20}$ & 0.01$^{+0.09}_{-0.00}$ & 0.1$^{+0.4}_{-0.1}$ & 1.46$^{+0.41}_{-0.24}$ & 0.0$^{+0.0}_{-0.0}$ & 0.89$^{+0.10}_{-0.20}$ & 1.55\\
 ... &  CB08 & 2.36$^{+0.00}_{-0.14}$ & 0.90$^{+0.10}_{-0.00}$ & 0.01$^{+0.07}_{-0.00}$ & 0.0$^{+0.0}_{-0.0}$ & 1.30$^{+0.11}_{-0.05}$ & 0.0$^{+0.0}_{-0.0}$ & 0.89$^{+0.10}_{-0.04}$ & 1.48\\
\hline
1030-101 &  BC03 (no irac) & 1.79$^{+0.07}_{-0.12}$ & 0.40$^{+1.85}_{-0.10}$ & 0.02$^{+0.27}_{-0.01}$ & 1.8$^{+0.3}_{-1.3}$ & 2.76$^{+0.62}_{-0.54}$ & 0.0$^{+13.5}_{-0.0}$ & 0.37$^{+1.32}_{-0.09}$ & 0.74\\
 ... &  BC03 & 1.79$^{+0.07}_{-0.09}$ & 0.40$^{+0.40}_{-0.10}$ & 0.04$^{+0.08}_{-0.03}$ & 1.9$^{+0.2}_{-0.4}$ & 2.91$^{+0.54}_{-0.49}$ & 0.4$^{+10.6}_{-0.4}$ & 0.36$^{+0.16}_{-0.08}$ & 0.82\\
 ... &  MAR05 (no irac) & 1.72$^{+0.20}_{-0.06}$ & 1.00$^{+1.75}_{-0.80}$ & 0.12$^{+0.28}_{-0.11}$ & 0.6$^{+1.7}_{-0.4}$ & 1.45$^{+1.22}_{-0.44}$ & 0.3$^{+11.5}_{-0.3}$ & 0.88$^{+1.47}_{-0.70}$ & 0.84\\
 ... &  MAR05 & 1.79$^{+0.08}_{-0.09}$ & 0.70$^{+0.30}_{-0.40}$ & 0.04$^{+0.11}_{-0.03}$ & 0.7$^{+1.1}_{-0.3}$ & 1.49$^{+0.81}_{-0.18}$ & 0.0$^{+2.3}_{-0.0}$ & 0.66$^{+0.33}_{-0.38}$ & 0.89\\
 ... &  CB08 (no irac) & 1.72$^{+0.12}_{-0.04}$ & 0.90$^{+2.60}_{-0.50}$ & 0.12$^{+0.38}_{-0.11}$ & 1.1$^{+0.7}_{-1.0}$ & 2.10$^{+0.93}_{-0.70}$ & 1.2$^{+6.5}_{-1.2}$ & 0.78$^{+2.22}_{-0.43}$ & 0.80\\
 ... &  CB08 & 1.72$^{+0.12}_{-0.04}$ & 0.80$^{+1.45}_{-0.30}$ & 0.10$^{+0.20}_{-0.09}$ & 1.2$^{+0.3}_{-0.9}$ & 2.14$^{+0.45}_{-0.82}$ & 0.9$^{+6.6}_{-0.9}$ & 0.70$^{+1.25}_{-0.22}$ & 0.81\\
\hline
1030-609 &  BC03 (no irac) & 1.800 & 0.90$^{+1.60}_{-0.85}$ & 10.0$^{+10.0}_{-9.99}$ & 1.5$^{+0.4}_{-0.3}$ & 0.88$^{+0.42}_{-0.47}$ & 118.2$^{+156.5}_{-105.2}$ & 0.45$^{+0.81}_{-0.41}$ & 1.09\\
 ... &  BC03 & 1.800 & 0.90$^{+1.35}_{-0.70}$ & 10.0$^{+10.0}_{-9.88}$ & 1.5$^{+0.3}_{-0.3}$ & 0.88$^{+0.36}_{-0.31}$ & 118.2$^{+100.6}_{-53.1}$ & 0.45$^{+0.71}_{-0.30}$ & 1.12\\
 ... &  MAR05 (no irac) & 1.800 & 0.60$^{+0.30}_{-0.55}$ & 20.0$^{+0.00}_{-19.9}$ & 1.5$^{+0.6}_{-0.2}$ & 0.70$^{+0.05}_{-0.39}$ & 140.6$^{+396.2}_{-57.2}$ & 0.30$^{+0.16}_{-0.26}$ & 1.12\\
 ... &  MAR05 & 1.800 & 0.10$^{+0.40}_{-0.05}$ & 0.12$^{+19.8}_{-0.09}$ & 1.9$^{+0.1}_{-0.4}$ & 0.41$^{+0.23}_{-0.08}$ & 303.8$^{+126.0}_{-169.3}$ & 0.05$^{+0.20}_{-0.02}$ & 1.19\\
 ... &  CB08 (no irac) & 1.800 & 0.80$^{+0.70}_{-0.60}$ & 5.00$^{+15.0}_{-4.90}$ & 1.5$^{+0.3}_{-0.3}$ & 0.82$^{+0.24}_{-0.24}$ & 118.4$^{+104.2}_{-56.4}$ & 0.41$^{+0.41}_{-0.28}$ & 1.02\\
 ... &  CB08 & 1.800 & 0.10$^{+0.00}_{-0.04}$ & 0.04$^{+0.01}_{-0.03}$ & 1.8$^{+0.0}_{-0.2}$ & 0.50$^{+0.01}_{-0.09}$ & 131.8$^{+32.6}_{-118.9}$ & 0.06$^{+0.00}_{-0.01}$ & 1.23\\
\hline
1030-807 &  BC03 (no irac) & 2.367 & 0.70$^{+0.20}_{-0.20}$ & 0.12$^{+0.03}_{-0.04}$ & 0.2$^{+0.3}_{-0.2}$ & 1.11$^{+0.17}_{-0.12}$ & 3.6$^{+1.8}_{-2.4}$ & 0.58$^{+0.17}_{-0.08}$ & 1.87\\
 ... &  BC03 & 2.367 & 0.50$^{+0.30}_{-0.30}$ & 0.10$^{+0.05}_{-0.09}$ & 0.8$^{+0.5}_{-0.4}$ & 1.47$^{+0.22}_{-0.25}$ & 12.9$^{+14.3}_{-12.9}$ & 0.40$^{+0.25}_{-0.23}$ & 1.88\\
 ... &  MAR05 (no irac) & 2.367 & 0.60$^{+0.10}_{-0.54}$ & 0.08$^{+0.04}_{-0.07}$ & 0.0$^{+1.8}_{-0.0}$ & 0.82$^{+0.30}_{-1.52}$ & 0.7$^{+28.4}_{-0.1}$ & 0.52$^{+0.06}_{-0.47}$ & 1.97\\
 ... &  MAR05 & 2.367 & 0.60$^{+0.10}_{-0.51}$ & 0.08$^{+0.04}_{-0.07}$ & 0.0$^{+1.5}_{-0.0}$ & 0.82$^{+0.30}_{-0.00}$ & 0.7$^{+2.1}_{-0.1}$ & 0.52$^{+0.06}_{-0.44}$ & 1.78\\
 ... &  CB08 (no irac) & 2.367 & 0.70$^{+0.20}_{-0.20}$ & 0.12$^{+0.03}_{-0.04}$ & 0.2$^{+0.6}_{-0.2}$ & 1.11$^{+0.34}_{-0.14}$ & 3.6$^{+9.1}_{-2.7}$ & 0.58$^{+0.17}_{-0.17}$ & 1.87\\
 ... &  CB08 & 2.367 & 0.70$^{+0.20}_{-0.10}$ & 0.12$^{+0.03}_{-0.04}$ & 0.2$^{+0.0}_{-0.2}$ & 1.10$^{+0.00}_{-0.13}$ & 3.6$^{+0.0}_{-2.6}$ & 0.58$^{+0.17}_{-0.06}$ & 1.74\\
\hline
1030-1531 &  BC03 (no irac) & 2.613 & 0.50$^{+0.50}_{-0.00}$ & 0.20$^{+0.80}_{-0.00}$ & 0.4$^{+0.2}_{-0.1}$ & 1.43$^{+0.70}_{-0.00}$ & 81.0$^{+69.7}_{-28.4}$ & 0.34$^{+0.24}_{-0.00}$ & 1.38\\
 ... &  BC03 & 2.613 & 2.50$^{+0.00}_{-0.75}$ & 10.0$^{+10.0}_{-7.50}$ & 0.6$^{+0.1}_{-0.0}$ & 3.29$^{+0.40}_{-0.48}$ & 153.5$^{+38.0}_{-3.7}$ & 1.30$^{+0.05}_{-0.33}$ & 1.81\\
 ... &  MAR05 (no irac) & 2.613 & 0.40$^{+0.60}_{-0.10}$ & 0.25$^{+9.75}_{-0.15}$ & 0.6$^{+0.1}_{-0.2}$ & 1.30$^{+0.57}_{-0.21}$ & 160.2$^{+61.4}_{-90.2}$ & 0.25$^{+0.25}_{-0.03}$ & 1.58\\
 ... &  MAR05 & 2.613 & 0.90$^{+0.10}_{-0.20}$ & 2.50$^{+17.5}_{-1.85}$ & 0.7$^{+0.0}_{-0.1}$ & 1.91$^{+0.17}_{-0.23}$ & 219.1$^{+3.3}_{-48.6}$ & 0.47$^{+0.05}_{-0.09}$ & 1.60\\
 ... &  CB08 (no irac) & 2.613 & 0.50$^{+0.50}_{-0.00}$ & 0.20$^{+0.80}_{-0.00}$ & 0.4$^{+0.2}_{-0.1}$ & 1.43$^{+0.70}_{-0.00}$ & 81.0$^{+69.7}_{-28.5}$ & 0.34$^{+0.24}_{-0.00}$ & 1.39\\
 ... &  CB08 & 2.613 & 1.00$^{+0.50}_{-0.30}$ & 0.90$^{+1.60}_{-0.50}$ & 0.6$^{+0.0}_{-0.1}$ & 2.14$^{+0.27}_{-0.40}$ & 149.7$^{+4.3}_{-33.3}$ & 0.59$^{+0.24}_{-0.16}$ & 1.36\\
\hline
1030-1611 &  BC03 (no irac) & 1.93$^{+0.14}_{-0.15}$ & 0.50$^{+1.50}_{-0.30}$ & 0.12$^{+0.28}_{-0.11}$ & 1.8$^{+0.5}_{-1.1}$ & 1.92$^{+0.57}_{-0.55}$ & 32.4$^{+81.9}_{-32.4}$ & 0.38$^{+0.98}_{-0.21}$ & 0.75\\
 ... &  BC03 & 1.93$^{+0.04}_{-0.25}$ & 0.90$^{+2.60}_{-0.40}$ & 0.20$^{+0.60}_{-0.12}$ & 1.2$^{+0.4}_{-0.9}$ & 1.72$^{+1.34}_{-0.39}$ & 12.6$^{+8.7}_{-10.6}$ & 0.71$^{+2.15}_{-0.30}$ & 0.81\\
 ... &  MAR05 (no irac) & 1.93$^{+0.14}_{-0.13}$ & 0.70$^{+0.80}_{-0.65}$ & 0.12$^{+0.18}_{-0.11}$ & 0.7$^{+1.9}_{-0.5}$ & 0.98$^{+0.82}_{-0.18}$ & 3.1$^{+123.6}_{-3.1}$ & 0.58$^{+0.67}_{-0.51}$ & 0.75\\
 ... &  MAR05 & 1.93$^{+0.14}_{-0.25}$ & 0.30$^{+3.20}_{-0.22}$ & 0.03$^{+0.87}_{-0.02}$ & 1.4$^{+0.8}_{-1.3}$ & 1.19$^{+1.67}_{-0.43}$ & 0.2$^{+67.3}_{-0.2}$ & 0.27$^{+2.47}_{-0.19}$ & 0.73\\
 ... &  CB08 (no irac) & 1.93$^{+0.14}_{-0.14}$ & 0.60$^{+1.65}_{-0.40}$ & 0.12$^{+0.28}_{-0.11}$ & 1.5$^{+0.7}_{-1.1}$ & 1.63$^{+0.58}_{-0.41}$ & 12.2$^{+60.9}_{-12.2}$ & 0.48$^{+0.72}_{-0.23}$ & 0.77\\
 ... &  CB08 & 1.69$^{+0.27}_{-0.00}$ & 3.50$^{+0.00}_{-2.50}$ & 0.65$^{+0.05}_{-0.45}$ & 0.3$^{+0.5}_{-0.3}$ & 1.62$^{+0.32}_{-0.68}$ & 1.5$^{+2.6}_{-0.9}$ & 2.86$^{+0.04}_{-2.05}$ & 0.96\\
\hline
1030-1839 &  BC03 (no irac) & 2.312 & 2.25$^{+0.50}_{-1.55}$ & 2.00$^{+18.0}_{-1.75}$ & 1.1$^{+0.3}_{-0.3}$ & 2.94$^{+0.77}_{-1.36}$ & 94.1$^{+93.1}_{-51.3}$ & 1.33$^{+0.44}_{-0.85}$ & 2.74\\
 ... &  BC03 & 2.312 & 0.80$^{+0.20}_{-0.40}$ & 0.25$^{+0.15}_{-0.15}$ & 0.7$^{+0.3}_{-0.2}$ & 1.50$^{+0.19}_{-0.35}$ & 32.9$^{+37.4}_{-13.5}$ & 0.58$^{+0.15}_{-0.28}$ & 2.73\\
 ... &  MAR05 (no irac) & 2.312 & 1.50$^{+1.25}_{-1.10}$ & 2.75$^{+17.2}_{-2.63}$ & 1.2$^{+0.3}_{-0.5}$ & 2.14$^{+0.91}_{-1.05}$ & 136.4$^{+133.7}_{-100.6}$ & 0.81$^{+0.58}_{-0.57}$ & 2.71\\
 ... &  MAR05 & 2.312 & 0.30$^{+0.10}_{-0.10}$ & 0.08$^{+0.04}_{-0.03}$ & 1.0$^{+0.3}_{-0.7}$ & 1.15$^{+0.00}_{-0.36}$ & 43.0$^{+88.5}_{-34.7}$ & 0.22$^{+0.09}_{-0.08}$ & 2.74\\
 ... &  CB08 (no irac) & 2.312 & 2.25$^{+0.50}_{-1.65}$ & 2.00$^{+18.0}_{-1.80}$ & 1.1$^{+0.3}_{-0.3}$ & 2.94$^{+0.78}_{-1.40}$ & 93.8$^{+94.5}_{-51.2}$ & 1.33$^{+0.44}_{-0.90}$ & 2.74\\
 ... &  CB08 & 2.312 & 0.60$^{+0.20}_{-0.10}$ & 0.12$^{+0.08}_{-0.02}$ & 0.3$^{+0.2}_{-0.3}$ & 0.92$^{+0.30}_{-0.17}$ & 6.8$^{+14.8}_{-4.3}$ & 0.48$^{+0.13}_{-0.09}$ & 3.01\\
\hline
1030-2026 &  BC03 (no irac) & 2.511 & 1.00$^{+0.50}_{-0.30}$ & 0.20$^{+0.05}_{-0.05}$ & 0.7$^{+0.4}_{-0.7}$ & 3.01$^{+0.50}_{-1.07}$ & 13.6$^{+14.2}_{-11.0}$ & 0.80$^{+0.44}_{-0.25}$ & 1.46\\
 ... &  BC03 & 2.511 & 0.80$^{+0.20}_{-0.30}$ & 0.20$^{+0.05}_{-0.08}$ & 1.2$^{+0.3}_{-0.2}$ & 3.70$^{+0.46}_{-0.41}$ & 44.8$^{+33.0}_{-19.1}$ & 0.61$^{+0.15}_{-0.22}$ & 1.77\\
 ... &  MAR05 (no irac) & 2.511 & 0.90$^{+0.00}_{-0.20}$ & 0.15$^{+0.00}_{-0.03}$ & 0.2$^{+0.3}_{-0.1}$ & 1.80$^{+0.31}_{-0.25}$ & 3.8$^{+4.9}_{-1.7}$ & 0.75$^{+0.00}_{-0.17}$ & 1.58\\
 ... &  MAR05 & 2.511 & 0.80$^{+0.10}_{-0.10}$ & 0.15$^{+0.00}_{-0.03}$ & 0.5$^{+0.0}_{-0.3}$ & 2.10$^{+6.10}_{-0.27}$ & 8.7$^{+0.0}_{-4.8}$ & 0.65$^{+0.09}_{-0.07}$ & 1.81\\
 ... &  CB08 (no irac) & 2.511 & 1.00$^{+0.50}_{-0.30}$ & 0.20$^{+0.05}_{-0.05}$ & 0.7$^{+0.3}_{-0.5}$ & 2.99$^{+0.50}_{-0.80}$ & 13.5$^{+12.1}_{-9.9}$ & 0.80$^{+0.44}_{-0.25}$ & 1.44\\
 ... &  CB08 & 2.511 & 0.80$^{+0.20}_{-0.10}$ & 0.15$^{+0.05}_{-0.00}$ & 0.7$^{+0.2}_{-0.1}$ & 2.55$^{+0.39}_{--3.0}$ & 10.8$^{+11.2}_{-0.0}$ & 0.65$^{+0.15}_{-0.09}$ & 1.66\\
\hline
1030-2329 &  BC03 (no irac) & 2.236 & 0.80$^{+0.20}_{-0.60}$ & 0.15$^{+0.05}_{-0.14}$ & 0.7$^{+1.2}_{-0.6}$ & 1.54$^{+0.38}_{-0.40}$ & 6.5$^{+32.0}_{-6.5}$ & 0.65$^{+0.19}_{-0.49}$ & 1.38\\
 ... &  BC03 & 2.236 & 0.80$^{+0.20}_{-0.50}$ & 0.15$^{+0.05}_{-0.14}$ & 0.7$^{+0.6}_{-0.1}$ & 1.55$^{+0.23}_{-0.24}$ & 6.5$^{+8.3}_{-6.5}$ & 0.65$^{+0.15}_{-0.38}$ & 1.29\\
 ... &  MAR05 (no irac) & 2.236 & 0.70$^{+0.10}_{-0.10}$ & 0.10$^{+0.02}_{-0.02}$ & 0.1$^{+0.3}_{-0.1}$ & 0.86$^{+0.10}_{-0.08}$ & 1.0$^{+0.2}_{-0.2}$ & 0.60$^{+0.08}_{-0.08}$ & 1.25\\
 ... &  MAR05 & 2.236 & 0.70$^{+0.10}_{-0.20}$ & 0.10$^{+0.02}_{-0.09}$ & 0.1$^{+0.2}_{-0.1}$ & 0.85$^{+0.05}_{-0.11}$ & 1.0$^{+0.2}_{-1.0}$ & 0.60$^{+0.08}_{-0.14}$ & 1.13\\
 ... &  CB08 (no irac) & 2.236 & 0.70$^{+0.30}_{-0.50}$ & 0.12$^{+0.08}_{-0.11}$ & 0.6$^{+1.2}_{-0.3}$ & 1.32$^{+0.59}_{-0.19}$ & 4.3$^{+27.9}_{-4.3}$ & 0.58$^{+0.22}_{-0.41}$ & 1.38\\
 ... &  CB08 & 2.236 & 0.80$^{+0.20}_{-0.10}$ & 0.12$^{+0.03}_{-0.02}$ & 0.3$^{+0.2}_{-0.3}$ & 1.12$^{+0.16}_{-0.10}$ & 1.5$^{+2.2}_{-0.3}$ & 0.68$^{+0.17}_{-0.09}$ & 1.33\\
\hline
1030-2559 &  BC03 (no irac) & 2.25$^{+0.19}_{-0.04}$ & 0.90$^{+0.60}_{-0.50}$ & 0.12$^{+0.08}_{-0.11}$ & 0.6$^{+0.7}_{-0.6}$ & 1.84$^{+0.70}_{-0.51}$ & 1.1$^{+6.8}_{-1.1}$ & 0.78$^{+0.52}_{-0.30}$ & 1.39\\
 ... &  BC03 & 2.25$^{+0.19}_{-0.04}$ & 0.70$^{+0.80}_{-0.20}$ & 0.08$^{+0.12}_{-0.07}$ & 0.8$^{+0.4}_{-0.6}$ & 1.92$^{+0.37}_{-0.37}$ & 0.5$^{+4.4}_{-0.5}$ & 0.62$^{+0.68}_{-0.16}$ & 1.26\\
 ... &  MAR05 (no irac) & 2.30$^{+0.08}_{-0.08}$ & 0.90$^{+0.60}_{-0.30}$ & 0.10$^{+0.10}_{-0.09}$ & 0.0$^{+0.5}_{-0.0}$ & 1.13$^{+0.40}_{-0.13}$ & 0.1$^{+0.6}_{-0.1}$ & 0.80$^{+0.50}_{-0.22}$ & 1.57\\
 ... &  MAR05 & 2.30$^{+0.14}_{-0.08}$ & 0.90$^{+0.60}_{-0.20}$ & 0.10$^{+0.10}_{-0.09}$ & 0.0$^{+0.2}_{-0.0}$ & 1.12$^{+0.43}_{-0.15}$ & 0.1$^{+0.5}_{-0.1}$ & 0.80$^{+0.50}_{-0.15}$ & 1.47\\
 ... &  CB08 (no irac) & 2.30$^{+0.14}_{-0.06}$ & 0.80$^{+0.70}_{-0.30}$ & 0.10$^{+0.10}_{-0.09}$ & 0.6$^{+0.6}_{-0.6}$ & 1.83$^{+0.80}_{-0.40}$ & 0.8$^{+4.6}_{-0.8}$ & 0.70$^{+0.60}_{-0.22}$ & 1.39\\
 ... &  CB08 & 2.30$^{+0.17}_{-0.06}$ & 1.00$^{+0.00}_{-0.30}$ & 0.12$^{+0.00}_{-0.11}$ & 0.2$^{+0.3}_{-0.2}$ & 1.49$^{+0.14}_{-0.33}$ & 0.4$^{+0.5}_{-0.4}$ & 0.88$^{+0.09}_{-0.23}$ & 1.32\\
\hline
1030-2728 &  BC03 (no irac) & 2.504 & 0.50$^{+0.10}_{-0.20}$ & 0.08$^{+0.02}_{-0.07}$ & 0.8$^{+0.4}_{-0.3}$ & 1.93$^{+0.47}_{-0.34}$ & 6.1$^{+15.0}_{-6.1}$ & 0.42$^{+0.09}_{-0.15}$ & 1.93\\
 ... &  BC03 & 2.504 & 0.20$^{+0.20}_{-0.00}$ & 0.02$^{+0.06}_{-0.01}$ & 1.6$^{+0.1}_{-0.3}$ & 2.48$^{+0.27}_{-0.11}$ & 0.7$^{+29.1}_{-0.7}$ & 0.18$^{+0.14}_{-0.00}$ & 1.94\\
 ... &  MAR05 (no irac) & 2.504 & 0.30$^{+0.30}_{-0.10}$ & 0.02$^{+0.05}_{-0.01}$ & 0.9$^{+0.5}_{-0.6}$ & 1.70$^{+0.48}_{-0.51}$ & 0.0$^{+10.9}_{-0.0}$ & 0.27$^{+0.24}_{-0.09}$ & 2.04\\
 ... &  MAR05 & 2.504 & 0.40$^{+0.10}_{-0.00}$ & 0.05$^{+0.03}_{-0.04}$ & 0.6$^{+0.1}_{-0.2}$ & 1.49$^{+0.15}_{-0.19}$ & 1.2$^{+3.8}_{-1.2}$ & 0.35$^{+0.07}_{-0.00}$ & 1.96\\
 ... &  CB08 (no irac) & 2.504 & 0.50$^{+0.10}_{-0.20}$ & 0.08$^{+0.02}_{-0.07}$ & 0.8$^{+0.4}_{-0.3}$ & 1.93$^{+0.22}_{-0.34}$ & 6.1$^{+0.7}_{-6.1}$ & 0.42$^{+0.09}_{-0.14}$ & 1.93\\
 ... &  CB08 & 2.504 & 0.50$^{+0.10}_{-0.10}$ & 0.08$^{+0.02}_{-0.06}$ & 0.8$^{+0.1}_{-0.1}$ & 1.92$^{+0.08}_{-0.32}$ & 6.1$^{+0.6}_{-6.1}$ & 0.42$^{+0.08}_{-0.07}$ & 1.76\\
\hline
1030-2927 &  BC03 (no irac) & 1.83$^{+0.14}_{-0.13}$ & 0.40$^{+0.85}_{-0.31}$ & 0.12$^{+0.28}_{-0.11}$ & 1.4$^{+0.4}_{-1.3}$ & 1.05$^{+0.37}_{-0.28}$ & 41.0$^{+52.3}_{-41.0}$ & 0.29$^{+0.71}_{-0.20}$ & 1.20\\
 ... &  BC03 & 1.83$^{+0.11}_{-0.14}$ & 0.40$^{+0.60}_{-0.30}$ & 0.12$^{+0.28}_{-0.11}$ & 1.4$^{+0.4}_{-0.7}$ & 1.05$^{+0.37}_{-0.26}$ & 41.0$^{+30.4}_{-40.8}$ & 0.29$^{+0.47}_{-0.20}$ & 1.14\\
 ... &  MAR05 (no irac) & 1.84$^{+0.09}_{-0.07}$ & 0.20$^{+0.10}_{-0.15}$ & 0.04$^{+0.04}_{-0.03}$ & 1.3$^{+0.6}_{-0.4}$ & 0.83$^{+0.12}_{-0.29}$ & 17.3$^{+77.8}_{-17.0}$ & 0.16$^{+0.08}_{-0.10}$ & 1.06\\
 ... &  MAR05 & 1.83$^{+0.12}_{-0.08}$ & 0.10$^{+0.20}_{-0.05}$ & 0.02$^{+0.07}_{-0.01}$ & 1.8$^{+0.2}_{-0.5}$ & 0.74$^{+0.22}_{-0.23}$ & 65.8$^{+58.7}_{-64.7}$ & 0.07$^{+0.15}_{-0.03}$ & 1.08\\
 ... &  CB08 (no irac) & 1.83$^{+0.14}_{-0.13}$ & 0.10$^{+0.90}_{-0.01}$ & 0.01$^{+0.24}_{-0.00}$ & 1.7$^{+0.2}_{-1.5}$ & 0.79$^{+0.32}_{-0.07}$ & 0.4$^{+39.6}_{-0.0}$ & 0.09$^{+0.71}_{-0.00}$ & 1.32\\
 ... &  CB08 & 1.83$^{+0.12}_{-0.14}$ & 0.10$^{+0.90}_{-0.01}$ & 0.01$^{+0.24}_{-0.00}$ & 1.7$^{+0.0}_{-1.5}$ & 0.79$^{+0.25}_{-0.04}$ & 0.4$^{+11.3}_{-0.0}$ & 0.09$^{+0.71}_{-0.00}$ & 1.23\\
\hline
1256-0 &  BC03 (no irac) & 2.29$^{+0.20}_{-0.18}$ & 0.20$^{+1.30}_{-0.00}$ & 0.01$^{+0.24}_{-0.00}$ & 1.9$^{+0.3}_{-1.8}$ & 4.12$^{+1.29}_{-1.51}$ & 0.0$^{+194.9}_{-0.0}$ & 0.19$^{+1.11}_{-0.02}$ & 1.12\\
 ... &  BC03 & 2.29$^{+0.15}_{-0.21}$ & 0.70$^{+1.30}_{-0.40}$ & 0.12$^{+0.18}_{-0.11}$ & 1.0$^{+0.5}_{-0.8}$ & 3.62$^{+1.84}_{-0.70}$ & 11.9$^{+23.4}_{-11.9}$ & 0.58$^{+0.87}_{-0.31}$ & 1.06\\
 ... &  MAR05 (no irac) & 2.29$^{+0.18}_{-0.03}$ & 0.70$^{+0.30}_{-0.61}$ & 0.10$^{+0.05}_{-0.09}$ & 0.4$^{+1.7}_{-0.4}$ & 2.13$^{+1.85}_{-0.55}$ & 2.6$^{+19.5}_{-2.6}$ & 0.60$^{+0.27}_{-0.41}$ & 0.98\\
 ... &  MAR05 & 2.29$^{+0.18}_{-0.03}$ & 0.90$^{+0.35}_{-0.80}$ & 0.12$^{+0.08}_{-0.11}$ & 0.1$^{+1.9}_{-0.1}$ & 1.93$^{+1.81}_{-0.42}$ & 1.1$^{+16.6}_{-1.1}$ & 0.78$^{+0.09}_{-0.69}$ & 0.93\\
 ... &  CB08 (no irac) & 2.29$^{+0.20}_{-0.03}$ & 0.60$^{+0.90}_{-0.40}$ & 0.10$^{+0.10}_{-0.09}$ & 1.2$^{+1.0}_{-1.1}$ & 3.60$^{+1.33}_{-1.03}$ & 12.1$^{+101.3}_{-12.1}$ & 0.50$^{+0.55}_{-0.34}$ & 1.11\\
 ... &  CB08 & 2.29$^{+0.20}_{-0.20}$ & 1.00$^{+1.75}_{-0.40}$ & 0.15$^{+0.25}_{-0.14}$ & 0.4$^{+0.3}_{-0.4}$ & 2.72$^{+1.94}_{-0.66}$ & 3.1$^{+3.1}_{-3.1}$ & 0.85$^{+1.50}_{-0.25}$ & 1.15\\
\hline
1256-142 &  BC03 (no irac) & 2.24$^{+0.12}_{-0.04}$ & 2.25$^{+0.75}_{-1.85}$ & 0.30$^{+0.10}_{-0.29}$ & 0.1$^{+1.3}_{-0.1}$ & 2.76$^{+1.25}_{-0.96}$ & 0.7$^{+9.2}_{-0.7}$ & 1.95$^{+0.65}_{-1.57}$ & 1.19\\
 ... &  BC03 & 2.24$^{+0.19}_{-0.04}$ & 1.75$^{+0.75}_{-1.05}$ & 0.25$^{+0.05}_{-0.24}$ & 0.2$^{+0.5}_{-0.2}$ & 2.47$^{+0.63}_{-0.90}$ & 1.2$^{+1.5}_{-1.2}$ & 1.50$^{+0.69}_{-0.81}$ & 1.20\\
 ... &  MAR05 (no irac) & 2.24$^{+0.03}_{-0.03}$ & 1.50$^{+1.25}_{-0.80}$ & 0.20$^{+0.20}_{-0.19}$ & 0.1$^{+0.5}_{-0.1}$ & 1.76$^{+0.96}_{-0.33}$ & 0.6$^{+1.9}_{-0.6}$ & 1.30$^{+1.05}_{-0.64}$ & 1.27\\
 ... &  MAR05 & 2.24$^{+0.02}_{-0.04}$ & 1.75$^{+1.25}_{-0.25}$ & 0.25$^{+0.15}_{-0.05}$ & 0.0$^{+0.0}_{-0.0}$ & 1.90$^{+1.00}_{-0.35}$ & 0.9$^{+0.1}_{-0.5}$ & 1.50$^{+1.10}_{-0.20}$ & 1.39\\
 ... &  CB08 (no irac) & 2.24$^{+0.03}_{-0.03}$ & 1.50$^{+1.25}_{-1.00}$ & 0.20$^{+0.20}_{-0.19}$ & 0.4$^{+0.9}_{-0.4}$ & 2.32$^{+1.35}_{-0.71}$ & 0.9$^{+8.9}_{-0.9}$ & 1.30$^{+1.05}_{-0.73}$ & 1.10\\
 ... &  CB08 & 2.42$^{+0.02}_{-0.19}$ & 1.00$^{+0.00}_{-0.20}$ & 0.10$^{+0.02}_{-0.09}$ & 0.0$^{+0.1}_{-0.0}$ & 1.61$^{+0.31}_{-0.04}$ & 0.1$^{+0.3}_{-0.1}$ & 0.90$^{+0.08}_{-0.05}$ & 1.62\\
\hline
1256-519 &  BC03 (no irac) & 1.857 & 0.30$^{+1.70}_{-0.23}$ & 0.12$^{+19.8}_{-0.11}$ & 3.2$^{+0.2}_{-0.7}$ & 4.80$^{+2.68}_{-1.76}$ & 444.7$^{+946.1}_{-442.6}$ & 0.20$^{+0.82}_{-0.14}$ & 1.25\\
 ... &  BC03 & 1.857 & 2.50$^{+1.00}_{-0.75}$ & 0.40$^{+0.20}_{-0.39}$ & 0.9$^{+0.3}_{-0.2}$ & 4.64$^{+0.78}_{-0.60}$ & 3.0$^{+2.7}_{-3.0}$ & 2.10$^{+0.89}_{-0.64}$ & 2.55\\
 ... &  MAR05 (no irac) & 1.857 & 0.60$^{+0.65}_{-0.56}$ & 0.15$^{+19.8}_{-0.14}$ & 2.2$^{+1.5}_{-0.6}$ & 3.51$^{+1.46}_{-1.40}$ & 55.9$^{+252}_{-55.9}$ & 0.46$^{+0.44}_{-0.43}$ & 1.25\\
 ... &  MAR05 & 1.857 & 3.50$^{+0.00}_{-3.30}$ & 0.65$^{+0.05}_{-0.64}$ & 0.9$^{+1.5}_{-0.1}$ & 5.09$^{+0.34}_{-1.69}$ & 4.8$^{+2.1}_{-4.8}$ & 2.86$^{+0.04}_{-2.67}$ & 2.33\\
 ... &  CB08 (no irac) & 1.857 & 0.30$^{+0.50}_{-0.23}$ & 0.10$^{+0.80}_{-0.09}$ & 3.0$^{+0.4}_{-0.3}$ & 4.24$^{+1.30}_{-1.20}$ & 278.2$^{+687.8}_{-276.0}$ & 0.21$^{+0.25}_{-0.15}$ & 1.25\\
 ... &  CB08 & 1.857 & 3.50$^{+0.00}_{-0.25}$ & 0.50$^{+0.00}_{-0.49}$ & 0.5$^{+0.0}_{-0.1}$ & 4.44$^{+0.43}_{-0.28}$ & 1.1$^{+0.0}_{-1.1}$ & 3.00$^{+0.48}_{-0.00}$ & 3.07\\
\hline
1256-1207 &  BC03 (no irac) & 1.80$^{+0.08}_{-0.16}$ & 0.80$^{+0.95}_{-0.40}$ & 0.10$^{+0.15}_{-0.09}$ & 0.7$^{+0.7}_{-0.5}$ & 1.40$^{+0.50}_{-0.22}$ & 0.6$^{+5.4}_{-0.6}$ & 0.70$^{+0.60}_{-0.32}$ & 1.06\\
 ... &  BC03 & 1.66$^{+0.19}_{-0.02}$ & 0.80$^{+0.45}_{-0.40}$ & 0.12$^{+0.08}_{-0.11}$ & 1.0$^{+0.4}_{-0.5}$ & 1.35$^{+0.18}_{-0.30}$ & 1.9$^{+4.2}_{-1.9}$ & 0.68$^{+0.37}_{-0.30}$ & 0.99\\
 ... &  MAR05 (no irac) & 1.80$^{+0.08}_{-0.15}$ & 0.90$^{+0.60}_{-0.30}$ & 0.10$^{+0.10}_{-0.09}$ & 0.0$^{+0.3}_{-0.0}$ & 0.78$^{+0.24}_{-0.13}$ & 0.1$^{+0.3}_{-0.1}$ & 0.80$^{+0.50}_{-0.18}$ & 1.11\\
 ... &  MAR05 & 1.80$^{+0.09}_{-0.14}$ & 0.90$^{+0.10}_{-0.70}$ & 0.10$^{+0.02}_{-0.09}$ & 0.0$^{+1.8}_{-0.0}$ & 0.78$^{+0.85}_{-0.13}$ & 0.1$^{+0.3}_{-0.1}$ & 0.80$^{+0.09}_{-0.61}$ & 1.18\\
 ... &  CB08 (no irac) & 1.80$^{+0.09}_{-0.15}$ & 0.80$^{+0.20}_{-0.40}$ & 0.10$^{+0.05}_{-0.09}$ & 0.6$^{+0.5}_{-0.5}$ & 1.24$^{+0.23}_{-0.28}$ & 0.5$^{+1.4}_{-0.5}$ & 0.70$^{+0.19}_{-0.25}$ & 0.97\\
 ... &  CB08 & 1.80$^{+0.08}_{-0.15}$ & 1.00$^{+0.00}_{-0.30}$ & 0.12$^{+0.03}_{-0.11}$ & 0.2$^{+0.4}_{-0.1}$ & 1.04$^{+0.41}_{-0.02}$ & 0.2$^{+1.3}_{-0.2}$ & 0.88$^{+0.10}_{-0.23}$ & 0.95\\
\hline
1256-1967 &  BC03 (no irac) & 2.02$^{+0.25}_{-0.18}$ & 0.10$^{+0.00}_{-0.02}$ & 0.01$^{+0.01}_{-0.00}$ & 1.8$^{+0.2}_{-0.0}$ & 2.43$^{+0.30}_{-0.25}$ & 1.4$^{+103.2}_{-0.1}$ & 0.09$^{+0.00}_{-0.01}$ & 3.24\\
 ... &  BC03 & 2.02$^{+0.25}_{-0.15}$ & 0.30$^{+0.30}_{-0.10}$ & 0.02$^{+0.06}_{-0.01}$ & 0.9$^{+0.4}_{-0.7}$ & 1.96$^{+0.28}_{-0.52}$ & 0.0$^{+6.2}_{-0.0}$ & 0.28$^{+0.24}_{-0.09}$ & 4.15\\
 ... &  MAR05 (no irac) & 2.27$^{+0.00}_{-0.43}$ & 0.60$^{+0.00}_{-0.55}$ & 0.08$^{+0.00}_{-0.07}$ & 0.0$^{+2.5}_{-0.0}$ & 1.85$^{+1.15}_{-0.18}$ & 1.7$^{+199}_{-1.5}$ & 0.52$^{+0.00}_{-0.48}$ & 3.40\\
 ... &  MAR05 & 1.99$^{+0.04}_{-0.02}$ & 0.50$^{+0.10}_{-0.20}$ & 0.05$^{+0.03}_{-0.04}$ & 0.0$^{+0.6}_{-0.0}$ & 1.22$^{+0.47}_{-0.01}$ & 0.1$^{+1.1}_{-0.1}$ & 0.45$^{+0.07}_{-0.16}$ & 3.23\\
 ... &  CB08 (no irac) & 2.02$^{+0.25}_{-0.18}$ & 0.10$^{+0.00}_{-0.02}$ & 0.01$^{+0.01}_{-0.00}$ & 1.8$^{+0.2}_{-0.0}$ & 2.43$^{+0.18}_{-0.35}$ & 1.4$^{+98.3}_{-0.2}$ & 0.09$^{+0.00}_{-0.01}$ & 3.24\\
 ... &  CB08 & 2.02$^{+0.25}_{-0.16}$ & 0.70$^{+0.10}_{-0.60}$ & 0.10$^{+0.02}_{-0.09}$ & 0.0$^{+1.7}_{-0.0}$ & 1.57$^{+0.62}_{-0.30}$ & 1.9$^{+0.6}_{-1.9}$ & 0.60$^{+0.00}_{-0.51}$ & 4.24\\
\hline
hdfs1-259 &  BC03 (no irac) & 2.249 & 1.00$^{+0.75}_{-0.70}$ & 5.00$^{+15.0}_{-4.75}$ & 1.4$^{+0.3}_{-0.2}$ & 2.10$^{+0.67}_{-0.69}$ & 241.3$^{+223.3}_{-94.6}$ & 0.51$^{+0.38}_{-0.34}$ & 1.79\\
 ... &  BC03 & 2.249 & 1.00$^{+0.75}_{-0.70}$ & 5.00$^{+15.0}_{-4.75}$ & 1.4$^{+0.2}_{-0.2}$ & 2.10$^{+0.53}_{-0.69}$ & 241.1$^{+140.0}_{-93.0}$ & 0.51$^{+0.40}_{-0.34}$ & 1.62\\
 ... &  MAR05 (no irac) & 2.249 & 0.30$^{+0.40}_{-0.24}$ & 0.60$^{+19.4}_{-0.57}$ & 1.6$^{+0.3}_{-0.3}$ & 1.36$^{+0.39}_{-0.58}$ & 417.0$^{+492.2}_{-246.3}$ & 0.16$^{+0.19}_{-0.12}$ & 1.78\\
 ... &  MAR05 & 2.249 & 0.30$^{+0.20}_{-0.25}$ & 20.0$^{+0.00}_{-19.9}$ & 1.7$^{+0.2}_{-0.3}$ & 1.38$^{+0.15}_{-0.67}$ & 542.8$^{+357.0}_{-280.1}$ & 0.15$^{+0.10}_{-0.11}$ & 1.63\\
 ... &  CB08 (no irac) & 2.249 & 1.00$^{+0.75}_{-0.70}$ & 5.00$^{+15.0}_{-4.75}$ & 1.4$^{+0.3}_{-0.2}$ & 2.10$^{+0.58}_{-0.69}$ & 241.0$^{+223.2}_{-93.3}$ & 0.51$^{+0.37}_{-0.34}$ & 1.80\\
 ... &  CB08 & 2.249 & 0.10$^{+0.00}_{-0.04}$ & 0.04$^{+0.01}_{-0.03}$ & 1.7$^{+0.0}_{-0.2}$ & 1.10$^{+0.04}_{-0.20}$ & 289.6$^{+158.5}_{-261.1}$ & 0.06$^{+0.00}_{-0.01}$ & 1.83\\
\hline
hdfs1-1849 &  BC03 (no irac) & 2.30$^{+0.06}_{-0.05}$ & 0.40$^{+0.60}_{-0.20}$ & 0.08$^{+0.17}_{-0.07}$ & 1.5$^{+0.6}_{-0.7}$ & 2.97$^{+1.41}_{-0.71}$ & 32.1$^{+109.6}_{-32.1}$ & 0.32$^{+0.48}_{-0.15}$ & 1.47\\
 ... &  BC03 & 2.30$^{+0.08}_{-0.05}$ & 0.40$^{+0.60}_{-0.20}$ & 0.08$^{+0.17}_{-0.07}$ & 1.6$^{+0.3}_{-0.4}$ & 3.28$^{+1.14}_{-0.63}$ & 35.5$^{+84.3}_{-35.5}$ & 0.32$^{+0.44}_{-0.15}$ & 1.34\\
 ... &  MAR05 (no irac) & 2.30$^{+0.06}_{-0.05}$ & 0.40$^{+0.50}_{-0.31}$ & 0.05$^{+0.10}_{-0.04}$ & 0.9$^{+1.1}_{-0.6}$ & 1.96$^{+1.26}_{-0.45}$ & 1.7$^{+64.9}_{-1.7}$ & 0.35$^{+0.40}_{-0.26}$ & 1.43\\
 ... &  MAR05 & 2.30$^{+0.06}_{-0.05}$ & 0.40$^{+0.50}_{-0.34}$ & 0.05$^{+0.10}_{-0.04}$ & 0.9$^{+1.4}_{-0.5}$ & 1.96$^{+1.37}_{-0.40}$ & 1.7$^{+157.7}_{-1.7}$ & 0.35$^{+0.40}_{-0.27}$ & 1.33\\
 ... &  CB08 (no irac) & 2.30$^{+0.06}_{-0.05}$ & 0.40$^{+0.85}_{-0.20}$ & 0.08$^{+0.17}_{-0.07}$ & 1.5$^{+0.6}_{-0.8}$ & 2.96$^{+1.52}_{-0.74}$ & 32.0$^{+110.3}_{-32.0}$ & 0.32$^{+0.68}_{-0.16}$ & 1.45\\
 ... &  CB08 & 2.26$^{+0.07}_{-0.03}$ & 0.80$^{+0.20}_{-0.30}$ & 0.15$^{+0.05}_{-0.07}$ & 0.9$^{+0.2}_{-0.2}$ & 2.67$^{+0.41}_{-0.45}$ & 11.3$^{+12.1}_{-9.1}$ & 0.65$^{+0.15}_{-0.23}$ & 1.39\\
\hline
hdfs2-509 &  BC03 (no irac) & 2.918 & 0.60$^{+0.00}_{-0.00}$ & 0.10$^{+0.00}_{-0.00}$ & 0.1$^{+0.0}_{-0.0}$ & 4.28$^{+-6.1}_{-6.10}$ & 14.4$^{+0.0}_{-0.0}$ & 0.50$^{+0.00}_{-0.00}$ & 3.84\\
 ... &  BC03 & 2.918 & 0.50$^{+0.10}_{-0.00}$ & 0.10$^{+0.02}_{-0.00}$ & 0.4$^{+0.3}_{-0.3}$ & 5.20$^{+1.44}_{-0.86}$ & 45.7$^{+66.3}_{-31.1}$ & 0.40$^{+0.09}_{-0.01}$ & 4.49\\
 ... &  MAR05 (no irac) & 2.918 & 0.40$^{+0.10}_{-0.00}$ & 0.08$^{+0.02}_{-0.00}$ & 0.4$^{+0.0}_{-0.1}$ & 4.53$^{+0.62}_{-0.00}$ & 48.0$^{+0.0}_{-3.5}$ & 0.32$^{+0.08}_{-0.00}$ & 5.13\\
 ... &  MAR05 & 2.918 & 0.40$^{+0.10}_{-0.00}$ & 0.08$^{+0.02}_{-0.00}$ & 0.4$^{+0.0}_{-0.2}$ & 4.51$^{+-6.1}_{-0.31}$ & 47.9$^{+0.0}_{-11.6}$ & 0.32$^{+0.08}_{-0.00}$ & 4.69\\
 ... &  CB08 (no irac) & 2.918 & 0.60$^{+0.00}_{-0.00}$ & 0.10$^{+0.00}_{-0.00}$ & 0.1$^{+0.0}_{-0.0}$ & 4.29$^{+0.00}_{-0.00}$ & 14.3$^{+0.0}_{-0.0}$ & 0.50$^{+0.00}_{-0.00}$ & 3.87\\
 ... &  CB08 & 2.918 & 0.60$^{+0.00}_{-0.00}$ & 0.10$^{+0.00}_{-0.00}$ & 0.1$^{+0.0}_{-0.0}$ & 4.30$^{+0.00}_{-0.00}$ & 14.4$^{+0.0}_{-0.0}$ & 0.50$^{+0.00}_{-0.00}$ & 3.56\\
\hline
hdfs2-1099 &  BC03 (no irac) & 2.72$^{+0.07}_{-0.24}$ & 0.80$^{+0.20}_{-0.10}$ & 0.12$^{+0.03}_{-0.02}$ & 0.2$^{+0.5}_{-0.2}$ & 2.64$^{+1.28}_{-0.58}$ & 3.7$^{+12.9}_{-2.2}$ & 0.68$^{+0.17}_{-0.09}$ & 1.46\\
 ... &  BC03 & 2.51$^{+0.24}_{-0.03}$ & 0.70$^{+0.30}_{-0.20}$ & 0.12$^{+0.08}_{-0.04}$ & 0.7$^{+0.3}_{-0.3}$ & 3.09$^{+0.62}_{-0.64}$ & 10.1$^{+6.6}_{-6.4}$ & 0.58$^{+0.22}_{-0.16}$ & 1.66\\
 ... &  MAR05 (no irac) & 2.72$^{+0.06}_{-0.21}$ & 0.70$^{+0.10}_{-0.00}$ & 0.10$^{+0.02}_{-0.00}$ & 0.0$^{+0.1}_{-0.0}$ & 2.04$^{+0.36}_{-0.07}$ & 2.5$^{+0.8}_{-0.0}$ & 0.60$^{+0.08}_{-0.00}$ & 1.44\\
 ... &  MAR05 & 2.55$^{+0.10}_{-0.06}$ & 0.70$^{+0.10}_{-0.00}$ & 0.10$^{+0.02}_{-0.00}$ & 0.0$^{+0.1}_{-0.0}$ & 1.68$^{+0.34}_{--0.0}$ & 2.0$^{+0.7}_{-0.0}$ & 0.60$^{+0.08}_{-0.00}$ & 1.54\\
 ... &  CB08 (no irac) & 2.72$^{+0.10}_{-0.24}$ & 0.80$^{+0.20}_{-0.10}$ & 0.12$^{+0.03}_{-0.02}$ & 0.2$^{+0.5}_{-0.2}$ & 2.63$^{+1.34}_{-0.54}$ & 3.6$^{+13.1}_{-2.1}$ & 0.68$^{+0.17}_{-0.09}$ & 1.45\\
 ... &  CB08 & 2.60$^{+0.12}_{-0.12}$ & 1.00$^{+0.00}_{-0.20}$ & 0.15$^{+0.00}_{-0.03}$ & 0.0$^{+0.3}_{-0.0}$ & 2.31$^{+0.22}_{-0.16}$ & 2.6$^{+0.8}_{-1.3}$ & 0.85$^{+0.00}_{-0.17}$ & 1.52\\
\hline
hdfs2-2046 &  BC03 (no irac) & 2.24$^{+0.04}_{-0.04}$ & 0.30$^{+0.30}_{-0.10}$ & 0.02$^{+0.06}_{-0.01}$ & 0.8$^{+0.4}_{-0.6}$ & 1.38$^{+0.34}_{-0.20}$ & 0.0$^{+4.6}_{-0.0}$ & 0.28$^{+0.24}_{-0.09}$ & 1.00\\
 ... &  BC03 & 2.24$^{+0.05}_{-0.05}$ & 0.40$^{+0.20}_{-0.10}$ & 0.05$^{+0.03}_{-0.04}$ & 0.5$^{+0.4}_{-0.3}$ & 1.27$^{+0.21}_{-0.05}$ & 1.1$^{+3.6}_{-1.1}$ & 0.35$^{+0.17}_{-0.08}$ & 0.97\\
 ... &  MAR05 (no irac) & 2.24$^{+0.03}_{-0.05}$ & 0.30$^{+0.30}_{-0.10}$ & 0.02$^{+0.05}_{-0.01}$ & 0.6$^{+0.4}_{-0.6}$ & 1.26$^{+0.35}_{-0.25}$ & 0.0$^{+2.6}_{-0.0}$ & 0.27$^{+0.24}_{-0.09}$ & 1.14\\
 ... &  MAR05 & 2.25$^{+0.02}_{-0.05}$ & 0.20$^{+0.10}_{-0.00}$ & 0.01$^{+0.02}_{-0.00}$ & 0.9$^{+0.1}_{-0.4}$ & 1.43$^{+0.16}_{-0.27}$ & 0.0$^{+2.5}_{-0.0}$ & 0.19$^{+0.09}_{-0.01}$ & 1.17\\
 ... &  CB08 (no irac) & 2.24$^{+0.07}_{-0.05}$ & 0.40$^{+0.20}_{-0.20}$ & 0.05$^{+0.03}_{-0.04}$ & 0.5$^{+0.7}_{-0.3}$ & 1.26$^{+0.47}_{-0.15}$ & 1.1$^{+3.5}_{-1.1}$ & 0.35$^{+0.17}_{-0.08}$ & 1.05\\
 ... &  CB08 & 2.24$^{+0.09}_{-0.05}$ & 0.60$^{+0.00}_{-0.10}$ & 0.08$^{+0.00}_{-0.03}$ & 0.1$^{+0.1}_{-0.1}$ & 1.04$^{+0.14}_{-0.11}$ & 1.0$^{+0.1}_{-0.8}$ & 0.52$^{+0.00}_{-0.07}$ & 1.33\\
\hline
ecdfs-4454 &  BC03 (no irac) & 2.351 & 0.90$^{+0.10}_{-0.20}$ & 0.20$^{+0.10}_{-0.05}$ & 0.2$^{+0.7}_{-0.0}$ & 1.46$^{+0.69}_{-0.20}$ & 10.7$^{+54.2}_{-0.2}$ & 0.71$^{+0.05}_{-0.21}$ & 1.54\\
 ... &  BC03 & 2.351 & 1.00$^{+0.25}_{-0.30}$ & 0.30$^{+0.00}_{-0.10}$ & 0.6$^{+0.3}_{-0.4}$ & 2.04$^{+0.10}_{-0.27}$ & 32.9$^{+31.7}_{-20.0}$ & 0.73$^{+0.23}_{-0.24}$ & 1.47\\
 ... &  MAR05 (no irac) & 2.351 & 0.60$^{+0.20}_{-0.10}$ & 0.12$^{+0.08}_{-0.02}$ & 0.2$^{+0.3}_{-0.2}$ & 1.08$^{+0.31}_{-0.06}$ & 7.9$^{+12.8}_{--1.}$ & 0.48$^{+0.13}_{-0.09}$ & 1.45\\
 ... &  MAR05 & 2.351 & 0.60$^{+0.10}_{-0.10}$ & 0.12$^{+0.03}_{-0.02}$ & 0.2$^{+0.0}_{-0.2}$ & 1.07$^{+0.07}_{-0.04}$ & 7.9$^{+1.5}_{-0.0}$ & 0.48$^{+0.07}_{-0.08}$ & 1.34\\
 ... &  CB08 (no irac) & 2.351 & 0.90$^{+0.10}_{-0.20}$ & 0.20$^{+0.10}_{-0.05}$ & 0.2$^{+0.7}_{-0.0}$ & 1.46$^{+0.69}_{-0.20}$ & 10.7$^{+54.1}_{-0.2}$ & 0.71$^{+0.05}_{-0.21}$ & 1.57\\
 ... &  CB08 & 2.351 & 0.90$^{+0.10}_{-0.20}$ & 0.20$^{+0.00}_{-0.05}$ & 0.2$^{+0.2}_{-0.2}$ & 1.44$^{+0.03}_{-0.19}$ & 10.6$^{+7.3}_{-4.2}$ & 0.71$^{+0.09}_{-0.15}$ & 1.47\\
\hline
ecdfs-4511 &  BC03 (no irac) & 2.122 & 3.00$^{+0.00}_{-1.25}$ & 10.0$^{+10.0}_{-7.75}$ & 1.2$^{+0.0}_{-0.1}$ & 4.31$^{+0.27}_{-0.96}$ & 164.8$^{+43.8}_{-35.8}$ & 1.57$^{+0.13}_{-0.55}$ & 1.98\\
 ... &  BC03 & 2.122 & 3.00$^{+0.00}_{-1.25}$ & 20.0$^{+0.00}_{-18.0}$ & 1.2$^{+0.0}_{-0.2}$ & 4.08$^{+0.47}_{-0.77}$ & 168.3$^{+41.2}_{-62.9}$ & 1.53$^{+0.25}_{-0.61}$ & 1.78\\
 ... &  MAR05 (no irac) & 2.122 & 1.75$^{+1.25}_{-0.95}$ & 20.0$^{+0.00}_{-19.0}$ & 1.2$^{+0.2}_{-0.1}$ & 2.76$^{+0.97}_{-0.66}$ & 192.3$^{+106.5}_{-46.8}$ & 0.88$^{+0.64}_{-0.47}$ & 1.98\\
 ... &  MAR05 & 2.122 & 1.50$^{+1.25}_{-0.90}$ & 5.00$^{+15.0}_{-4.50}$ & 1.2$^{+0.3}_{-0.2}$ & 2.60$^{+0.83}_{-0.71}$ & 188.5$^{+186.8}_{-70.1}$ & 0.78$^{+0.71}_{-0.48}$ & 1.85\\
 ... &  CB08 (no irac) & 2.122 & 3.00$^{+0.00}_{-1.25}$ & 20.0$^{+0.00}_{-17.7}$ & 1.2$^{+0.0}_{-0.1}$ & 4.07$^{+0.50}_{-0.99}$ & 167.3$^{+41.9}_{-39.2}$ & 1.53$^{+0.17}_{-0.52}$ & 2.01\\
 ... &  CB08 & 2.122 & 1.00$^{+0.00}_{-0.40}$ & 0.65$^{+1.85}_{-0.40}$ & 1.1$^{+0.2}_{-0.3}$ & 2.37$^{+0.10}_{-0.43}$ & 128.0$^{+87.7}_{-63.2}$ & 0.62$^{+0.06}_{-0.17}$ & 2.14\\
\hline
ecdfs-4937 &  BC03 (no irac) & 2.309 & 0.50$^{+0.00}_{-0.20}$ & 0.20$^{+0.00}_{-0.12}$ & 1.0$^{+0.0}_{-0.5}$ & 3.08$^{+0.00}_{-1.00}$ & 174.1$^{+0.0}_{-133.1}$ & 0.34$^{+0.04}_{-0.11}$ & 1.28\\
 ... &  BC03 & 2.309 & 0.40$^{+0.10}_{-0.10}$ & 0.10$^{+0.05}_{-0.02}$ & 0.7$^{+0.3}_{-0.0}$ & 2.47$^{+0.12}_{-0.10}$ & 58.3$^{+103.6}_{-0.0}$ & 0.30$^{+0.06}_{-0.08}$ & 1.30\\
 ... &  MAR05 (no irac) & 2.309 & 0.20$^{+0.10}_{-0.11}$ & 0.05$^{+0.05}_{-0.02}$ & 0.9$^{+0.4}_{-0.1}$ & 2.04$^{+0.32}_{-0.37}$ & 92.3$^{+195.8}_{-12.8}$ & 0.15$^{+0.07}_{-0.08}$ & 1.25\\
 ... &  MAR05 & 2.309 & 0.10$^{+0.20}_{-0.02}$ & 0.03$^{+0.07}_{-0.01}$ & 1.3$^{+0.0}_{-0.4}$ & 1.75$^{+0.76}_{-0.11}$ & 252.9$^{+36.4}_{-159.0}$ & 0.07$^{+0.15}_{-0.01}$ & 1.26\\
 ... &  CB08 (no irac) & 2.309 & 0.50$^{+0.00}_{-0.20}$ & 0.15$^{+0.05}_{-0.07}$ & 0.7$^{+0.3}_{-0.2}$ & 2.54$^{+0.54}_{-0.46}$ & 79.8$^{+94.2}_{-38.8}$ & 0.36$^{+0.01}_{-0.14}$ & 1.29\\
 ... &  CB08 & 2.309 & 0.50$^{+0.10}_{-0.00}$ & 0.10$^{+0.02}_{-0.00}$ & 0.2$^{+0.0}_{-0.0}$ & 1.89$^{+0.08}_{-6.10}$ & 16.6$^{+0.0}_{-1.8}$ & 0.40$^{+0.08}_{-0.00}$ & 1.43\\
\hline
ecdfs-5856 &  BC03 (no irac) & 2.56$^{+0.09}_{-0.03}$ & 0.50$^{+0.40}_{-0.20}$ & 0.08$^{+0.07}_{-0.07}$ & 0.7$^{+0.4}_{-0.6}$ & 1.93$^{+0.56}_{-0.49}$ & 6.1$^{+15.2}_{-6.1}$ & 0.42$^{+0.33}_{-0.14}$ & 0.92\\
 ... &  BC03 & 2.55$^{+0.11}_{-0.04}$ & 0.60$^{+0.20}_{-0.30}$ & 0.10$^{+0.05}_{-0.09}$ & 0.8$^{+0.3}_{-0.3}$ & 2.23$^{+0.18}_{-0.52}$ & 7.5$^{+13.2}_{-7.5}$ & 0.50$^{+0.15}_{-0.22}$ & 1.05\\
 ... &  MAR05 (no irac) & 2.56$^{+0.13}_{-0.03}$ & 0.60$^{+0.20}_{-0.40}$ & 0.08$^{+0.04}_{-0.07}$ & 0.2$^{+1.1}_{-0.2}$ & 1.36$^{+0.84}_{-0.26}$ & 1.2$^{+9.9}_{-1.2}$ & 0.52$^{+0.16}_{-0.33}$ & 1.09\\
 ... &  MAR05 & 2.56$^{+0.09}_{-0.05}$ & 0.30$^{+0.40}_{-0.10}$ & 0.03$^{+0.07}_{-0.02}$ & 0.8$^{+0.5}_{-0.8}$ & 1.69$^{+0.67}_{-0.58}$ & 0.3$^{+10.9}_{-0.3}$ & 0.27$^{+0.33}_{-0.09}$ & 1.08\\
 ... &  CB08 (no irac) & 2.56$^{+0.09}_{-0.03}$ & 0.50$^{+0.40}_{-0.20}$ & 0.08$^{+0.07}_{-0.07}$ & 0.7$^{+0.4}_{-0.6}$ & 1.93$^{+0.65}_{-0.44}$ & 6.1$^{+15.9}_{-6.1}$ & 0.42$^{+0.26}_{-0.07}$ & 0.91\\
 ... &  CB08 & 2.56$^{+0.09}_{-0.03}$ & 0.80$^{+0.20}_{-0.20}$ & 0.12$^{+0.03}_{-0.04}$ & 0.1$^{+0.2}_{-0.1}$ & 1.55$^{+0.23}_{-0.17}$ & 2.1$^{+1.7}_{-1.2}$ & 0.68$^{+0.17}_{-0.16}$ & 1.05\\
\hline
ecdfs-6842 &  BC03 (no irac) & 2.40$^{+0.07}_{-0.11}$ & 0.80$^{+0.20}_{-0.40}$ & 0.25$^{+0.35}_{-0.15}$ & 1.2$^{+0.5}_{-0.4}$ & 3.74$^{+1.07}_{-0.96}$ & 82.0$^{+207.2}_{-49.0}$ & 0.58$^{+0.15}_{-0.30}$ & 2.64\\
 ... &  BC03 & 2.40$^{+0.07}_{-0.03}$ & 0.60$^{+0.40}_{-0.20}$ & 0.12$^{+0.08}_{-0.04}$ & 0.8$^{+0.1}_{-0.4}$ & 2.32$^{+0.35}_{-0.29}$ & 17.4$^{+4.7}_{-5.9}$ & 0.48$^{+0.32}_{-0.16}$ & 2.81\\
 ... &  MAR05 (no irac) & 2.40$^{+0.06}_{-0.03}$ & 0.40$^{+0.30}_{-0.10}$ & 0.10$^{+0.10}_{-0.02}$ & 1.1$^{+0.5}_{-0.5}$ & 2.44$^{+0.68}_{-0.56}$ & 56.5$^{+127.3}_{-42.0}$ & 0.30$^{+0.17}_{-0.08}$ & 2.60\\
 ... &  MAR05 & 2.40$^{+0.07}_{-0.02}$ & 0.20$^{+0.10}_{-0.00}$ & 0.04$^{+0.01}_{-0.01}$ & 1.3$^{+0.0}_{-0.5}$ & 2.15$^{+0.03}_{-0.34}$ & 44.4$^{+0.4}_{-41.1}$ & 0.16$^{+0.09}_{-0.00}$ & 2.60\\
 ... &  CB08 (no irac) & 2.40$^{+0.06}_{-0.10}$ & 0.80$^{+0.20}_{-0.40}$ & 0.25$^{+0.25}_{-0.15}$ & 1.2$^{+0.5}_{-0.4}$ & 3.73$^{+0.96}_{-0.98}$ & 81.6$^{+203.3}_{-48.9}$ & 0.58$^{+0.15}_{-0.27}$ & 2.62\\
 ... &  CB08 & 2.43$^{+0.05}_{-0.05}$ & 0.70$^{+0.20}_{-0.00}$ & 0.12$^{+0.03}_{-0.00}$ & 0.2$^{+0.1}_{-0.2}$ & 1.70$^{+0.27}_{-0.14}$ & 5.5$^{+2.7}_{-1.8}$ & 0.58$^{+0.17}_{-0.00}$ & 3.52\\
\hline
ecdfs-6956 &  BC03 (no irac) & 2.037 & 1.50$^{+1.00}_{-0.90}$ & 20.0$^{+0.00}_{-18.7}$ & 1.0$^{+0.2}_{-0.1}$ & 1.43$^{+0.45}_{-0.44}$ & 118.8$^{+65.9}_{-29.9}$ & 0.75$^{+0.51}_{-0.40}$ & 2.01\\
 ... &  BC03 & 2.037 & 1.25$^{+1.00}_{-0.65}$ & 10.0$^{+10.0}_{-8.75}$ & 1.0$^{+0.2}_{-0.1}$ & 1.25$^{+0.45}_{-0.25}$ & 120.3$^{+62.6}_{-30.2}$ & 0.63$^{+0.50}_{-0.32}$ & 1.97\\
 ... &  MAR05 (no irac) & 2.037 & 0.60$^{+0.40}_{-0.20}$ & 2.50$^{+17.5}_{-1.70}$ & 1.1$^{+0.1}_{-0.1}$ & 0.93$^{+0.17}_{-0.14}$ & 168.8$^{+47.6}_{-40.3}$ & 0.31$^{+0.19}_{-0.10}$ & 1.99\\
 ... &  MAR05 & 2.037 & 0.50$^{+0.10}_{-0.10}$ & 1.50$^{+18.5}_{-1.20}$ & 1.1$^{+0.1}_{-0.1}$ & 0.83$^{+0.10}_{-0.08}$ & 171.3$^{+50.9}_{-46.7}$ & 0.26$^{+0.06}_{-0.06}$ & 1.89\\
 ... &  CB08 (no irac) & 2.037 & 1.50$^{+0.75}_{-1.10}$ & 20.0$^{+0.00}_{-19.2}$ & 1.0$^{+0.3}_{-0.1}$ & 1.43$^{+0.35}_{-0.58}$ & 118.1$^{+112.2}_{-28.2}$ & 0.75$^{+0.40}_{-0.55}$ & 2.01\\
 ... &  CB08 & 2.037 & 0.09$^{+0.01}_{-0.03}$ & 0.03$^{+0.02}_{-0.02}$ & 1.3$^{+0.1}_{-0.0}$ & 0.57$^{+0.01}_{-0.09}$ & 118.7$^{+94.7}_{-103.9}$ & 0.06$^{+0.00}_{-0.01}$ & 2.22\\
\hline
ecdfs-9822 &  BC03 (no irac) & 1.612 & 0.60$^{+2.90}_{-0.55}$ & 20.0$^{+0.00}_{-19.9}$ & 1.9$^{+0.4}_{-0.6}$ & 0.96$^{+0.84}_{-0.54}$ & 196.0$^{+316.2}_{-189.5}$ & 0.30$^{+1.36}_{-0.26}$ & 1.30\\
 ... &  BC03 & 1.612 & 0.40$^{+0.85}_{-0.35}$ & 20.0$^{+0.00}_{-19.9}$ & 2.0$^{+0.2}_{-0.3}$ & 0.81$^{+0.42}_{-0.38}$ & 245.4$^{+147.4}_{-230.1}$ & 0.20$^{+0.43}_{-0.16}$ & 1.21\\
 ... &  MAR05 (no irac) & 1.612 & 0.50$^{+1.00}_{-0.46}$ & 1.50$^{+18.5}_{-1.48}$ & 1.7$^{+0.6}_{-0.6}$ & 0.75$^{+0.29}_{-0.44}$ & 154.9$^{+509.9}_{-117.3}$ & 0.26$^{+0.50}_{-0.24}$ & 1.30\\
 ... &  MAR05 & 1.612 & 0.08$^{+0.02}_{-0.04}$ & 0.12$^{+19.8}_{-0.10}$ & 2.2$^{+0.0}_{-0.0}$ & 0.44$^{+0.06}_{-0.11}$ & 443.1$^{+217.9}_{-147.6}$ & 0.04$^{+0.01}_{-0.01}$ & 1.25\\
 ... &  CB08 (no irac) & 1.612 & 0.06$^{+1.69}_{-0.01}$ & 0.01$^{+19.9}_{-0.00}$ & 1.9$^{+0.4}_{-0.7}$ & 0.49$^{+0.68}_{-0.07}$ & 15.0$^{+533.3}_{-8.5}$ & 0.05$^{+0.74}_{-0.00}$ & 1.32\\
 ... &  CB08 & 1.612 & 0.10$^{+0.90}_{-0.05}$ & 0.05$^{+0.60}_{-0.04}$ & 2.1$^{+0.1}_{-0.8}$ & 0.58$^{+0.43}_{-0.15}$ & 214.6$^{+213.5}_{-199.4}$ & 0.06$^{+0.55}_{-0.02}$ & 1.22\\
\hline
ecdfs-11490 &  BC03 (no irac) & 2.37$^{+0.09}_{-0.05}$ & 0.30$^{+0.10}_{-0.10}$ & 0.05$^{+0.03}_{-0.04}$ & 0.5$^{+0.2}_{-0.5}$ & 0.70$^{+0.12}_{-0.20}$ & 4.5$^{+4.3}_{-4.5}$ & 0.25$^{+0.07}_{-0.00}$ & 2.12\\
 ... &  BC03 & 2.33$^{+0.08}_{-0.04}$ & 0.20$^{+0.20}_{-0.00}$ & 0.04$^{+0.08}_{-0.01}$ & 0.9$^{+0.0}_{-0.4}$ & 0.80$^{+0.16}_{-0.11}$ & 16.9$^{+20.2}_{-13.0}$ & 0.16$^{+0.14}_{-0.00}$ & 2.43\\
 ... &  MAR05 (no irac) & 2.33$^{+0.05}_{-0.02}$ & 0.30$^{+0.00}_{-0.10}$ & 0.04$^{+0.01}_{-0.03}$ & 0.1$^{+0.4}_{-0.1}$ & 0.54$^{+0.10}_{-0.06}$ & 0.9$^{+3.0}_{-0.9}$ & 0.26$^{+0.00}_{-0.08}$ & 2.06\\
 ... &  MAR05 & 2.31$^{+0.11}_{-0.03}$ & 0.40$^{+0.00}_{-0.35}$ & 0.08$^{+0.00}_{-0.07}$ & 0.2$^{+1.1}_{-0.1}$ & 0.65$^{+0.20}_{-0.18}$ & 6.9$^{+80.3}_{-2.4}$ & 0.32$^{+0.00}_{-0.28}$ & 2.38\\
 ... &  CB08 (no irac) & 2.33$^{+0.13}_{-0.04}$ & 0.30$^{+0.10}_{-0.10}$ & 0.05$^{+0.03}_{-0.04}$ & 0.5$^{+0.2}_{-0.5}$ & 0.67$^{+0.08}_{-0.21}$ & 4.4$^{+3.8}_{-4.4}$ & 0.25$^{+0.07}_{-0.08}$ & 2.12\\
 ... &  CB08 & 2.33$^{+0.11}_{-0.04}$ & 0.30$^{+0.20}_{-0.00}$ & 0.02$^{+0.08}_{-0.01}$ & 0.1$^{+0.1}_{-0.1}$ & 0.50$^{+0.17}_{-0.04}$ & 0.0$^{+5.9}_{-0.0}$ & 0.28$^{+0.12}_{-0.00}$ & 2.31\\
\hline
ecdfs-12514 &  BC03 (no irac) & 2.024 & 0.60$^{+0.40}_{-0.30}$ & 2.75$^{+17.2}_{-2.60}$ & 1.2$^{+0.0}_{-0.2}$ & 1.43$^{+0.34}_{-0.33}$ & 266.3$^{+77.4}_{-113.5}$ & 0.31$^{+0.19}_{-0.14}$ & 2.10\\
 ... &  BC03 & 2.024 & 0.04$^{+0.00}_{-0.01}$ & 0.01$^{+0.00}_{-0.00}$ & 1.5$^{+0.1}_{-0.0}$ & 0.68$^{+0.00}_{-0.16}$ & 145.2$^{+160.2}_{-0.0}$ & 0.03$^{+0.00}_{-0.00}$ & 3.36\\
 ... &  MAR05 (no irac) & 2.024 & 0.30$^{+0.10}_{-0.22}$ & 1.50$^{+18.5}_{-1.47}$ & 1.3$^{+0.1}_{-0.2}$ & 1.08$^{+0.15}_{-0.42}$ & 388.7$^{+117.7}_{-191.8}$ & 0.15$^{+0.06}_{-0.10}$ & 2.12\\
 ... &  MAR05 & 2.024 & 0.02$^{+0.00}_{-0.00}$ & 0.01$^{+0.00}_{-0.00}$ & 1.7$^{+0.0}_{-0.0}$ & 0.39$^{+0.07}_{-0.00}$ & 660.3$^{+124.8}_{-0.0}$ & 0.01$^{+0.00}_{-0.00}$ & 3.15\\
 ... &  CB08 (no irac) & 2.024 & 0.60$^{+0.40}_{-0.30}$ & 3.50$^{+16.5}_{-3.30}$ & 1.2$^{+0.0}_{-0.2}$ & 1.41$^{+0.33}_{-0.31}$ & 268.1$^{+75.4}_{-115.7}$ & 0.30$^{+0.19}_{-0.13}$ & 2.09\\
 ... &  CB08 & 2.024 & 0.70$^{+0.80}_{-0.30}$ & 20.0$^{+0.00}_{-19.2}$ & 1.2$^{+0.0}_{-0.2}$ & 1.55$^{+0.60}_{-0.26}$ & 271.1$^{+66.6}_{-99.4}$ & 0.35$^{+0.41}_{-0.14}$ & 2.79\\
\hline
ecdfs-13532 &  BC03 (no irac) & 2.336 & 0.80$^{+1.45}_{-0.70}$ & 0.40$^{+19.6}_{-0.37}$ & 1.4$^{+0.8}_{-0.8}$ & 1.75$^{+1.02}_{-0.71}$ & 87.9$^{+465.1}_{-74.8}$ & 0.52$^{+0.61}_{-0.38}$ & 0.97\\
 ... &  BC03 & 2.336 & 0.80$^{+0.70}_{-0.50}$ & 0.30$^{+0.40}_{-0.22}$ & 1.2$^{+0.4}_{-0.5}$ & 1.64$^{+0.27}_{-0.39}$ & 52.6$^{+86.9}_{-38.1}$ & 0.55$^{+0.57}_{-0.34}$ & 0.88\\
 ... &  MAR05 (no irac) & 2.336 & 0.20$^{+0.70}_{-0.17}$ & 0.10$^{+19.9}_{-0.09}$ & 1.8$^{+0.7}_{-1.0}$ & 1.14$^{+0.52}_{-0.53}$ & 213.0$^{+1228.}_{-196.3}$ & 0.13$^{+0.27}_{-0.10}$ & 0.96\\
 ... &  MAR05 & 2.336 & 0.20$^{+0.20}_{-0.15}$ & 0.10$^{+0.20}_{-0.09}$ & 1.8$^{+0.2}_{-0.9}$ & 1.14$^{+0.20}_{-0.46}$ & 212.2$^{+131.8}_{-189.8}$ & 0.13$^{+0.17}_{-0.09}$ & 0.87\\
 ... &  CB08 (no irac) & 2.336 & 0.80$^{+1.20}_{-0.70}$ & 0.40$^{+19.6}_{-0.37}$ & 1.4$^{+0.8}_{-0.7}$ & 1.75$^{+0.82}_{-0.71}$ & 87.6$^{+464.9}_{-73.2}$ & 0.52$^{+0.49}_{-0.45}$ & 0.98\\
 ... &  CB08 & 2.336 & 0.09$^{+0.91}_{-0.02}$ & 0.01$^{+0.24}_{-0.00}$ & 1.7$^{+0.0}_{-1.3}$ & 0.96$^{+0.33}_{-0.16}$ & 1.4$^{+80.4}_{-0.8}$ & 0.08$^{+0.72}_{-0.01}$ & 1.02\\
\hline
ecdfs-16671 &  BC03 (no irac) & 2.61$^{+0.02}_{-0.01}$ & 0.03$^{+0.02}_{-0.00}$ & 0.01$^{+0.02}_{-0.00}$ & 1.1$^{+0.0}_{-0.0}$ & 0.68$^{+0.09}_{-0.01}$ & 396.0$^{+282.5}_{-6.4}$ & 0.02$^{+0.00}_{-0.00}$ & 4.71\\
 ... &  BC03 & 2.61$^{+0.00}_{-0.01}$ & 0.03$^{+0.00}_{-0.00}$ & 0.01$^{+0.00}_{-0.00}$ & 1.1$^{+0.0}_{-0.0}$ & 0.68$^{+-0.0}_{-0.00}$ & 396.2$^{+0.0}_{-1.0}$ & 0.02$^{+0.00}_{-0.00}$ & 4.46\\
 ... &  MAR05 (no irac) & 2.63$^{+0.04}_{-0.02}$ & 0.02$^{+0.01}_{-0.00}$ & 0.01$^{+0.29}_{-0.00}$ & 1.2$^{+0.0}_{-0.0}$ & 0.51$^{+0.05}_{-0.00}$ & 865.5$^{+1056.}_{-10.2}$ & 0.01$^{+0.00}_{-0.00}$ & 4.90\\
 ... &  MAR05 & 2.63$^{+0.04}_{-0.02}$ & 0.02$^{+0.00}_{-0.00}$ & 0.01$^{+0.01}_{-0.00}$ & 1.2$^{+0.0}_{-0.0}$ & 0.51$^{+-0.0}_{-0.01}$ & 865.4$^{+682.9}_{-24.4}$ & 0.01$^{+0.00}_{-0.00}$ & 4.64\\
 ... &  CB08 (no irac) & 2.61$^{+0.04}_{-0.01}$ & 0.03$^{+0.02}_{-0.00}$ & 0.01$^{+0.02}_{-0.00}$ & 1.1$^{+0.0}_{-0.0}$ & 0.68$^{+0.06}_{-0.03}$ & 396.2$^{+256.1}_{-22.8}$ & 0.02$^{+0.00}_{-0.00}$ & 4.70\\
 ... &  CB08 & 2.61$^{+0.00}_{-0.01}$ & 0.03$^{+0.00}_{-0.00}$ & 0.01$^{+0.00}_{-0.00}$ & 1.1$^{+0.0}_{-0.0}$ & 0.68$^{+0.00}_{-0.00}$ & 396.3$^{+0.5}_{-0.0}$ & 0.02$^{+0.00}_{-0.00}$ & 4.45\\
\hline
cdfs-6202 &  BC03 (no irac) & 2.225 & 0.80$^{+0.20}_{-0.40}$ & 20.0$^{+0.00}_{-19.6}$ & 1.3$^{+0.1}_{-0.2}$ & 2.32$^{+0.38}_{-0.55}$ & 356.8$^{+94.5}_{-164.1}$ & 0.40$^{+0.12}_{-0.15}$ & 1.30\\
 ... &  BC03 & 2.225 & 0.60$^{+0.40}_{-0.30}$ & 0.65$^{+19.3}_{-0.50}$ & 1.2$^{+0.0}_{-0.1}$ & 2.04$^{+0.32}_{-0.42}$ & 259.8$^{+96.8}_{-68.0}$ & 0.34$^{+0.15}_{-0.15}$ & 1.87\\
 ... &  MAR05 (no irac) & 2.225 & 0.40$^{+0.20}_{-0.30}$ & 20.0$^{+0.00}_{-19.9}$ & 1.4$^{+0.2}_{-0.2}$ & 1.74$^{+0.37}_{-0.67}$ & 515.9$^{+266.7}_{-204.9}$ & 0.20$^{+0.10}_{-0.14}$ & 1.35\\
 ... &  MAR05 & 2.225 & 0.20$^{+0.10}_{-0.14}$ & 0.65$^{+3.35}_{-0.62}$ & 1.5$^{+0.0}_{-0.2}$ & 1.30$^{+0.30}_{-0.44}$ & 649.3$^{+27.1}_{-280.2}$ & 0.10$^{+0.06}_{-0.06}$ & 1.97\\
 ... &  CB08 (no irac) & 2.225 & 0.70$^{+0.30}_{-0.30}$ & 5.00$^{+15.0}_{-4.60}$ & 1.3$^{+0.1}_{-0.2}$ & 2.15$^{+0.51}_{-0.38}$ & 357.8$^{+93.7}_{-165.3}$ & 0.35$^{+0.16}_{-0.12}$ & 1.31\\
 ... &  CB08 & 2.225 & 0.10$^{+0.00}_{-0.01}$ & 0.03$^{+0.01}_{-0.00}$ & 1.3$^{+0.1}_{-0.0}$ & 1.14$^{+0.00}_{-0.04}$ & 168.0$^{+121.4}_{-0.0}$ & 0.07$^{+0.00}_{-0.00}$ & 2.39\\
\hline
\enddata
\end{deluxetable}
\begin{deluxetable}{lcc}
\tabletypesize{\footnotesize}
\scriptsize
\tablecaption{Comparison of Photometric Redshifts}
\tablewidth{3.3in}
\tablehead{\colhead{Data Set} & \colhead{Offset} & \colhead{ Scatter } 
  \\ 
\colhead{} & \colhead{($z_{\mbox{\scriptsize
      phot}}$-$z_{\mbox{\scriptsize spec}}$)/(1 + $z$)} &
\colhead{$\sigma$/(1 + $z$)} 
}
\startdata
All U$\rightarrow$K & 0.17 & 0.12 \\
All U$\rightarrow$4.5$\micron$ & 0.18 & 0.12 \\
All U$\rightarrow$8.0$\micron$ & 0.14 & 0.11 \\
Deep U$\rightarrow$K & 0.00 & 0.05 \\
Deep U$\rightarrow$4.5$\micron$ & 0.01 & 0.05 \\
Deep U$\rightarrow$8.0$\micron$ & 0.02 & 0.05
 \\
Wide U$\rightarrow$K & 0.46 & 0.19 \\
Wide U$\rightarrow$4.5$\micron$ & 0.46 & 0.18 \\
Wide U$\rightarrow$8.0$\micron$ & 0.35 & 0.17

\enddata
\end{deluxetable}
\begin{deluxetable}{lccccc}
\tabletypesize{\footnotesize}
\scriptsize
\tablecaption{Systematic Offsets in SED Parameters Compared to U$\rightarrow$8$\micron$+NIRspec Fits}
\tablewidth{6.4in}
\tablehead{\colhead{Data} & \colhead{M$_{star}$} & \colhead{$\tau$} & \colhead{$\langle t\rangle_{\mbox{\scriptsize SFR}}$} & \colhead{A$_{v}$} & \colhead{SFR} \\
\colhead{} & \colhead{Log(M$_{\odot}$)} & \colhead{Log(Gyr)} & \colhead{Log(Gyr)} & \colhead{mag} & \colhead{Log(M$_{\odot}$ yr$^{-1}$)}
}
\startdata
U$\rightarrow$K & 0.111 $\pm$ 0.045 & 0.091 $\pm$ 0.138 & -0.009 $\pm$ 0.078 & 0.070 $\pm$ 0.137 & -0.003 $\pm$ 0.187 \\  
U$\rightarrow$z$^{\prime}$+NIRspec & 0.022 $\pm$ 0.020 & 0.039 $\pm$ 0.114 & 0.002 $\pm$ 0.054 & 0.032 $\pm$ 0.094 & -0.024 $\pm$ 0.118 \\  
U$\rightarrow$8$\micron$ & -0.012 $\pm$ 0.019 & -0.055 $\pm$ 0.154 & 0.015 $\pm$ 0.062 & 0.026 $\pm$ 0.055 & -0.080 $\pm$ 0.076
\enddata
\tablecomments{Quoted values are the mean of Log(Data/U$\rightarrow$8$\micron$+NIRspec), except for A$_{v}$ which is mean (Data - U$\rightarrow$8$\micron$+NIRspec).  Errors are standard errors of the mean.}
\end{deluxetable}
\begin{deluxetable}{lcccccc}
\tabletypesize{\footnotesize}
\scriptsize
\tablecolumns{7}
\tablecaption{Effect of Stellar Synthesis Code on SED Parameters}
\tablewidth{5.6in}
\tablehead{\colhead{Parameter} & \multicolumn{3}{c}{MAR05/BC03} & \multicolumn{3}{c}{CB08/BC03} \\
\colhead{} & \multicolumn{3}{c}{----------------------} & \multicolumn{3}{c}{---------------------} \\
\colhead{} & \colhead{EL} & \colhead{NEL} & \colhead{All} & \colhead{EL} &
\colhead{NEL} & \colhead{All}
}
\startdata
$\tau$ & 0.66 (0.84) & 0.83 (0.84) & 0.75 (0.84) & 0.66 (111.) & 1.25 (1.48) & 1.00 (59.8) \\
$\langle t\rangle_{\mbox{\scriptsize SFR}}$ & 0.42 (0.64) & 1.03 (1.00) & 0.65 (0.81) & 0.96 (1.39) & 1.46 (1.53) & 1.24 (1.45) \\
A$_{v}$ & 0.10 (-0.0) & -0.4 (-0.3) & 0.00 (-0.1) & -0.2 (-0.1) & -0.6 (-0.5) & -0.4 (-0.3) \\
M$_{star}$ & 0.63 (0.65) & 0.69 (0.69) & 0.63 (0.67) & 0.70 (0.76) & 0.77 (0.77) & 0.74 (0.76) \\
SFR & 1.58 (1.79) & 0.33 (0.69) & 1.12 (1.27) & 0.64 (1.04) & 0.31 (1.64) & 0.32 (1.33) 
\enddata
\tablecomments{Quoted values are the median ratio, with the mean ratio
  in brackets}
\end{deluxetable}
\begin{deluxetable}{lccccccccc}
\tabletypesize{\footnotesize}
\scriptsize
\tablecaption{Effect of Dust Law on SED Parameters}
\tablewidth{8.4in}
\tablehead{\colhead{Parameter} & \multicolumn{3}{c}{BC03+SMC/BC03+Calz} & \multicolumn{3}{c}{BC03+LMC/BC03+Calz} & \multicolumn{3}{c}{BC03+MW/BC03+Calz}\\
\colhead{} & \multicolumn{3}{c}{----------------------} & \multicolumn{3}{c}{---------------------} & \multicolumn{3}{c}{---------------------}\\
\colhead{} & \colhead{EL} & \colhead{NEL} & \colhead{All} & \colhead{EL} &\colhead{NEL} & \colhead{All} & \colhead{EL} & \colhead{NEL} & \colhead{All}
}
\startdata
$\tau$ & 2.00 (16.7) & 3.33 (131.) & 2.50 (70.7) & 1.00 (2.73) & 1.25 (1.33) & 1.00 (2.07) & 1.00 (1.60) & 1.00 (1.14) & 1.00 (1.38) \\
$\langle t\rangle_{\mbox{\scriptsize SFR}}$ & 2.81 (6.62) & 2.35 (5.56) & 2.80 (6.12) & 1.11 (1.07) & 1.00 (1.09) & 1.09 (1.08) & 1.00 (0.93) & 1.00 (1.06) & 1.00 (0.99) \\
A$_{v}$ & -0.6 (-0.6) & -0.1 (-0.3) & -0.4 (-0.5) & -0.2 (-0.1) & -0.1 (-0.1) & -0.2 (-0.1) & -0.1 (-0.1) & -0.1 (-0.1) & -0.1 (-0.1) \\
M$_{star}$ & 1.20 (1.37) & 1.07 (1.34) & 1.15 (1.36) & 0.91 (0.80) & 0.94 (0.91) & 0.92 (0.85) & 0.83 (0.73) & 0.93 (0.91) & 0.90 (0.82) \\
SFR & 0.55 (3.82) & 2.36 (58.5) & 1.00 (29.5) & 0.77 (210.) & 1.28 (2.75) & 1.00 (112.) & 0.73 (208.) & 0.93 (2.35) & 0.83 (111.)
\enddata
\tablecomments{Quoted values are the median ratio, with the mean ratio
  in brackets}
\end{deluxetable}
\begin{deluxetable}{lcccccc}
\tabletypesize{\footnotesize}
\scriptsize
\tablecaption{Effect of Metallicity on SED Parameters}
\tablewidth{5.7in}
\tablehead{\colhead{Parameter} & \multicolumn{3}{c}{BC03\_m42/BC03\_m62} & \multicolumn{3}{c}{BC03\_m72/BC03\_m62} \\
\colhead{} & \multicolumn{3}{c}{---------------------} & \multicolumn{3}{c}{---------------------}\\
\colhead{} & \colhead{EL} &\colhead{NEL} & \colhead{All} & \colhead{EL} & \colhead{NEL} & \colhead{All}
}
\startdata
$\tau$ & 0.92 (0.84) & 0.39 (0.97) & 0.50 (0.90) & 0.53 (28.3) & 1.25 (126.) & 0.79 (74.3) \\
$\langle t\rangle_{\mbox{\scriptsize SFR}}$ & 0.47 (0.68) & 1.17 (1.36) & 0.94 (1.00) & 0.86 (1.75) & 1.17 (1.61) & 1.03 (1.68) \\
A$_{v}$ & 0.39 (0.43) & 0.19 (0.06) & 0.20 (0.26) & -0.3 (-0.3) & -0.4 (-0.4) & -0.3 (-0.3) \\
M$_{star}$ & 0.74 (0.71) & 1.03 (1.01) & 0.91 (0.85) & 0.83 (0.95) & 0.81 (0.91) & 0.82 (0.93) \\
SFR & 1.04 (180.) & 0.39 (1.43) & 0.65 (96.4) & 0.59 (1.60) & 0.67 (1.53) & 0.59 (1.57)
\enddata
\tablecomments{Quoted values are the median ratio, with the mean ratio
  in brackets}
\end{deluxetable}

\end{document}